\pdfoutput=1
\documentclass[cernpreprint, PAPER, atlasdraft=false, texlive=2023, UKenglish, texmf, orcidlogo]{atlasdoc}
 
\usepackage[block=none]{atlaspackage}
\usepackage{atlasbiblatex}
 
\usepackage{atlasphysics}
 
\usepackage{atlascover}
\usepackage{multirow}
\usepackage{float}
\usepackage{atlasparticle}
 
\graphicspath{{logos/}{figures/}}
 
\usepackage{ANA-TOPQ-2019-17-PAPER-defs}

\AtlasTitle{Search for flavour-changing neutral current interactions of the top quark and the Higgs boson in events with a pair of $\tau$-leptons in $pp$ collisions at $\sqrt{s}=13$~TeV with the ATLAS detector}
\AtlasAbstract{
A search for flavour-changing neutral current (FCNC) $tqH$ interactions involving a top quark, another up-type quark ($q=u,\, c$), and a
Standard Model (SM) Higgs boson decaying into a $\tau$-lepton pair ($H\rightarrow \tau^+\tau^-$) is presented.
The search is based on a dataset of $pp$ collisions at $\sqrt{s}=13~\tev$ that corresponds to an integrated luminosity of 139 fb$^{-1}$ recorded with the ATLAS detector at the
Large Hadron Collider.
Two processes are considered:  single top quark FCNC production in association with a Higgs boson ($pp\rightarrow tH$), and top quark pair production in
which one of top quarks decays into $Wb$ and the other decays into $qH$ through the FCNC interactions.
The search selects events with two hadronically decaying $\tau$-lepton candidates ($\ensuremath{\tau_{\text{had}}}$) or at least one $\ensuremath{\tau_{\text{had}}}$
with an additional lepton ($e$, $\mu$),
as well as multiple jets. Event kinematics is used to separate signal from the background through a multivariate discriminant.
A slight excess of data is observed with a significance of 2.3$\sigma$ above the expected SM background,
and 95\% CL upper limits on the $t\to qH$ branching ratios are derived.
The observed (expected) 95\% CL upper limits set on the $t\to cH$ and $t\to uH$ branching ratios are $9.4 \times 10^{-4}\, (4.8^{+2.2}_{-1.4}\times 10^{-4})$
and $6.9\times 10^{-4}\, (3.5^{+1.5}_{-1.0}\times 10^{-4})$, respectively.
The corresponding combined observed (expected) upper limits on the dimension-6 operator Wilson coefficients in
the effective $tqH$ couplings are $C_{c\phi} <1.35\, (0.97)$ and $C_{u\phi} <1.16\, (0.82)$.}
\author{The ATLAS Collaboration}
 
\AtlasRefCode{TOPQ-2019-17}

\AtlasJournalRef{JHEP 06 (2023) 155}
\AtlasDOI{10.1007/JHEP06(2023)155}
 
\PreprintIdNumber{CERN-EP-2022-123}

\AtlasJournal{JHEP}

\hypersetup{pdftitle={ATLAS document},pdfauthor={The ATLAS Collaboration}}

\begin{document}
 
\maketitle
 
\tableofcontents

\section{Introduction}
\label{sec:intro}
Since the discovery of the Higgs boson in 2012 at the Large Hadron Collider (LHC) by the ATLAS~\cite{Aad:2012tfa} and
CMS~\cite{Chatrchyan:2012ufa} collaborations,
a comprehensive programme of measurements 
has been conducted to explore this particle. Measurements so far have proved to be consistent with the Standard Model (SM) predictions.
The programme is ongoing and precision measurements as well as searches for rare new-physics processes beyond the Standard Model (BSM)
are underway. One such possibility is flavour-changing neutral current (FCNC) interactions between the Higgs boson,
the top quark, and an up-type quark, $tqH$ ($q=u, c$), which have been searched for by the ATLAS and CMS collaborations.
Since the Higgs boson is lighter than the top quark~\cite{Aad:2015zhl},
such interactions could manifest themselves as FCNC top-quark decays~($t\to qH$)~\cite{Agashe:2013hma}.
 
In the Standard Model (SM), the FCNC interaction is forbidden at tree level and suppressed at higher orders through the Glashow--Iliopoulos--Maiani (GIM) mechanism~\cite{Glashow:1970gm}. An observation of an enhanced rate of this decay would be clear evidence of new physics.
Furthermore, if the $tqH$ interaction exists, the associated single-top and  Higgs  production process through this interaction would enhance
the total production cross section of $pp\rightarrow tH$.
The $t\to qH$ branching fraction in the SM is calculated to be exceedingly small, $\BR(t\to qH)\approx10^{-15}$~\cite{Eilam:1990zc,Mele:1998ag,AguilarSaavedra:2004wm,Zhang:2013xya}.
However, these branching ratios can be large enough to be observed at LHC
when processes beyond the SM are included. Examples of these processes include: quark-singlet models~\cite{AguilarSaavedra:2002kr}, two-Higgs-doublet models (2HDMs)~\cite{ Branco:2hdm2012} with or without flavour violation,
the minimal supersymmetric SM (MSSM)~\cite{Bejar:2000ub, Guasch:1999jp,Cao:2007dk,Cao:2014udj},
supersymmetric models with R-parity violation~\cite{Eilam:2001dh}, composite Higgs models with partial  compositeness~\cite{Azatov:2014lha},
and warped extra dimensions models with SM fermions in the bulk~\cite{Azatov:2009na}.
An even larger branching ratio of  $\BR(t\to cH) \sim 10^{-3}$ can be reached in 2HDMs without explicit flavour conservation, since in these models the tree-level FCNC coupling is no longer forbidden by any symmetry~\cite{Cheng:1987rs,Baum:2008qm,Chen:2013qta,Chiang:2015cba,Crivellin:2015hha,Botella:2015hoa, Gori:2017tvg,Chiang:2017fjr}.
The study of $pp\rightarrow tH$ processes will also contribute to the FCNC interaction searches~\cite{Greljo:2014dka}.
In the SM, associated production of $tH$ in $pp$ collisions is expected to have a cross section of $\sigma_{tH}=92^{+7}_{-12}$ fb at a centre-of-mass energy of $\sqrt{s}= 13~\tev$~\cite{deFlorian:2016spz}.
 
Searches for $t \to qH$ decays have been performed by the ATLAS and CMS collaborations, taking advantage of the large samples
of top-quark pair ($\ttbar$) events collected in proton--proton ($pp$) collisions at centre-of-mass energies of $\sqrt{s}=7~\tev$ and $8~\tev$~\cite{Aad:2014dya,Aad:2015pja,Khachatryan:2016atv} during Run~1 of the LHC, as well as at $\sqrt{s}=13~\tev$~\cite{TOPQ-2017-07} using early Run~2 data.
In these searches, one of the top quarks is required to decay into $Wb$, while the other top quark decays into $qH$ with a small branching ratio
$\BR(t\to qH)$, a process denoted by $\ttbar \to WbHq$.\footnote{In the following, $WbHq$ is used to denote both $W^+b H\bar{q}$ and its charge conjugate, $HqW^- \bar{b}$. Similarly,
$WbWb$ is used to denote $W^+b W^- \bar{b}$.}  The Higgs boson is assumed to have a mass of $m_H=125~\gev$ and to decay as predicted by
the SM.
Compared to Run~1, the Run~2 searches, summarised in Table~\ref{tab:limits_summary_ref}, benefit from the increased $\ttbar$ cross section at $\sqrt{s}=13~\tev$, as well as the larger integrated luminosity.
Using 36.1~fb$^{-1}$ of data at $\sqrt{s}=13~\tev$, the ATLAS Collaboration has derived upper limits at 95\% confidence level (CL) on the
$t\to cH$ branching ratio: $\BR(t\to cH)<0.22\%$ using $H\to \gamma\gamma$ decays~\cite{Aaboud:2017mfd} and $\BR(t\to cH)<0.16\%$ based on
multi-lepton (electron or muon) signatures resulting from
$H \to  WW^*, ZZ^*$, $\tau^+\tau^-$ in which both $\tau$-leptons decay leptonically~\cite{Aaboud:2018pob}.
ATLAS also set upper limits of $\BR(t\to cH)<0.42\%$ using $H\to b\bar{b}$ decay~\cite{TOPQ-2017-07} and $\BR(t\to cH)<0.19\%$ using $H\to \tau^+\tau^-$ decays in which at least
one of the $\tau$-leptons decays hadronically~\cite{TOPQ-2017-07}.
These upper limits are derived assuming that the branching ratio $\BR(t\to uH)=0$. Similar upper limits are obtained for $\BR(t\to uH)$ when assuming $\BR(t\to cH)=0$.
Combining all of the ATLAS searches using 36.1~fb$^{-1}$ of Run~2 data, upper limits at 95\% CL on the branching fractions are
set at $\BR(t\to cH)<0.11\%$ assuming $\BR(t\to uH)=0$, and at $\BR(t\to uH)<0.12\%$ assuming $\BR(t\to cH)=0$~\cite{TOPQ-2017-07}.
 
The CMS Collaboration performed a similar search using
$H\to b\bar{b}$ decays~\cite{Sirunyan:2017uae} with 35.9 fb$^{-1}$ of data at $\sqrt{s}=13~\tev$, resulting
in upper limits of $\BR(t\to cH)<0.47\%$ and $\BR(t\to uH)<0.47\%$, in each case neglecting the other decay mode.
The search in Ref.~\cite{Sirunyan:2017uae} also considers the contribution to the signal from
$pp \to tH$ production~\cite{Greljo:2014dka}.
CMS subsequently updated their search for $tqH$ in the $H\to b\bar{b}$ channel using
the full Run~2 dataset, obtaining observed (expected) upper limits of  $\BR(t\to cH) < 9.4\times10^{-4}\, (8.6\times10^{-4})$
and $\BR(t\to uH) < 7.9\times10^{-4}\, (1.1\times10^{-3})$~\cite{CMS:2021gfa}.
CMS also recently reported a search for $tqH$ in the $H\rightarrow \gamma\gamma$ mode using the full Run~2 data and set observed (expected) upper limits on $\mathcal{B}(t\to Hu)$
of $1.9\times10^{-4}\, (3.1\times10^{-4})$ and $\mathcal{B}(t\to Hc)$ of $7.3\times10^{-4}\, (5.1\times10^{-4})$~\cite{CMS-PAS-TOP-20-007}.

\begin{table}[t!]
\caption{\small{Summary of 95\% CL upper limits on $\BR(t \to cH)$ and $\BR(t \to uH)$ obtained by the ATLAS and CMS collaborations with Run~2 data. Each limit is obtained assuming the other branching ratio is zero.}}
\begin{center}
\small
\begin{tabular}{ccccc}
\toprule\toprule
& &\multirow{2}{*}{$\mathcal{L}$ [fb$^{-1}$]} & \multicolumn{2}{c}{95\% CL observed upper limits}  \\
& & 										    & \multicolumn{1}{c}{on $\BR(t \to cH)$}            & \multicolumn{1}{c}{on $\BR(t \to uH)$} \\
\midrule
\multirow{5}{*}{ATLAS}
& $H \to b\bar{b}$~\cite{TOPQ-2017-07}                                    & 36.1         & $4.2 \times 10^{-3}$ & $5.2 \times 10^{-3}$ \\
& $H \to \gamma\gamma$~\cite{Aaboud:2017mfd}                              & 36.1         & $2.2 \times 10^{-3}$  & $2.4 \times 10^{-3}$  \\
& $H \to \tau\tau$ ($\lephad$, $\hadhad$)~\cite{TOPQ-2017-07}             & 36.1         & $1.9 \times 10^{-3}$  & $1.7 \times 10^{-3}$  \\
& $H \to WW^*, \tau\tau, ZZ^*$ ($2\ell$SS, $3\ell$)~\cite{Aaboud:2018pob} & 36.1         & $1.6 \times 10^{-3}$  & $1.9 \times 10^{-3}$\\
& Combination~\cite{TOPQ-2017-07}                                         & 36.1         & $1.1 \times 10^{-3}$  & $1.2 \times 10^{-3}$  \\\midrule
\multirow{2}{*}{CMS}
& $H \to b\bar{b}$~\cite{Sirunyan:2017uae}                                & 35.9         & $4.7 \times 10^{-3}$  & $4.7 \times 10^{-3}$  \\
& $H \to b\bar{b}$~\cite{CMS:2021gfa}                                     & 137          &$9.4 \times 10^{-4}$  & $7.9 \times 10^{-4}$  \\
& $H \to \gamma\gamma$~\cite{CMS-PAS-TOP-20-007}                          & 137          & $7.3 \times 10^{-4}$  & $1.9 \times 10^{-4}$  \\
\bottomrule\bottomrule
\end{tabular}
\label{tab:limits_summary_ref}
\end{center}
\end{table}

The analysis reported here targets Higgs boson decays into
$\tau$-leptons in the complete Run~2 dataset collected by ATLAS in 2015--2018. The corresponding integrated luminosity is 139 fb$^{-1}$. Both $\ttbar \to WbHq$ decays and $pp \to tH$ production are sought. The dataset is divided into several final states depending on the production mode and
the  $W$ boson and $\tau$-lepton decays. Top quark pair production events where the $W$ boson decays hadronically (leptonically) are denoted by $t_h$ ($t_{\ell}$).
The decays $\tau \to \ell \nu_{\ell} \nu_{\tau}$ are denoted by $\lep$, while the decays $\tau \to \text{hadrons} + \nu_{\tau}$ are denoted by $\thad$.
The contribution of
$W\rightarrow\tau\nu$ is included as $\lep$ in $t_{\ell}$ when the $\tau$-lepton decays into a light lepton (electron or muon) or as $\thad$
in $t_h$ when the $\tau$-lepton decays hadronically.
The $H\rightarrow \tau\tau$ decay is detected in
either the $\tlhad$ or $\thadhad$ final states when the top quark decays hadronically ($t_h$), but only the $\thadhad$ final state is considered
when the top quark decays leptonically ($t_{\ell}$), in order to avoid overlaps with
other ATLAS searches~\cite{Aaboud:2018pob}.
In this search, events with two hadronically decaying $\tau$-leptons and no electron or muon define the hadronic channel. Events with at least one \tauhad and an additional electron or muon (corresponding to $t_{\ell}$ or $\lep$ events) are assigned to leptonic channels. More signal regions (with the top quark decaying leptonically) are exploited here than in the previous FCNC $\Htautau$ search, which was conducted using the partial Run~2 dataset~\cite{TOPQ-2017-07}.
In addition, an improved treatment of misidentified $\tau$-leptons (`fakes') in simulation and in data-driven
estimations of fakes from multi-jet background is implemented.
Finally, a multivariate technique based on boosted decision trees is used to discriminate between the signal and the background on the basis of their different kinematical distributions.


\section{ATLAS detector}
 
The ATLAS detector~\cite{PERF-2007-01} at the LHC covers almost the entire solid angle around the collision point,\footnote{ATLAS
uses a right-handed coordinate system with its origin at the nominal interaction point (IP) in the
centre of the detector.
The $x$-axis points from
the IP to the centre of the LHC ring, 
the $y$-axis points upward,
and the $z$-axis coincides with the axis of the beam pipe.
Cylindrical coordinates ($r$,$\phi$) are used
in the transverse plane, $\phi$ being the azimuthal angle around the beam pipe. The pseudorapidity is defined in
terms of the polar angle $\theta$ as $\eta = - \ln \tan(\theta/2)$.
Angular separation is measured in units of $\Delta R\equiv \sqrt{(\Delta\eta)^2+(\Delta\phi)^2}$.} and it consists of an inner tracking detector surrounded by a thin superconducting solenoid producing a
2~T axial magnetic field, electromagnetic and hadronic calorimeters, and a muon spectrometer incorporating three large toroid magnet assemblies with eight coils each. The inner detector contains a high-granularity silicon pixel detector, including the 
insertable B-layer~\cite{IBL1,IBL2,Abbott:2018ikt} added as a new innermost layer in 2014, and a silicon microstrip tracker, together providing precise reconstruction of tracks of charged particles in the pseudorapidity range $|\eta|<2.5$.
The inner detector also includes a transition radiation tracker that provides tracking and electron identification for $|\eta|<2.0$.
The calorimeter system covers the pseudorapidity range $|\eta| < 4.9$. Within the region $|\eta|< 3.2$, electromagnetic (EM) calorimetry is provided by barrel and endcap high-granularity lead/liquid-argon (LAr) sampling calorimeters, with an additional thin LAr presampler covering $|\eta| < 1.8$ to correct for energy loss in material upstream of the calorimeters. Hadronic calorimetry is provided by 
a steel/scintillator-tile calorimeter, segmented into three barrel structures within $|\eta| < 1.7$, and two copper/LAr hadronic endcap calorimeters.
The solid angle coverage is completed with forward copper/LAr and tungsten/LAr calorimeter modules optimised for electromagnetic and hadronic measurements, respectively.
The calorimeters are surrounded by a muon spectrometer within a magnetic field provided by air-core toroid magnets with a bending integral of about \SI{2.5}{\tesla\metre} in the barrel and up to \SI{6.0}{\tesla\metre} in the endcaps.
The muon spectrometer measures the trajectories of muons with $|\eta|<2.7$ using multiple layers of high-precision tracking chambers, and it is instrumented with separate trigger chambers covering $|\eta|<2.4$. A two-level trigger system~\cite{Aaboud:2016leb}, consisting of a hardware-based level-1 trigger followed by a software-based high-level trigger, is used to reduce the event rate to a maximum of around \SI{1}{\kHz} for offline storage.
An extensive software suite~\cite{ATL-SOFT-PUB-2021-001} is used in the reconstruction and analysis of real and simulated data, in detector operations, and in the trigger and data acquisition systems of the experiment.


\section{Event reconstruction}
\label{sec:objects}
 
Events are selected from $pp$ collisions at $\sqrt{s}=13$~\TeV\ recorded by the ATLAS detector during 2015--2018~\cite{DAPR-2018-01}.
Only events for which all relevant subsystems were operational are considered.
Events are required to have at least one primary vertex with two or more tracks with transverse momentum
($\pt$) larger than $500~\mev$ that are consistent with originating from the
beam collision region in the $x$--$y$ plane. If more than one primary vertex candidate is found, the
candidate whose associated tracks form the largest sum of squared $\pt$~\cite{ATL-PHYS-PUB-2015-026}
is selected as the hard-scatter primary vertex.
 
Electron candidates~\cite{EGAM-2018-01} are reconstructed from energy
clusters in the EM calorimeter that are matched to reconstructed tracks in the inner detector;
electron candidates in the transition region between the EM barrel and endcap calorimeters
($1.37 < |\eta_{\textrm{cluster}}| < 1.52$) are excluded.
Electron candidates are required to have $|\eta_{\textrm{cluster}}| < 2.47$, and to satisfy `tight' likelihood-based identification
criteria~\cite{PERF-2017-01}
based on calorimeter, tracking and combined variables that provide
separation between electrons and jets.
 
Muon candidates~\cite{MUON-2018-03} are reconstructed by matching track segments in 
different layers of the muon spectrometer to tracks found in the inner detector;
the resulting muon candidates are re-fitted using the complete track information from both detector systems.
Muon candidates are required to have $|\eta|<2.5$, and to satisfy `medium' identification criteria~\cite{Aad:2016jkr}.
 
Electron (muon) candidates are matched to the primary vertex by requiring that the significance of their transverse impact parameter, $d_0$,
satisfies $|d_0/\sigma(d_0)|<5\,(3)$, where $\sigma(d_0)$ is the measured uncertainty in $d_0$,
and by requiring that their longitudinal impact parameter, $z_0$, satisfies $|z_0 \sin\theta|<0.5$~mm.
To further reduce the background from non-prompt leptons, which originate mostly from heavy-flavour hadron decays, photon conversions and misidentified hadrons,
lepton candidates are also required to be isolated in the tracker and in the calorimeter.
A track- and cluster-based lepton isolation criterion is defined by placing requirements on the quantities $I_R = \sum \pt^{\textrm{trk}}$ and $E_R = \sum E^{\textrm{clst}}$, where
the scalar sum runs over all tracks $\pt^{\textrm{trk}}$ or cluster energy deposits $E^{\textrm{clst}}$ (excluding the lepton candidate itself)
within the cone defined by $\Delta R<0.2~(0.3)$ around the 
direction of the electron (muon). The electron (muon) candidates are required to satisfy both
$I_R/\pt^\ell < 0.2~(0.3)$ and $E_R/\pt^\ell < 0.2~(0.3)$, where $\ell=e$ or $\mu$.
 
`Tight' isolation working points are also applied in some channels to reduce fake and non-prompt lepton contributions by using a trained isolation boosted decision tree (BDT)
\texttt{PromptLeptonVeto} (PLIV), which identifies non-prompt light leptons by using lifetime information associated with a track-jet that matches the selected light lepton.
These additional reconstructed charged-particle tracks inside the jet can be
used to increase the efficiency for identifying the displaced decay vertices of heavy-flavour ($b$, $c$) hadrons that produced non-prompt leptons.
The `tight' working points are used for leptons with high $\pt$ ($>20~\GeV$).
Simulation-to-data scale factors for the efficiencies of the `tight' PLIV working points are measured using the tag-and-probe method~\cite{PERF-2017-01}
with $Z\rightarrow \ell^+\ell^-$ events. These scale factors were checked for electrons or muons from the $\tau$-lepton decays
using $Z\rightarrow\tau\tau\rightarrow e\mu4\nu$ samples, and are consistent at the 2\% level. To be conservative, an
additional uncertainty of $\pm 2\%$ is considered for the PLIV efficiency for the $\tau$-lepton in the lepton+$\thad$ channels.
 
Candidate jets are reconstructed using the anti-$k_t$ algorithm~\cite{Cacciari:2008gp,Cacciari:2011ma}
with a radius parameter $R=0.4$ applied to topological energy clusters~\cite{Aad:2016upy} and charged-particle tracks, processed using
a particle-flow algorithm~\cite{PERF-2015-09}. 
The reconstructed jets are then calibrated to the particle level by the application of a jet energy scale
derived from simulation and in situ corrections based on $\sqrt{s}=13~\tev$ data~\cite{Aaboud:2017jcu}. 
After being calibrated, the jets are required to have $\pt > 25~\gev$ and $|\eta| < 2.5$.
The four-momentum of each jet is corrected for pile-up effects using the jet-area method~\cite{Cacciari:2008gn}.
 
Quality criteria are imposed to reject events that contain any jets arising from non-collision sources
or detector noise~\cite{ATLAS-CONF-2015-029}.  To reduce the contamination due to jets originating from pile-up interactions,
additional requirements are imposed on the jet vertex tagger (JVT)~\cite{Aad:2015ina} output for jets with $\pt<60~\gev$ and $|\eta| < 2.4$.
 
Jets containing $b$-hadrons are identified ($b$-tagged) via the DL1r tagger~\cite{FTAG-2018-01,ATL-PHYS-PUB-2017-013}, 
which uses multivariate techniques to combine information about the impact parameters of displaced tracks and the  topological properties
of secondary and tertiary decay vertices reconstructed within the jet. For each jet, a value for the multivariate $b$-tagging discriminant is
calculated. A jet is considered $b$-tagged if this value is above the threshold corresponding to
an average 70\% efficiency to tag a $b$-quark jet, with a light-jet\footnote{A `light jet' refers to a jet originating from the hadronisation of a light quark ($u$, $d$, $s$) or a gluon.} rejection factor of about 385 and a charm-jet rejection factor of about 12, as determined for jets with
$\pt >20~\gev$ and $|\eta|<2.5$ in simulated $\ttbar$ events~\cite{FTAG-2018-01}.
 
Hadronically decaying $\tau$-lepton ($\had$) candidates are reconstructed from calorimeter energy clusters and
associated inner-detector tracks~\cite{ATL-PHYS-PUB-2019-033}. 
Candidates are required to have either one or three associated tracks,
with a total charge of $\pm 1$, and to have $\pt>25$ GeV with $|\eta|<2.5$, excluding the EM calorimeter's transition region.
A recurrent neural network (RNN)~\cite{Graves:2012SCI}
using calorimeter- and tracking-based variables is used to identify $\had$ candidates and reject jet backgrounds.
Three working points, labelled `loose', `medium' and `tight', are defined and correspond to different $\had$ identification efficiency values, with the efficiency designed to be independent of $\pt$. The $\Htautau$ search uses the `medium'
working point for the $\had$ selection.
The `medium' working point has a combined reconstruction and identification efficiency of 75\% (60\%) for one-prong (three-prong) $\had$
decays, and an expected rejection factor of 35 (240) against light jets~\cite{ATL-PHYS-PUB-2019-033}.
Electrons that are reconstructed as one-prong $\had$ candidates are removed via a BDT trained to reject electrons.
Events are rejected if the jet seeding the $\had$ candidate is $b$-tagged.
 
To avoid double-counting of reconstructed objects, an overlap removal procedure is applied.
Electron candidates that lie
within $\Delta R = 0.01$ of a muon candidate are removed to suppress contributions from muon bremsstrahlung.
Energy clusters from identified electrons are not excluded during jet reconstruction.
In order to avoid double-counting of electrons as jets, the closest jet whose axis is within ${\Delta}R = 0.2$ of an electron
is discarded if the jet is not $b$-tagged; otherwise the electron is removed.
If the electron is within ${\Delta}R = 0.4$ of the axis of any jet after this initial removal, the jet is retained and  the electron is removed.
The overlap removal procedure between the remaining jet candidates and muon candidates is designed to remove those muons
that are likely to have arisen in the decay of hadrons and to retain the overlapping jet instead.
Jets and muons may also appear in close proximity when the jet results from high-$\pt$ muon bremsstrahlung,
and in such cases the jet is removed and the muon retained. Such jets are characterised by having very
few matching inner-detector tracks. Selected muons that satisfy $\Delta R(\mu,{\textrm{jet}}) < 0.2$ are rejected
if the jet is either $b$-tagged or has at least three tracks originating from the primary vertex; otherwise the jet is removed and the muon is kept.
The $\tauhad$ within a $\Delta R=0.2$ cone around an electron or muon are removed.
In order to avoid double-counting of $\tauhad$ as jets, the closest jet whose axis is
within ${\Delta}R = 0.2$ of a $\tauhad$ is discarded if the jet is not $b$-tagged; otherwise the $\tauhad$ is removed.

The missing transverse momentum $\mpt$ (with magnitude $\met$) is defined as the negative vector sum of the
$\pt$ of all selected and calibrated objects in the event, including a term to account for momentum from soft particles
in the event which are not associated with any of the selected objects.
This soft term is calculated from inner-detector tracks matched to the selected primary vertex to make it more resilient to
contamination from pile-up interactions~\cite{Aaboud:2018tkc}.


\section{Data sample and event preselection}
 
The search is based on a dataset of $pp$ collisions at $\sqrt{s}=13~\tev$ with 25 ns bunch spacing collected from 2015 to 2018, corresponding to an integrated luminosity of $139~\ifb$.
Only events recorded with a single-electron trigger, a single-muon trigger, or a di-$\tau$-lepton trigger~\cite{TRIG-2018-05,TRIG-2018-01,id_trigger,l1topo_trigger} under stable beam conditions
and for which all detector subsystems were operational are considered for analysis. The events recorded by dilepton triggers that fall into the control regions are used for fake-$\tau$-lepton background
estimation as discussed in Section~\ref{sec:background_model}. The number of $pp$ interactions per bunch crossing in this dataset ranges from about 8 to 45, with an average of 24.
 
Single-electron and single-muon triggers with low $\pt$ thresholds and lepton isolation requirements are combined in a logical OR
with higher-threshold triggers that have a looser identification criterion and no isolation requirement.
The lowest $\pt$ threshold used for muons is 20 (26)~\gev\ in 2015 (2016--2018), while for electrons the threshold is 24 (26)~\gev.
For di-$\tau$ triggers, the $\pt$ threshold for the leading (subleading) $\had$ candidate is 35 (25)~\gev.
To reduce the impact of the trigger efficiency uncertainty around the threshold, the leptons are required to have a \pt that is at least 1~\GeV\ above the threshold.
The reconstructed $\tau$-leptons are required to have a \pt at least 5~\GeV\ higher than the trigger threshold.
The events in the leptonic channels are recorded by a single-electron or single-muon trigger, and are required to have exactly one electron or muon that matches, within $\Delta R < 0.15$, the lepton reconstructed by at least one of the possible triggers. The following additional requirements are applied.
\begin{itemize}
\item $t_h\lephad$: To enhance selection of the $t_hH$ and $t_ht(qH)$ final states with a $H\to\lephad$ decay, exactly one $\had$ with opposite-sign charge to $\lep$ is required, plus at least three jets with exactly one $b$-jet.
\item $t_{\ell}\hadhad$: To enhance selection of $t_{\ell}H$ and $t_{\ell}t(qH)$ final states with a $H\to\hadhad$ decay, exactly one light lepton and two opposite-sign $\had$ are required,
plus jets with exactly one $b$-jet.
\item $t_{\ell}\had$: This channel enhances the selection of $t_{\ell}H$ and $t_{\ell}t(qH)$ final states with a $H\to\hadhad$ decay where one $\had$ fails the applied reconstruction or identification
criterion so that there is only one reconstructed $\had$ candidate.
To reduce the SM backgrounds and avoid overlaps with the final states used in other ATLAS searches, exactly one $\had$ with the same charge as that assigned to the light lepton is required.
In addition, at least two jets including exactly one $b$-jet are required.
\end{itemize}
 
The events in the hadronic channel are selected by a di-$\tau$ trigger. Further requirements are:
\begin{itemize}
\item $t_h\hadhad$: The $t_hH$ and $t_ht(qH)$ final states with $H\to\hadhad$ decay are targeted. Exactly two $\had$ with opposite-sign charge and at least three jets, including exactly one $b$-jet, are required.
\end{itemize}
 
The above requirements apply to the reconstructed objects defined in Section~\ref{sec:objects}.
These requirements are referred to as the preselection and are summarised in Table~\ref{tab:preselection}.
 
\begin{table*}[t!]
\caption{\small{Summary of the preselection requirements.
The leading and subleading $\had$ candidates are denoted by $\tau_{\mathrm{had1}}$ and $\tau_{\mathrm{had2}}$ respectively.}}
\begin{center}
\begin{tabular}{c|ccc|c}
\toprule\toprule
\multirow{2}{*}{Requirement} &  \multicolumn{3}{c|}{Leptonic channels}  & \multicolumn{1}{c}{Hadronic channel} \\
& $t_h\lephad$ & $t_{\ell}\hadhad$ &  $t_{\ell}\had$ & $t_h\hadhad$\\
\midrule
Trigger & \multicolumn{3}{c|}{single-lepton trigger} & di-$\tau$ trigger  \\
Leptons  & \multicolumn{3}{c|}{=1 isolated $e$ or $\mu$}  & =0 isolated $e$ or $\mu$ \\
$\had$  & $=$1 $\had$ & $=$2 $\had$ & $=$1 $\had$ & $=$2 $\had$ \\
Electric charge ($Q$) & $Q_\ell \times Q_{\tau_{\mathrm{had1}}} = -1$ & $Q_{\tau_{\mathrm{had1}}} \times Q_{\tau_{\mathrm{had2}}} = -1$ & $Q_\ell \times Q_{\tau_{\mathrm{had1}}} = 1$ & $Q_{\tau_{\mathrm{had1}}} \times Q_{\tau_{\mathrm{had2}}} = -1$ \\
Jets  &   $\geq$3 jets & $\geq$1 jets & $\geq$2 jets & $\geq$3 jets \\
$b$-tagging & \multicolumn{3}{c|}{=1 $b$-jets} & =1 $b$-jets\\
\bottomrule\bottomrule
\end{tabular}
\label{tab:preselection}
\end{center}
\end{table*}


\section{Simulated events}
 
An overview of the Monte Carlo (MC) generators used for the main signal and background samples is summarised in Table~\ref{mob}.
Samples of simulated $\ttbar \to WbHq$ ($tt(qH)$) events were generated with the next-to-leading-order (NLO) generator\footnote{In the following,
the order of a generator should be understood as referring to the order in the strong coupling constant at which the matrix-element (ME) calculation
is performed.}{\powheg}\,v2 \cite{Frixione:2007nw,Nason:2004rx,Frixione:2007vw,Alioli:2010xd}
with the \textsc{NNPDF3.0nlo}~\cite{Ball:2014uwa} parton distribution function (PDF) set and interfaced to {\pythia}\,8.212~\cite{Sjostrand:2007gs} with the \textsc{NNPDF2.3lo}~\cite{Ball:2012cx} PDF set for the modelling of the parton showers (PS), hadronisation, and underlying event.
A set of tuned parameters called the A14 tune~\cite{ATLASUETune4} was used in {\pythia} to control the modelling of multi-parton interactions and
initial- and final-state radiation.
The signal sample is normalised to the same total cross section as is used for the inclusive $t\bar{t}\to WbWb$ sample (see discussion below)
assuming a benchmark branching ratio of $\BR_{\mathrm{ref}}(t\to qH)=0.1\%$.
The case of both top quarks decaying into $qH$ is neglected in the analysis given the existing upper limits on $\BR(t \to qH)$ (Section~\ref{sec:intro}).
 
The $tH$ signal events were generated by {\amcatnlolong}\,2.6.2~\cite{Alwall:2014hca}  (referred to in the following as {\amcatnlo})
with the \textsc{NNPDF3.0nlo} parton distribution function (PDF) set. The parton showers, hadronisation, and underlying event were modelled by {\pythia}\,8.212 with the \textsc{NNPDF2.3lo} PDF set
in combination with the A14 tune.
Depending on whether the up quark or the charm quark is involved in the FCNC production process, the effective Lagrangian of the $tqH$ interaction is parameterised using
dimension-6 operators~\cite{fcnc_production_theory}. The cross sections $\sigma(ug\to tH) = 0.711$~pb and $\sigma(cg\to tH) = 0.103$~pb were obtained using $\BR_{\mathrm{ref}}(t\to qH)=0.1\%$ as the benchmark.
 
The sample used to model the $\ttbar$ background was generated with the NLO generator {\powheg}\,v2
using the \textsc{NNPDF3.0nlo} PDF set. The {\powheg} model parameter $h_{\textrm{damp}}$, which controls
matrix element to parton shower matching and regulates the high-$\pt$ radiation, was set to 1.5 times the top-quark mass.
The parton showers, hadronisation, and underlying event were modelled by {\pythia}\,8.210 with the \textsc{NNPDF2.3lo} PDF set in combination with the A14 tune.
Alternative $\ttbar$ simulation samples used to derive parton shower systematic uncertainties are described in Section~\ref{sec:syst_bkgmodeling}.
The generated $\ttbar$ samples are normalised to a theoretical cross section of $\sigma_{\ttbar}=832^{+46}_{-51}$~pb,
computed using \textsc{Top++}\,2.0~\cite{Czakon:2011xx} at next-to-next-to-leading order (NNLO),
including resummation of next-to-next-to-leading logarithmic (NNLL) soft gluon
terms~\cite{Cacciari:2011hy,Baernreuther:2012ws,Czakon:2012zr,Czakon:2012pz,Czakon:2013goa}.
 
Samples of single-top-quark events corresponding to the $t$-channel production mechanism were generated with the
{\powheg}\,v2 generator~\cite{Frederix:2012dh}, using the four-flavour scheme  for the NLO matrix-element calculations
and the corresponding \textsc{NNPDF3.0nlo} set of PDFs.
Samples corresponding to the $tW$- and $s$-channel production mechanisms were generated
with {\powheg}\,v2 using the five-flavour scheme. Overlaps between the $\ttbar$ and $tW$ final states were avoided by using
the diagram removal scheme~\cite{Frixione:2005vw}.
The parton showers, hadronisation and underlying event were modelled using {\pythia}\,8.230~\cite{Sjostrand:2006za} with the A14 tune and the \textsc{NNPDF2.3lo} PDF set.
The single-top-quark samples are normalised to the approximate NNLO theoretical cross
sections~\cite{Kidonakis:2011wy,Kidonakis:2010ux,Kidonakis:2010tc}.
 
Samples of $W/Z$+jets events were generated with the {\sherpa}\,2.2.1~\cite{Gleisberg:2008ta} generator.
The matrix element was calculated for up to two partons at NLO and up to four partons at LO using
\textsc{Comix}~\cite{Gleisberg:2008fv} and \textsc{OpenLoops}~\cite{Cascioli:2011va}. The matrix-element calculation
was merged with the {\sherpa} parton shower~\cite{Schumann:2007mg} using the MEPS@NLO prescription~\cite{Hoeche:2012yf}.
The PDF set used for the matrix-element calculation is \textsc{NNPDF3.0nnlo}~\cite{Ball:2014uwa} with a dedicated parton shower tune developed for {\sherpa}.
Separate samples were generated for different $W/Z$+jets categories using filters for a $b$-jet
($W/Z$+$\geq$1$b$+jets), a $c$-jet and no $b$-jet ($W/Z$+$\geq$1$c$+jets), and with a veto on $b$- and $c$-jets
($W/Z$+light-jets), and were combined into the inclusive $W/Z$+jets samples.
The $W$+jets and $Z$+jets samples are normalised to their respective inclusive NNLO theoretical
cross sections calculated with \textsc{FEWZ}~\cite{Anastasiou:2003ds}.
 
Samples of $WW/WZ/ZZ$+jets events were generated with {\sherpa}\,2.2.1 using the CT10 PDF set
and include processes containing up to four electroweak vertices.
In the case of $WW/WZ$+jets ($ZZ$+jets) the matrix element was calculated for zero (up to one) additional partons
at NLO and up to three partons at LO using the same procedure as for the $W/Z$+jets samples.
The final states that were simulated require one of the bosons to decay leptonically and the other hadronically.
All diboson samples are normalised to their NLO theoretical cross sections provided by {\sherpa}.
 
Samples of $\ttbar V$ ($V=W, Z$ boson) and $\ttbar H$ events were generated with {\amcatnlo}\,2.2.1, using NLO matrix elements and the \textsc{NNPDF3.0nlo} PDF set,
and interfaced to {\pythia}\,8.210 with the \textsc{NNPDF2.3lo} PDF set and the A14 tune.
The $\ttbar V$ samples are normalised to the NLO cross section computed with {\amcatnlo}, while the $\ttbar H$ sample is normalised using
the NLO cross section recommended in Ref.~\cite{deFlorian:2016spz}.
 
Samples of $WH$ and $ZH$, collectively referred to as $VH$, were generated using {\powheg}\,v2~\cite{Frixione:2007nw,Nason:2004rx,Frixione:2007vw,Alioli:2010xd}
and interfaced to {\pythia}\,8.210 with the PDF4LHC15 PDF set and the AZNLO tune.
The contribution of $tH$ associated production is also considered as part of the SM Higgs background.
The sample was generated using {\amcatnlolong}\,2.6.2~\cite{Alwall:2014hca} and interfaced to {\pythia}\,8.210 with the CT10 PDF set
and the A14 tune. The MC predictions are normalised using the NLO cross section recommended in Ref.~\cite{deFlorian:2016spz}.
The contribution of triboson production is found to be negligible.
 
All generated samples, except those produced with the {\sherpa}~\cite{Gleisberg:2008ta} event generator,
utilise \textsc{EvtGen}\,1.2.0~\cite{Lange:2001uf} to model the decays of heavy-flavour hadrons.
The effects of multiple interactions in the same and nearby bunch crossings (pile-up) were modelled by overlaying minimum-bias events, simulated using the soft QCD processes of {\pythia}\,8.186~\cite{Sjostrand:2006za} with the A3 tune~\cite{ATL-PHYS-PUB-2016-017} and \textsc{NNPDF2.3lo} PDF set~\cite{Ball:2012cx}.
 
The generated events were processed through a simulation~\cite{Aad:2010ah} of the ATLAS detector geometry and response
using \textsc{Geant4}~\cite{Agostinelli:2002hh}. A faster simulation, where the full \textsc{Geant4} simulation of
the calorimeter response is replaced by a detailed parameterisation of the shower shapes~\cite{FastCaloSim},
was adopted for some of the samples used to estimate systematic uncertainties in background modelling.
Simulated events were processed through the same reconstruction software as the data, and corrections were applied so that the object identification
efficiencies, energy scales and energy resolutions match those determined from data control samples.

\begin{table}
\footnotesize
\caption{Overview of the MC generators used for the main signal and background samples, including the matrix element (ME), parton shower (PS), parton distribution function set (PDF), and cross-section calculation order (Order).}
\begin{center}
\begin{tabular}[h]{l|c|c|c|c|c|c}
\hline \hline
\multirow{2}{*}{Process} & \multicolumn{2}{c|}{Generator} & \multicolumn{2}{c|}{PDF set} & \multirow{2}{*}{Tune} & \multirow{2}{*}{Order} \\ \cline{2-5}
&  ME   &  PS    &  ME  & PS &   &  \\\hline
$tt(qH)$ signal & {\powheg} & {\pythia}\,8 & \textsc{NNPDF3.0nlo} & \textsc{NNPDF2.3lo} & A14 & NLO \\ \hline
$tH$ signal & {\amcatnlolong} & {\pythia}~8 & \textsc{NNPDF3.0nlo} & \textsc{NNPDF2.3lo} & A14 & NLO \\ \hline
$W/Z$+jets & \multicolumn{2}{c|}{{\sherpa}\,2.2.1} & \multicolumn{2}{c|}{\textsc{NNPDF3.0nnlo}} & {\sherpa} & NLO/LO \\ \hline
\ttbar & {\powheg} & {\pythia}\,8 & \textsc{NNPDF3.0nlo} & \textsc{NNPDF2.3lo} & A14 & NLO \\ \hline
Single top & {\powheg} & {\pythia}\,8 & \textsc{NNPDF3.0nlo} & \textsc{NNPDF2.3lo} & A14 & NLO \\ \hline
$t\bar{t}X$ & {\amcatnlolong} & {\pythia}\,8 & \textsc{NNPDF3.0nlo} & \textsc{NNPDF2.3lo} & A14 & NLO \\ \hline
$VH$ & {\powheg} & {\pythia}\,8 & PDF4LHC15&CTEQ6L1 & AZNLO & NLO \\ \hline
$tH$ & {\amcatnlolong} & {\pythia}\,8 & \multicolumn{2}{c|}{CT10} & A14 & NLO \\ \hline
Diboson & \multicolumn{2}{c|}{{\sherpa}\,2.2.1} & \multicolumn{2}{c|}{\textsc{NNPDF3.0nnlo}} & {\sherpa} & NLO/LO \\ \hline\hline
\end{tabular}
\end{center}
\label{mob}
\end{table}


\section{Analysis strategy}
\label{sec:strategy_Htautau}
 
The analysis strategy adopted in this FCNC $\Htautau$ search is similar to the one used in Refs.~\cite{TOPQ-2017-07,Chen:2015nta} but extended to more search channels.
The $tt(qH)$ and $tH$ signal being probed is characterised by the presence of $\tau$-leptons from the decay of
the Higgs boson, where the remaining top quark decays into $Wb$. There is an additional $q$-jet from the FCNC $t\to qH$ decay in the top pair production.
If the $W$ boson or one of the $\tau$-leptons decays leptonically, an isolated electron or muon, together with significant $\met$, is also expected.
In a significant fraction of the events, the lowest-$\pt$ jet from the hadronic $W$ boson decay fails the minimum $\pt$ requirement of $25~\gev$,
resulting in only three reconstructed jets where the production mode is dominant.
In order to optimise the sensitivity of the search, the selected events are categorised into seven signal regions (SRs) based on the  numbers of light leptons,
$\had$ candidates, and light-flavour jets:
$t_{\ell}\hadhad$, $t_{\ell}\had$ (1j and 2j), $t_h\lephad$ (2j and 3j), $t_h\hadhad$ (2j and 3j), as shown in Table~\ref{tab:srcr}.

\begin{table}
\caption{Overview of the signal regions (SR), validation region (VR), and $t\bar{t}$ control regions (CRtt) used for the fake-$\tau$-lepton scale factor derivation in the leptonic channels. Leptons are required to have either same-sign (SS) or opposite-sign (OS) charges in each region.}
\label{tab:srcr}
\begin{center}
\begin{tabular}[h]{c|c|c|c|c|c|c}
\hline \hline
\multicolumn{2}{c|}{Regions} & $b$-jets & Light-flavour jets        & Leptons & Hadronic $\tau$ decays & Charge\\ \hline
\multirow{7}{*}{SR}&$t_{\ell}\thadhad$     & 1     & $\ge0$~~~~                                & 1      & 2             & $\thadhad$ OS\\ \cline{2-7}
&$t_{\ell}\tauhad$-1j  & 1     & 1                                   & 1      & 1                     & $t_{\ell}\tauhad$ SS\\ \cline{2-7}
&$t_{\ell}\tauhad$-2j  & 1     & 2                                        & 1      & 1                     & $t_{\ell}\tauhad$ SS\\ \cline{2-7}
&$t_h\tlhad$-2j   & 1     & 2                           & 1      & 1             & $\tlhad$ OS\\ \cline{2-7}
&$t_h\tlhad$-3j   & 1     & $\ge3$~~~~                      & 1      & 1             & $\tlhad$ OS\\ \cline{2-7}
&$t_h\thadhad$-2j & 1     & 2                            & 0      & 2             & $\thadhad$ OS\\ \cline{2-7}
&$t_h\thadhad$-3j & 1     & $\ge3$~~~~                       & 0      & 2             & $\thadhad$ OS\\ \hline
\multirow{2}{*}{VR}&$t_{\ell}\thadhad$-SS     & 1     & $\ge0$~~~~                                & 1      & 2             & $\thadhad$ SS\\ \cline{2-7}
&$t_h\thadhad$-3j SS & 1     & $\ge3$~~~~                       & 0      & 2             & $\thadhad$ SS\\ \hline
\multirow{6}{*}{CRtt}&$t_{\ell}t_{\ell}1b\tauhad$ & 1     & $\ge0$~~~~                            & 2      & 1                     & $t_{\ell}t_{\ell}$ OS\\ \cline{2-7}
&$t_{\ell}t_{\ell}2b\tauhad$      & 2     & $\ge0$~~~~                            & 2      & 1                     & $t_{\ell}t_{\ell}$ OS\\ \cline{2-7}
&$t_{\ell}t_h2b\tauhad$-2jSS & 2     & 2                             & 1      & 1             & $t_{\ell}\tauhad$ SS\\ \cline{2-7}
&$t_{\ell}t_h2b\tauhad$-2jOS & 2     & 2                             & 1      & 1             & $t_{\ell}\tauhad$ OS\\ \cline{2-7}
&$t_{\ell}t_h2b\tauhad$-3jSS & 2     & $\ge3$~~~~                        & 1      & 1             & $t_{\ell}\tauhad$ SS\\ \cline{2-7}
&$t_{\ell}t_h2b\tauhad$-3jOS & 2     & $\ge3$~~~~                & 1      & 1             & $t_{\ell}\tauhad$ OS\\ \hline
\end{tabular}
\end{center}
\end{table}

This event categorisation is primarily used to optimise the sensitivity in each signal region that targets either leptonic or hadronic top-quark
decays as well as the Higgs boson decays into either the $\lephad$ or $\hadhad$ final state.
The contribution of background to the signal regions of the $t_h\hadhad$ and $t_h\lephad$ channels is reduced by placing kinematic constraints on the di-$\tau$ mass and the $\met$ in the event~\cite{Chen:2015nta}.

For the $t_ht(qH)$ events, the jet from $t\to qH$, referred to as the FCNC jet ($q$-jet), should be a high-$\pt$ jet from the
decay chain $t\to qH\to q\tau\tau$, with $\tau$-leptons reconstructed as $\lephad$ or $\hadhad$.
Events should contain four jets, with the one having the smallest angular separation from the visible di-$\tau$ system being labelled the $q$-jet since the FCNC top-quark decay products are likely to be boosted closer together.
If there are more than two jets besides the $q$-jet and $b$-jet, the jets from the $W$ boson decay are chosen to be those from the combination
with an invariant mass closest to the $W$ boson mass. It is possible that one of the jets fails the $\pt$ requirement and is not reconstructed.
Events of this kind produce the $t_hH$ final state.
When both the $t_h$ and $H$ can be reconstructed, three jets come from the top quark's hadronic decay, including a $b$-jet, and a pair of opposite-sign $\had$ come from the Higgs boson decay.
 
The four-momenta of the invisible decay products from the decay of the $\tau$-leptons
are estimated using a kinematic fit and assuming a collinear approximation for the $\tau$ decay products.
The fit is done by minimising a $\chi^2$ function based on the Gaussian constraints placed on the Higgs boson mass ($m_H=125~\gev$) and the
measured $E_{x,y}^{\text{miss}}$  within their expected resolutions ($\sigma_{E_{x,y}^{\text{miss}}}$), defined as
\begin{eqnarray}
\begin{array}{ll}
\chi^2 =
\left( \frac{m_{\tau\tau,\text{fit}} - m_{H}}{\sigma_{\tau\tau}} \right)^2 +
\left( \frac{E_{x,\text{fit}}^{\text{miss}} - E_{x}^{\text{miss}}}{\sigma_{E_{x}^{\text{miss}}}} \right)^2
+\left( \frac{E_{y,\text{fit}}^{\text{miss}} - E_{y}^{\text{miss}}}{\sigma_{E_{y}^{\text{miss}}}} \right)^2.
\end{array}
\label{eq:eq1}
\end{eqnarray}
 
The Higgs boson mass resolution ($\sigma_{\tau\tau}$) is estimated to be $20~\gev$ from a fit of the mass
distribution of the simulated  $tqH$ signal events, while the $\met$ measurement's resolution is parameterised as a linear function of
$\sqrt{\sum E_{\text{T}}}$, where $\sum E_{\text{T}}$ is the scalar sum of the $E_{\text{T}}$ values of all physics objects contributing to the $\met$ reconstruction~\cite{Aaboud:2018tkc}.
After the $\chi^2$ minimisation, both the Higgs boson $\pt$ and the
$\pt$ values of the parent top quarks are determined with better resolution in the signal events.
 
For the  $t_{\ell}t(qH)$ and $t_{\ell}H$ events where the $W$ boson from $t\to W b$ decay decays leptonically,
the kinematic fit is no longer feasible due to the neutrino from the $W\rightarrow \ell\nu$ decay. The kinematic variables are calculated using the visible
objects only. After the event reconstruction, a number of these variables are used in a multivariate analysis to discriminate the signal from the background, as described in Section~\ref{sec:tmva}.


\FloatBarrier

\section{Background estimation}
\label{sec:background_model}
 
Most background processes are modelled using MC simulation.
After the event preselection, the main background is $\ttbar$ production, often in association with jets, denoted by $\ttbar$+jets in the following.
Small contributions arise from single-top-quark, $W/Z$+jets, multi-jet and diboson ($WW,WZ,ZZ$) production, as well as from the associated
production of a vector boson $V$ ($V=W,Z$) or a Higgs boson and a $\ttbar$ pair ($\ttbar V$ and $\ttbar H$). All backgrounds
with prompt leptons, i.e.\ those originating from the decay of a $W$ boson, a $Z$ boson, or a $\tau$-lepton,
are estimated using samples of simulated events and are initially normalised to their theoretical cross sections.
In the simulation, the top-quark and SM Higgs boson masses are set to $172.5~\gev$ and $125~\gev$, respectively,
and the Higgs boson is forced to decay via $H\to \tau\tau$ with a branching ratio calculated using \textsc{Hdecay}~\cite{Djouadi:1997yw}.
Backgrounds with non-prompt light leptons (electron or muon), with photons or jets misidentified as electrons, or with jets misidentified as $\tau$-lepton candidates,
generically referred to as fake leptons, are estimated using data-driven methods.
The background prediction is improved during the statistical analysis by performing a likelihood
fit to data using several signal-depleted control regions as shown in Table~\ref{tab:srcr}.
The events with one light lepton, two same-sign $\had$, and one $b$-jet, are also selected, providing a validation region (VR) for the background estimation in the leptonic channels.
\subsection{Backgrounds with fake $\tau$-leptons}
The background with one or more fake $\tau$ candidates arises mainly from $\ttbar$ or
multi-jet production, depending on the search channel.
Studies based on simulation show that, for all the above processes, fake $\tau$ candidates primarily result from the
misidentification of light jets and $b$-quark jets.
It is also found that the fake rate decreases for all jet flavours as the $\tau$ candidate's $\pt$ increases.
 
In the leptonic channels, the events with a prompt electron or muon and fake $\tau$-leptons are modelled by calibrating the MC samples with scale factors (SF)
derived from the dedicated $t\bar t$
control regions (CRtt) using dileptonic decays of $t\bar t$ pairs and semileptonic decays of $t\bar t$ pairs with two
$b$-jets, as is summarised in Table~\ref{tab:srcr}. The control regions are defined similarly to the signal regions but with an additional $b$-jet or light lepton.
There are four kinds of fake $\tau$-leptons that need to be calibrated: Type-1 fake $\tau$-leptons from hadronic $W$ boson decay ($\tau_{W}$) 
with opposite-sign (OS) charge relative to the light lepton;
Type-2 $\tau_{W}$'s with same-sign (SS) charge relative to the light lepton; Type-3 fake $\tau$-leptons originating from $b$-hadron decays; and Type-4 fake $\tau$-leptons from light-flavour hadron decays.
The dilepton regions ($t_{\ell}t_{\ell}1b\thad$ and $t_{\ell}t_{\ell}2b\thad$) are used to calibrate Type-3 and Type-4 fake $\tau$-leptons. The semileptonic
regions ($t_{\ell}t_h2b\thad$-2jOS and $t_{\ell}t_h2b\thad$-3jOS), where the $\thad$ and light lepton have opposite charges, are used to calibrate Type-1 fake $\tau$-leptons.
Similarly for Type-2, the semileptonic regions ($t_{\ell}t_h2b\thad$-2jSS and $t_{\ell}t_h2b\thad$-3jSS), where the $\thad$ and light lepton have same charges, are used.
A simultaneous fit to data is made to derive the scale factors for fake $\tau$-leptons in the MC samples. These consist of a total of 24 parameters
depending on four types of fake $\tau$-leptons, three $\pt$ bins, and two bins for 1- and 3-prong $\tau$ decays separately. The post-fit distributions of the
leading $\thad$ $\pt$ and measured $\tau$ scale factors are presented in Appendix~\ref{sec:tauFF_appendix}.
The values of these scale factors range from 0.3 to 1.28 and are used to correct the MC-estimated fakes in the corresponding  signal regions with a single $b$-jet.
In the $t_{\ell}\thadhad$ channel, both $\tau$ candidates can be misidentified, so the calibration is applied to each $\tau$ candidate separately, following the same procedure used in the $\tlhad$ channel.
The central values of the scale factors vary according to their uncertainties in the final fit. A closure test is made for the fake-$\tau$ estimates, using the same procedure as
in the $t_{\ell}\thadhad$-SS validation region, and it shows good agreement between data and the background prediction.
 
In the hadronic channels, the contribution of fakes is estimated from data using the fake-factor (FF) method to derive the transfer functions from the background-enriched control
regions (CR) to the SR~\cite{HIGG-2019-09}.
These CRs do not overlap with the main signal regions discussed in Section~\ref{sec:strategy_Htautau}.
The CR selection requirements are analogous to those used to define signal regions, except
that the subleading $\tau$ candidate is required to fail the `medium' $\tau$ identification while still passing a loose requirement.
The contribution of fakes with subleading $\tau$ candidates can be calculated by rescaling the templates for `loose' $\tau$ candidates in the CR
with the FFs.
The templates are produced by subtracting all MC background contributions with real subleading $\tau$-leptons from the data.
The FFs are computed as the ratio of misidentified $\thad$ candidates that either pass or fail the `medium' $\tau$ ID selection in regions enriched in events originating from
$W$+jets processes~\cite{HIGG-2019-09}.
FFs obtained from the $W$+jets events are compared with those from the same-sign $\thadhad$ control regions and the differences are treated as systematic uncertainties.
 
\subsection{Background with fake light leptons}
The background originating from non-prompt or misidentified light leptons is primarily from multi-jet production.
The contribution from these events is estimated with a data-driven method called ABCD.
The estimate is based on the numbers of `tight' and `loose' light leptons that passed and failed the PLIV cut in the low ($<25~\gev$) and high
($>25~\gev$) $\met$ regions. The number of
fake light leptons in the signal region is evaluated by scaling the number of `loose' leptons in the high $\met$ region by the ratio of `tight' leptons to `loose' leptons in the
low $\met$ region. The contributions of prompt leptons and calibrated fake $\tau$-leptons to the numbers of `tight' and `loose' leptons are subtracted using MC simulation.
A closure test is made for the background estimates in the signal-depleted low-BDT-score ($<-0.6$) regions, defined in Section~\ref{sec:tmva}.
The data are in good agreement with the background prediction in the
$t_{\ell}\had$ channels, while the fake light leptons are negligible in other leptonic channels.


\section{Multivariate discriminant}
\label{sec:tmva}
 
\begin{table*}[t!]
\caption{\small{Discriminating variables (n) used in the training of the BDT of each SR.
The ranking of the input variables according to their importance in the training is reported from highest (1) to
lowest (n). Variables whose ranking is missing are not included in the training of that SR. The description of each variable is provided in the text.}}
\label{tab:importance}
\centering
\vspace{0.2cm}
\begin{tabular}{cccccccc} \toprule\toprule
& $t_{\ell}\tauhad$-1j                                  &  $t_{h}\tlhad$-2j   &  $t_{\ell}\tauhad$-2j & $t_{h}\tlhad$-3j & $t_{\ell}2\tauhad$     & $t_h2\tauhad$-2j & $t_h2\tauhad$-3j       \\\midrule
Total variables~(n)                           & $12$~~ & $15$~~ & $12$~~ & $17$~~ & $15$~~ & $12$~~ & $12$~~ \\\midrule
$\met$                                              & $5$       &  $11$~~               &  $10$~~             & $13$~~     & $6$           & $7$    & $13$~~          \\
${\pt}_{\tau1}$                                 & $1$       &  $4$                &  $1$              & $1$      & $5$           & $11$~~   & $10$~~           \\
\text{max}($\eta_{\tau}$)                           & $4$       &             &  $4$              &  & $10$~~          &  &        \\
${\pt}^{\ell}$                                 & $12$~~      &  $15$~~               &  $12$~~             & $17$~~     &       &  &         \\
$\chi^{2}$                                          &   &             &           & $14$~~     &       &  &       \\
$m_{\text{bjj}}$                            &   &  $1$                &           & $2$      &       & $3$      & $4$          \\
$m^{W}_{\text{T}}$                           & $11$~~      &             &  $8$              &  & $13$~~          &  &         \\
$m_{\tau\tau,\text{fit}}$                                     &   &  $2$                &           & $3$      &       & $1$      & $1$          \\
$m_{\text{jj}}$                                      &   &             &           & $9$      &       & $6$      & $7$\\
$m_{\tau\tau\text{q},\text{fit}}$                          &   &             &           &  &       & $10$~~     & $6$\\
$m_{\tau\tau}$                           & $10$~~      &  $14$~~               &  $11$~~             & $6$      & $1$           & $2$    & $2$          \\
${\pt}_{\tautau} $                          &   &             &           &  & $9$           &  &         \\
$m_{\tau\tau\text{q}}$               &   &             &           &  & $3$           &  &        \\
$m_{\ell \text{b}}$                 & $3$       &             &  $5$              &  & $4$           &  &         \\
\text{min}($m_{\tau\tau \text{j}}$)             & $9$       &             &  $3$              &  & $14$~~          &  &         \\
\text{min}($m_{\text{jj}}$) &   &  $12$~~               &           & $11$~~     &       &  &         \\
$\met\text{\,centrality}$                             &   &  $13$~~               &           & $15$~~     &       & $12$~~   & $9$         \\
$E_{\tau1}/E_{\tau1,\text{fit}}$                  &   &  $10$~~               &           & $12$~~     &       & $8$    & $8$          \\
$E_{\tau2}/E_{\tau2,\text{fit}}$                  &   &  $7$                &           & $4$      &       & $9$    & $11$~~         \\
$\Delta\phi(\tau\tau,\met)$                         &   &  $6$                            &           & $16$~~     &       & $13$~~   & $12$~~         \\
\text{min}($\Delta R(\ell,\tau)$)                               & $8$       &  $9$                &  $9$              & $10$~~     & $15$~~          &  &         \\
$\Delta R(\tau,\tau)$                               &   &             &           &  & $2$           & $4$    & $3$             \\
$\Delta R(\ell,\text{$b$-jet})$                       & $2$       &  $3$                &  $2$              & $8$      & $12$~~          &  &         \\
$\Delta R(\tau1,\text{$b$-jet})$                       & $6$       &  $5$                &  $6$              & $7$      & $11$~~          &  &        \\
$\Delta R(\ell+\text{$b$-jet},\tau\tau )$             &   &             &           &  & $7$           &  &         \\
$\Delta R(\tau1,\text{light-jet})$                   & $7$       &  $8$                &  $7$              & $5$      & $8$           & $5$    & $5$    \\
\bottomrule\bottomrule\\
\end{tabular}
\end{table*}

Boosted decision trees implemented in the TMVA framework~\cite{Hocker:2007ht} are used in each SR to improve the separation between signal and background.
In the training process, all signal events from $tt(qH)$ and $tH$ are combined for $tuH$ and $tcH$. All background sources from SM processes
(including both the real and fake $\tau$ contributions) are also used in the training.
 
Many potential variables were investigated in each SR separately. The discrimination of a given variable is quantified by the `separation' (which measures the degree of overlap between background and signal distribution shape) and `importance' (which ranks the power of the variable in the classification of the events) provided by the TMVA package. 
The BDT discriminant is trained for each SR starting from a large list of variables; then the least important variables are removed sequentially, and the BDT is retrained until the
receiver operating characteristic (ROC) score, the area under the ROC curve, drops by more than 2\%.
The final BDT input variables in each SR and their importance ranking are listed in Table~\ref{tab:importance}. The discriminating variables are as follows:
\begin{itemize}
\item $\met$ is the missing transverse momentum.
\item ${\pt}_{\tau1} $ is the transverse momentum of the leading $\tau$-lepton candidate.
\item \text{max}($\eta_{\tau}$) is the maximum $\eta$ of the $\tauhad$ candidate(s).
\item ${\pt}^{\ell}$ is the transverse momentum of the leading light lepton.
\item $\chi^2$ of the kinematic fit of the momentum of the invisible decay products of the $\tau$-leptons as defined in Eq.~(\ref{eq:eq1}).
\item $m_{\text{bjj}}$ is the invariant mass of the $b$-jet and the two jets from the $W$ boson decay, and reflects the top-quark mass in the decay $t\to Wb \to j_1j_2b$. This variable is only defined for the 4-jet $t_hH$ and $t_ht(qH)$ events.
\item $m^{W}_{\text{T}}$ is the transverse mass calculated from the lepton and $\met$ in the leptonic channels, defined as
\begin{equation*}
m^{W}_{\text{T}} = \sqrt{2 {\pt}_{\ell} E_{\text{T}}^{\text{miss}} \left(1-\cos\Delta\phi_{\ell,\text{miss}} \right)},
\end{equation*}
where $\Delta\phi_{\ell,\text{miss}}$ is the azimuthal angle between the light lepton and $\met$.
\item $m_{\tau\tau,\text{fit}}$ is the fitted invariant mass of the $\tau$ candidates and reconstructed neutrinos for the $t_hH$ and $t_ht(qH)$ events.
\item $m_{\text{jj}}$ is the reconstructed invariant mass of two light jets from the $W$ decay with a mass closest to the known $W$ boson mass.
\item $m_{\tau\tau\text{q},\text{fit}}$ is the fitted invariant mass of the FCNC-decaying top quark reconstructed from the di-$\tau$ candidates, $q$-jet and reconstructed neutrinos.
\item $m_{\tau\tau}$ is the visible invariant mass of the di-$\tau$ system (including the light lepton when there is only one $\had$ candidate).
\item ${\pt}_{\tau\tau}$ is the visible $\pT$ of the di-$\tau$ system.
\item $m_{\tau\tau\text{q}}$ is the reconstructed visible mass of the FCNC-decaying top quark.
\item $m_{\ell \text{b}}$ is the invariant mass of the lepton and the $b$-jet, which reflects the visible top-quark mass.
\item \text{min}($m_{\tau\tau \text{j}}$) is the minimum value of the visible mass of the di-$\tau$ candidates (including leptonically decaying $\tau$ candidates) and the light-flavour jet,
reflecting the invariant mass of the visible FCNC top-quark decay products.
\item \text{min}($m_{\text{jj}}$) is the minimum value of the invariant mass of all the light-flavour jet pairs, reflecting the invariant mass of the $W$ boson candidate.
\item $\met$ centrality is a measure of how centrally the $\met$ lies between the two $\tau$ candidates in the transverse plane, and is defined as
\begin{eqnarray*}
\begin{array}{l}
\met~\text{centrality} = {(x+y)}/{\sqrt{x^2+y^2}}, \\
\text{with}~x=\frac{\sin(\phi_{\text{miss}}-\phi_{\tau_1})}{\sin(\phi_{\tau_2}-\phi_{\tau_1})}, \quad  y=\frac{\sin(\phi_{\tau_2}-\phi_{\text{miss}})}{\sin(\phi_{\tau_2}-\phi_{\tau_1})} ,
\end{array}
\end{eqnarray*}
\item $E_{\tau{i}}/E_{\tau{i},\text{fit}}$ $({i}=1,2)$ is the momentum fraction carried by the visible decay products from the leading and subleading $\tau$ decays. It is based on the best fit of the four-momentum of the neutrino(s) according to the event reconstruction algorithm in this section. For the $\tauhad$ decay mode, the visible decay products carry most of the $\tau$-lepton's energy since there is only a single neutrino in the final state.
\item $\Delta\phi(\tau\tau,\met)$ is the azimuthal angle between the $\met$ and di-$\tau$ system's $\pT$.
\item $\Delta R(a,b)$ is the angular distance between the $a$ and $b$ objects in the event.
\end{itemize}
 
Comparisons between the data and the predicted background for the distributions of leading $\had$ $\pt$  and visible invariant mass of the di-$\tau$ system ($m_{\tau\tau}$)
in SRs and VRs, after applying fake factors,
are shown in Figures~\ref{fig:taupt_prefit} and~\ref{fig:mtt}, respectively.
The data are well described by the background model in all cases.
The final observable used to extract the signal contribution is the BDT distribution in each SR corresponding to either the $tuH$ or $tcH$ signal.
 
\begin{figure}[H]
\centering
\begin{tabular}{@{}ccc@{}}
\includegraphics[page=1,width=0.28\textwidth]{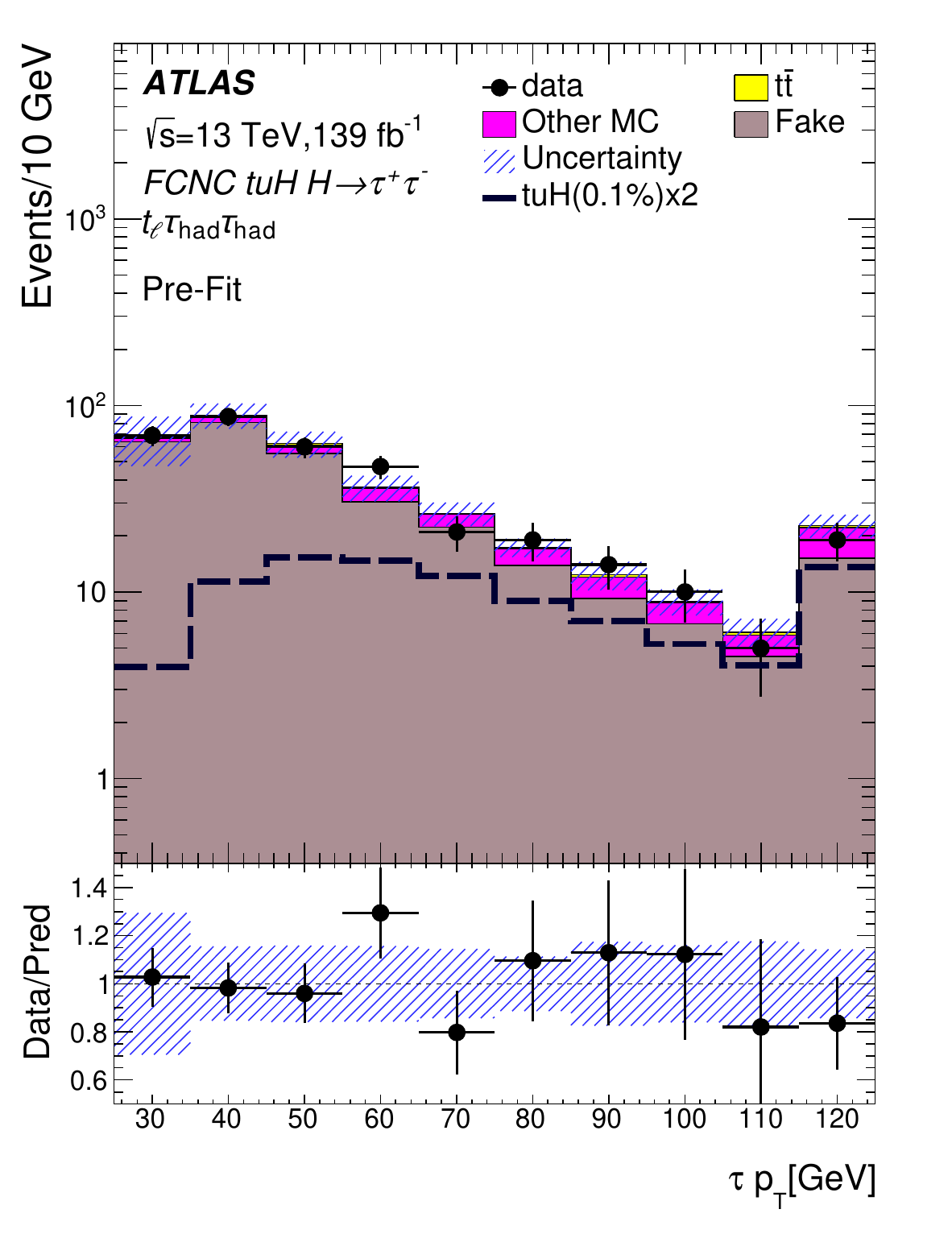} &
\includegraphics[page=1,width=0.28\textwidth]{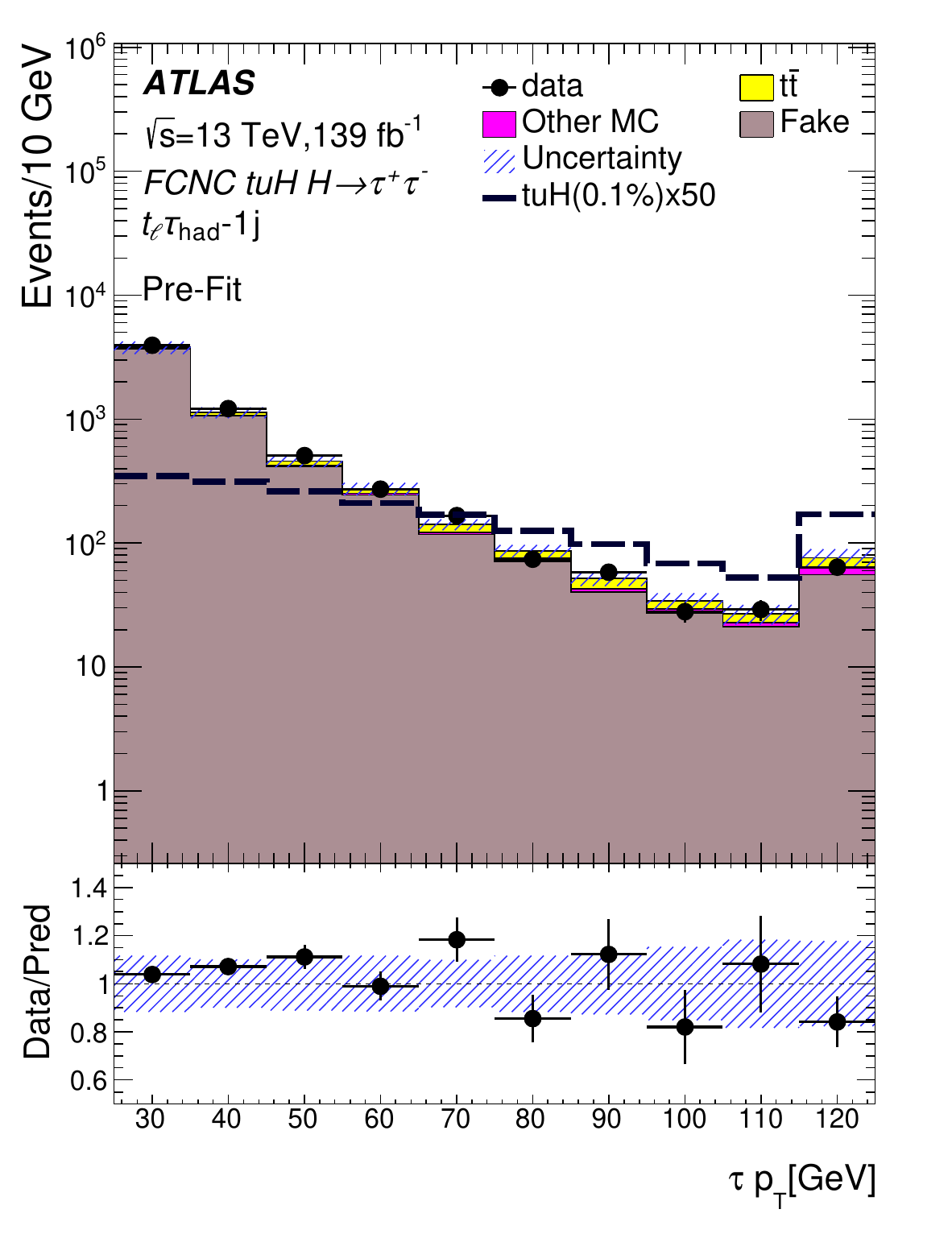}&
\includegraphics[page=1,width=0.28\textwidth]{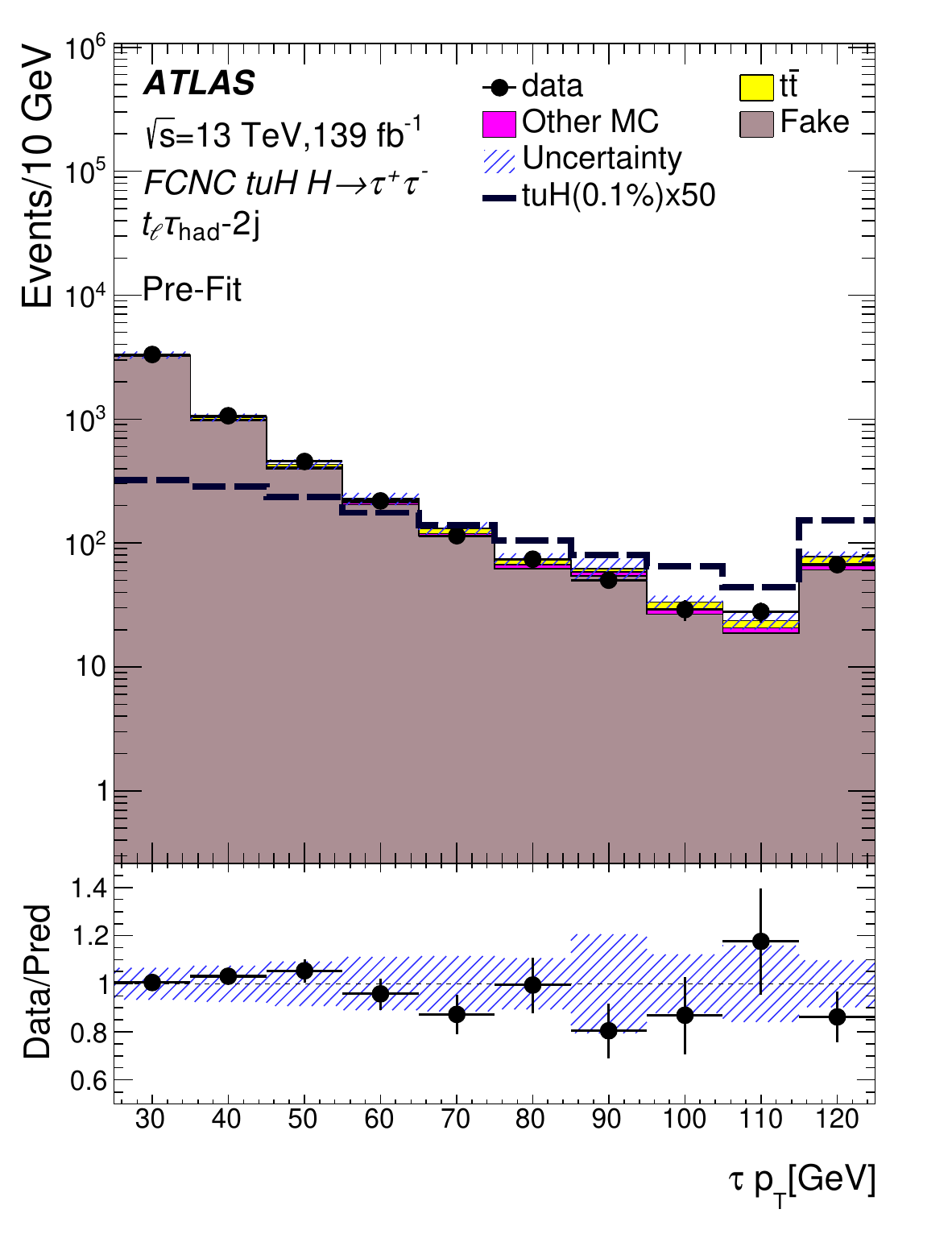}\\
(a) & (b) & (c) \\
\includegraphics[page=1,width=0.28\textwidth]{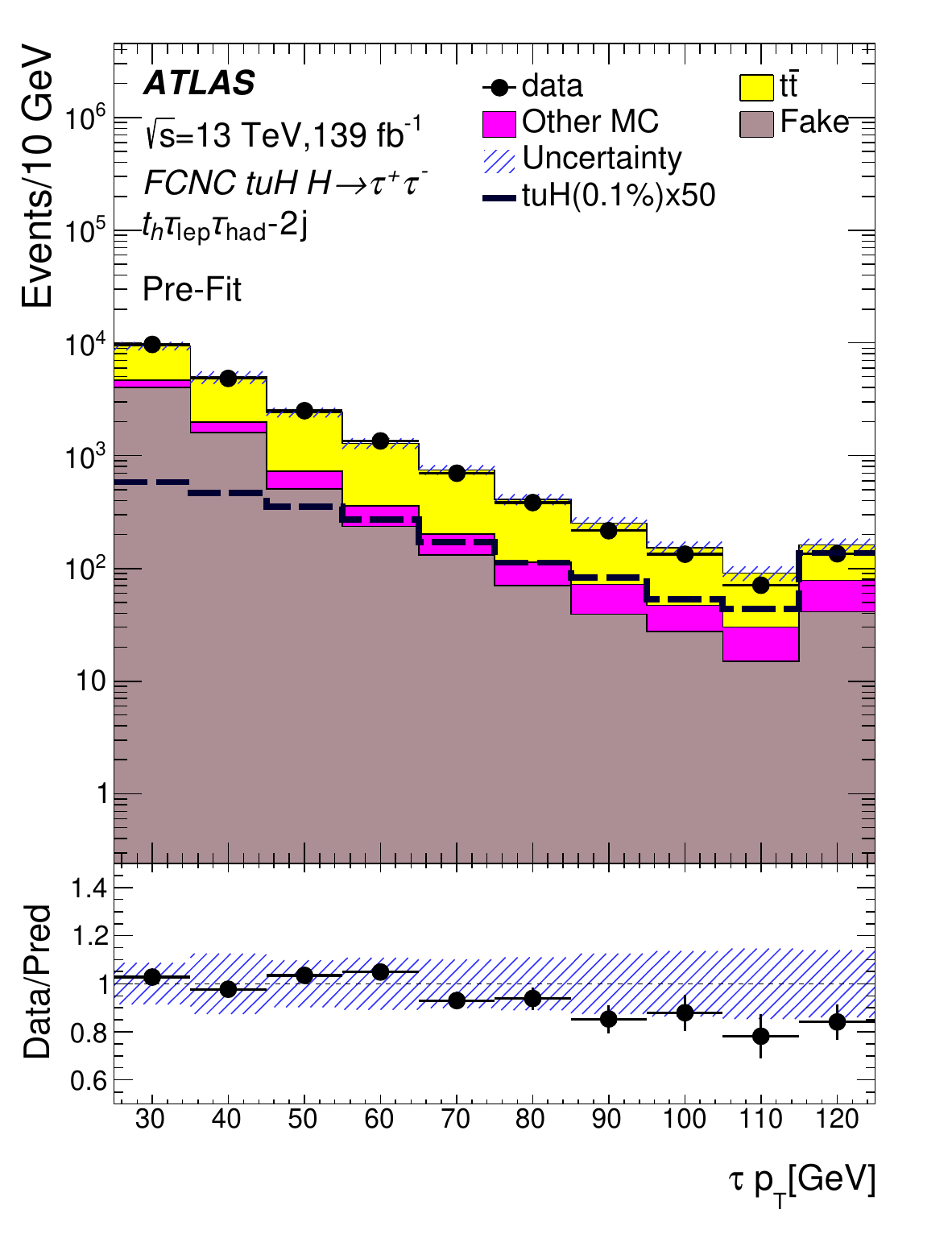}&
\includegraphics[page=1,width=0.28\textwidth]{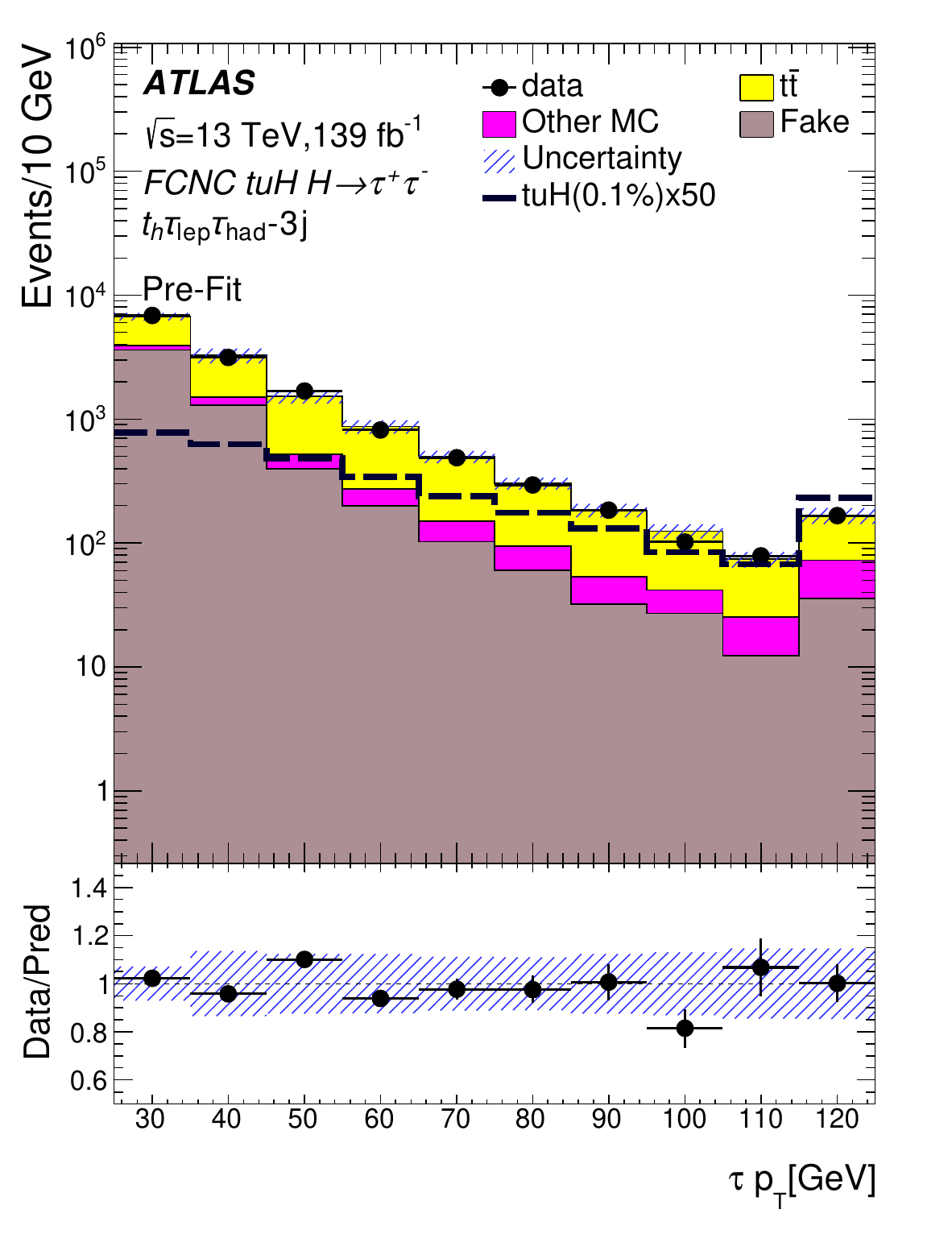} &
\includegraphics[page=1,width=0.28\textwidth]{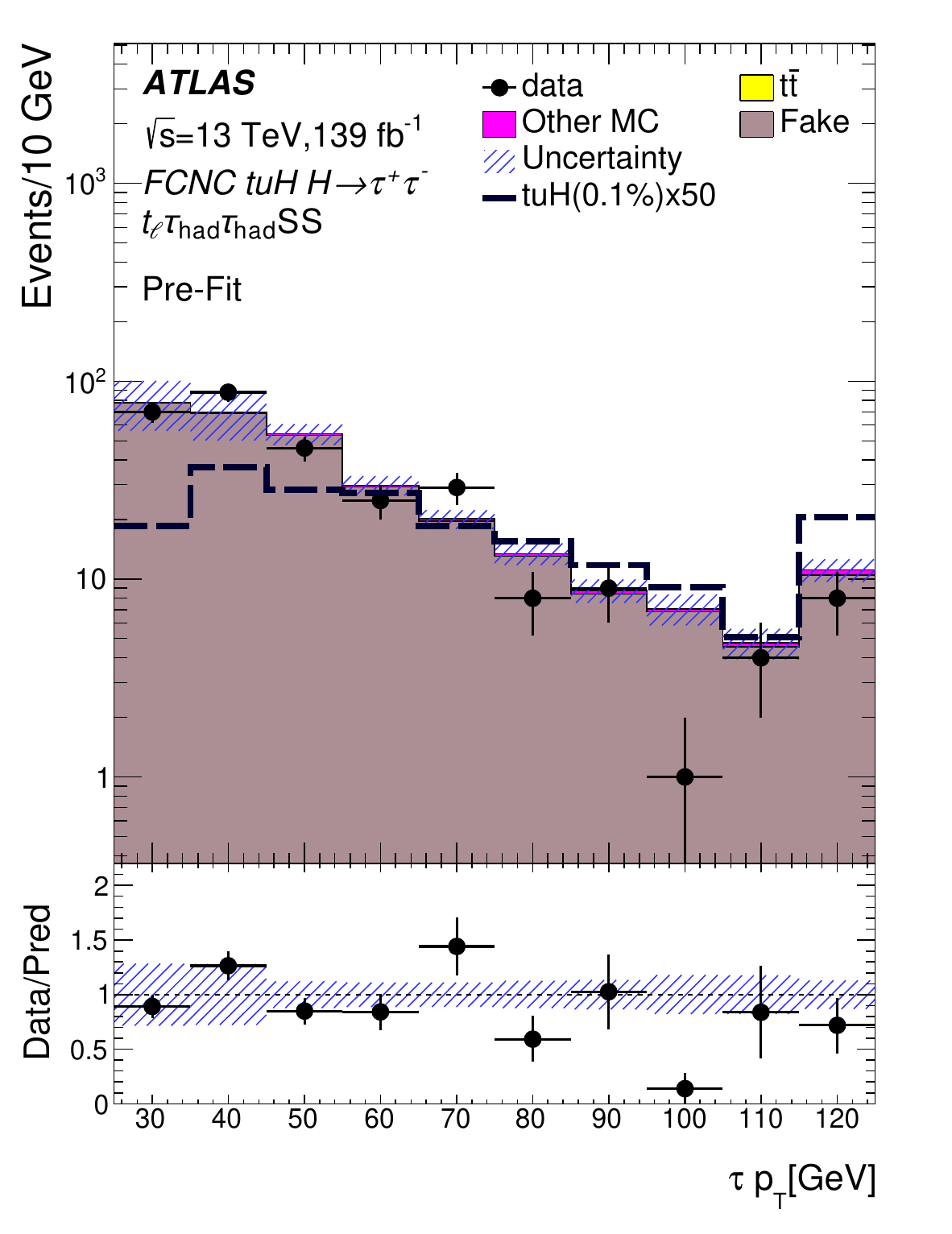} \\
(d) & (e) & (f)\\
\includegraphics[page=1,width=0.28\textwidth]{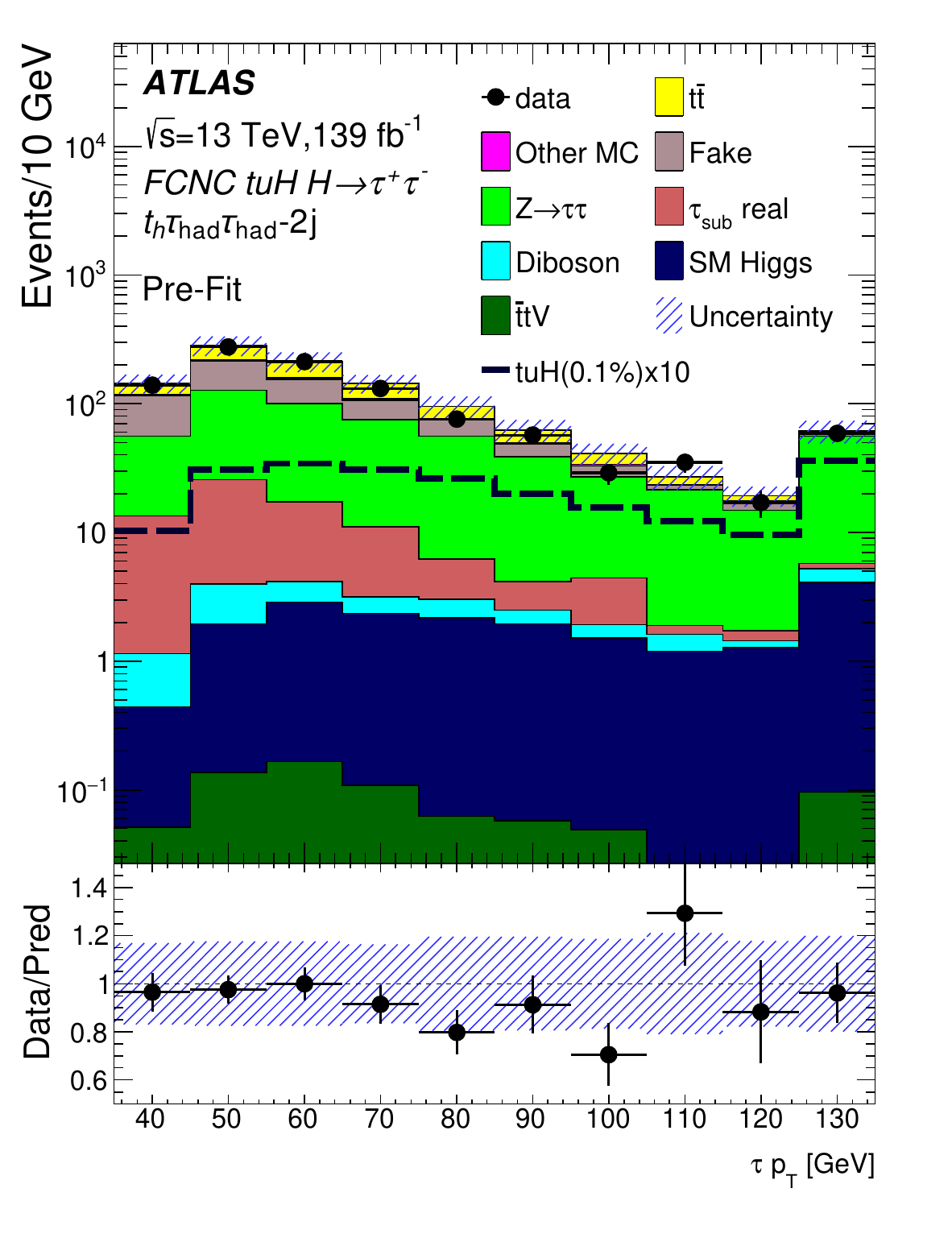}&
\includegraphics[page=1,width=0.28\textwidth]{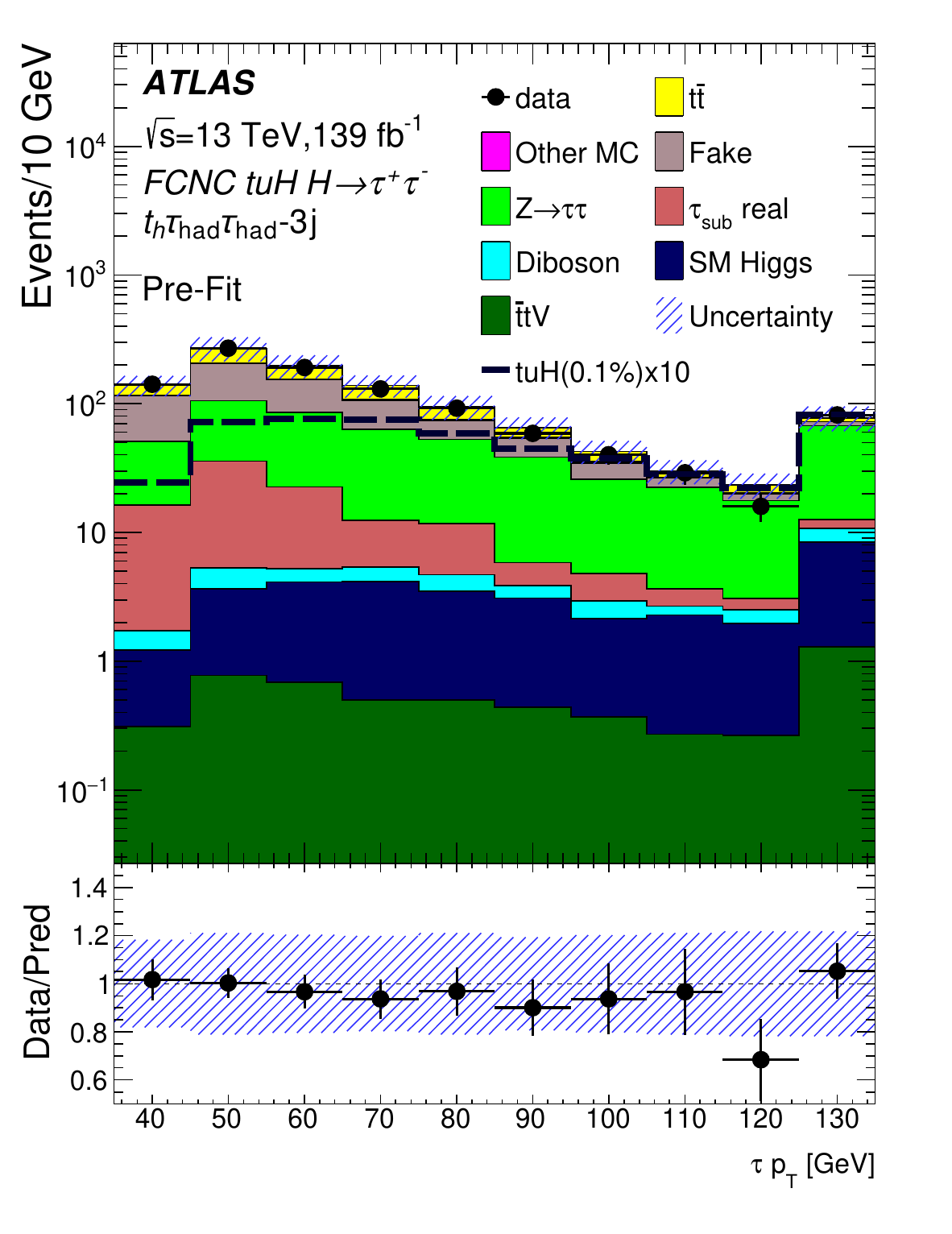}&
\includegraphics[page=1,width=0.28\textwidth]{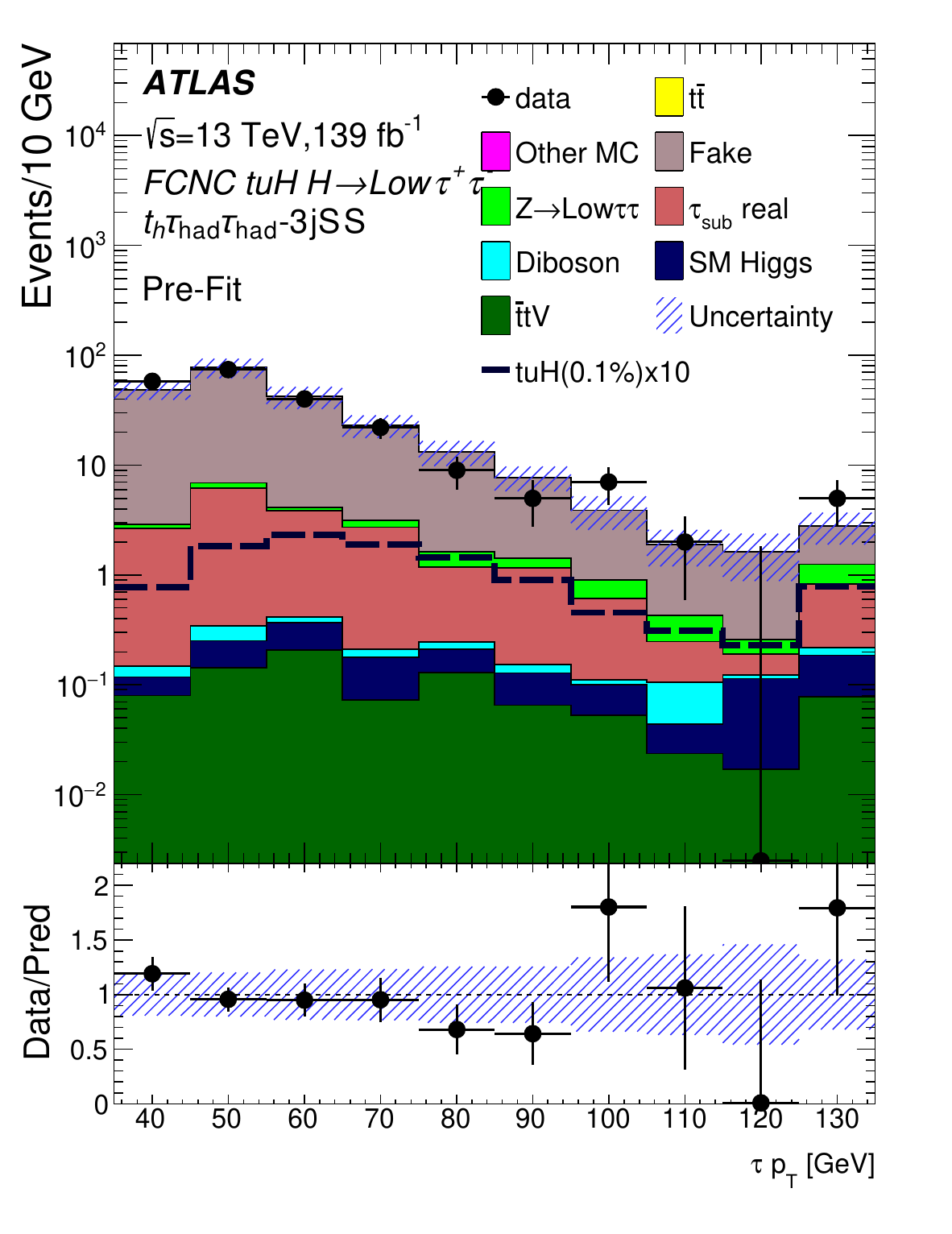}\\
(g) & (h) & (i)\\
\end{tabular}
\caption{Leading $\thad$ $\pt$  distributions obtained before the fit to data (`Pre-Fit') showing
the expected background and $tuH$ signals after applying fake factors in the following regions: (a) $t_{\ell}\thadhad$,
(b) $t_{\ell}\tauhad$-1j,  (c) $t_{\ell}\tauhad$-2j, (d) $t_h\tlhad$-2j, (e) $t_h\tlhad$-3j, (f) $t_{\ell}\thadhad$-SS, (g) $t_h\thadhad$-2j, (h) $t_h\thadhad$-3j and (i) $t_h\thadhad$-3j SS.
The total statistical and systematic uncertainty of the background prediction is indicated by the hatched band.
Overflow events are included in the last bin. `Other MC' includes single-top, $V$+jets, and other small backgrounds in the leptonic and hadronic channel. The $tuH$ signal is scaled by a
normalisation factor of either 2, 10, or 50.
The lower panels show the ratio of data to prediction.}
\label{fig:taupt_prefit}
\end{figure}

\begin{figure}[H]
\centering
\begin{tabular}{@{}ccc@{}}
\includegraphics[page=1,width=0.28\textwidth]{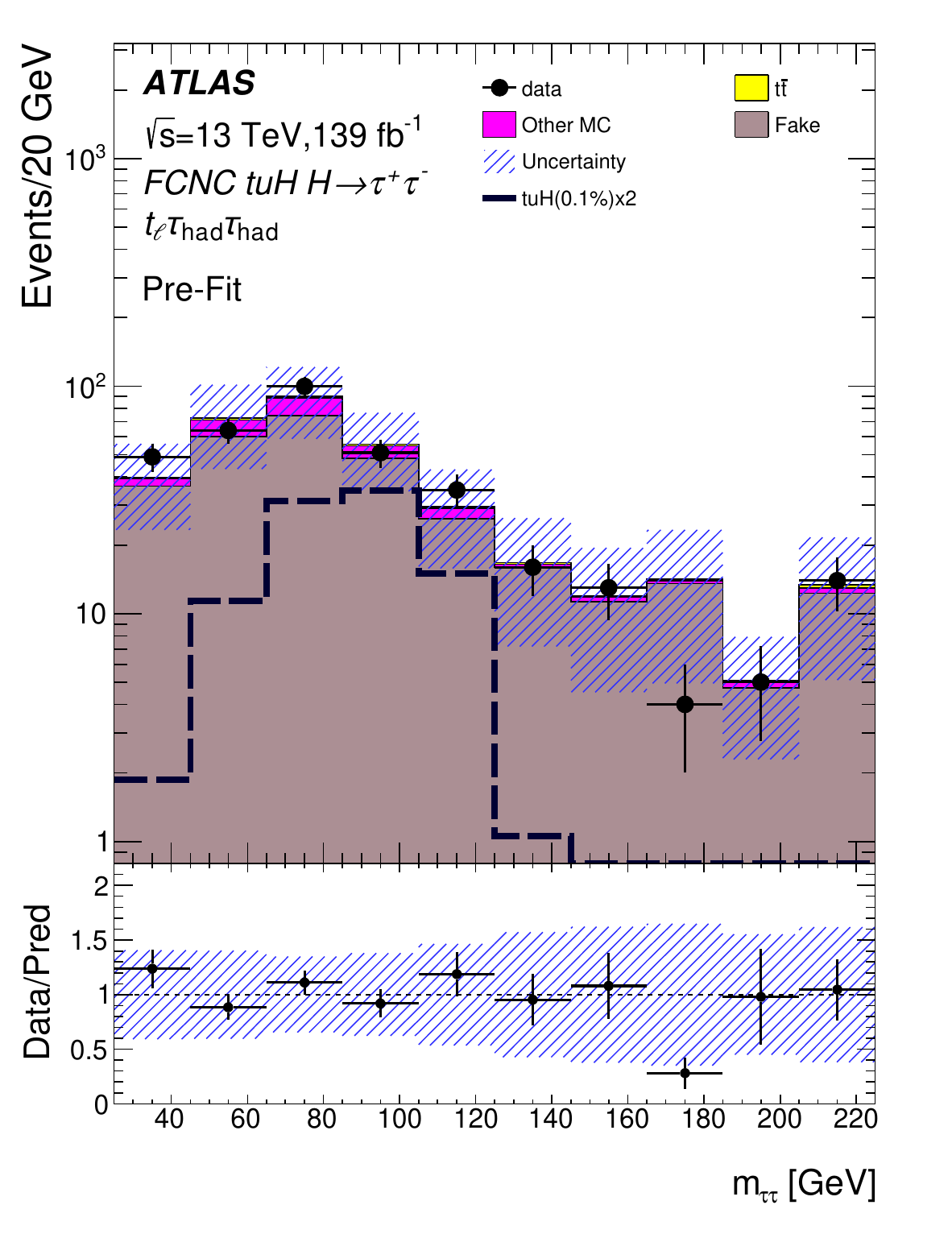} &
\includegraphics[page=1,width=0.28\textwidth]{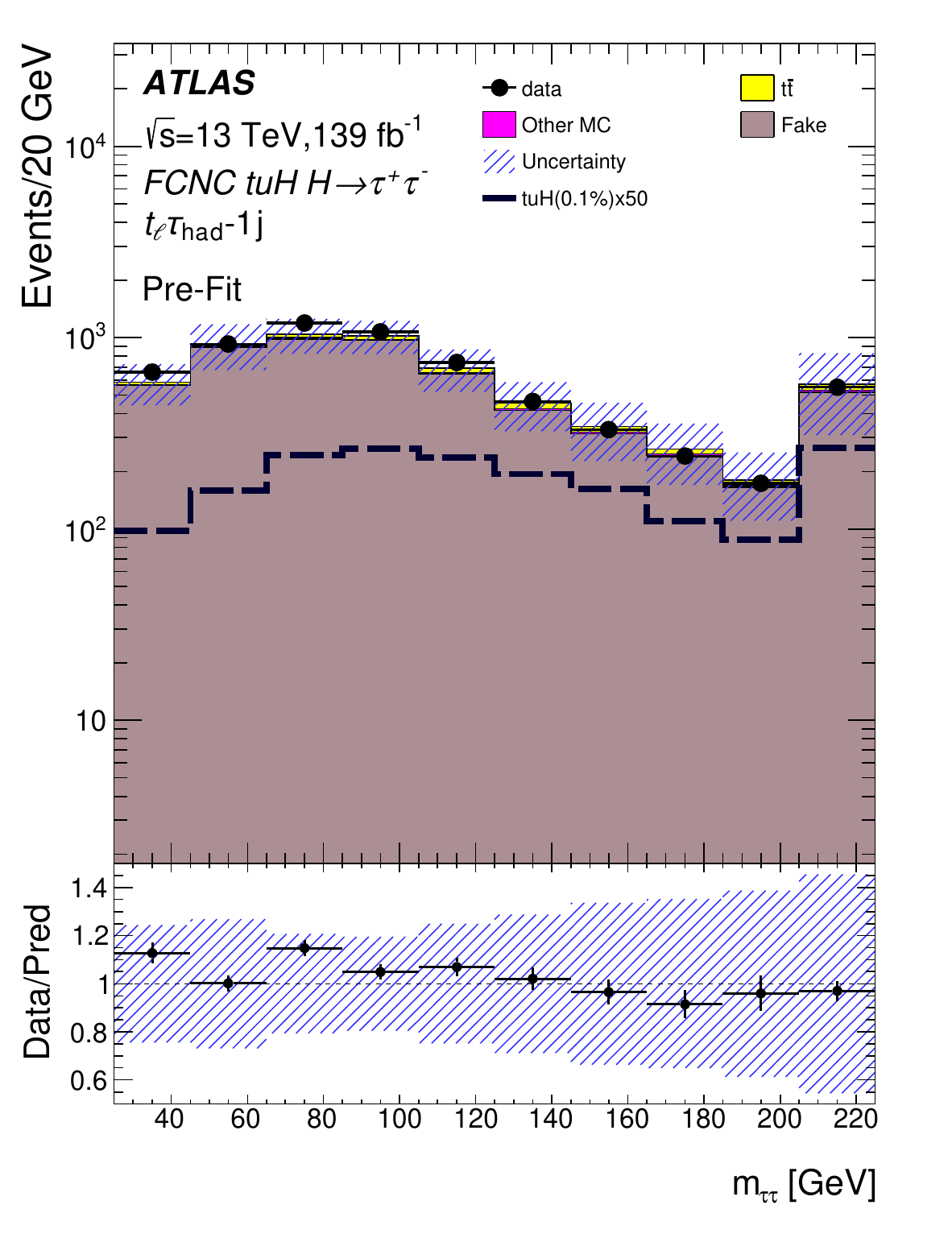}&
\includegraphics[page=1,width=0.28\textwidth]{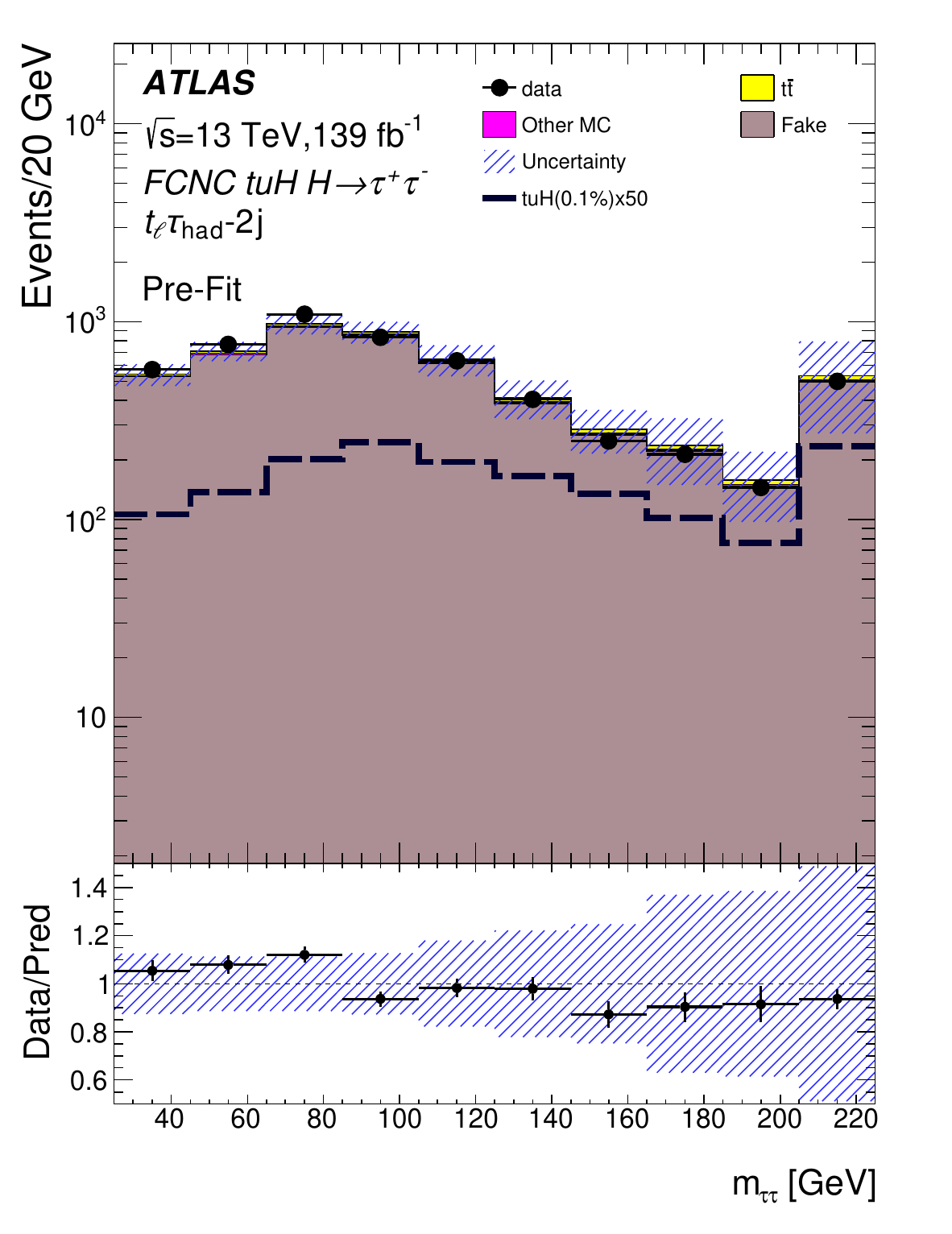}\\
(a) & (b) & (c) \\
\includegraphics[page=1,width=0.28\textwidth]{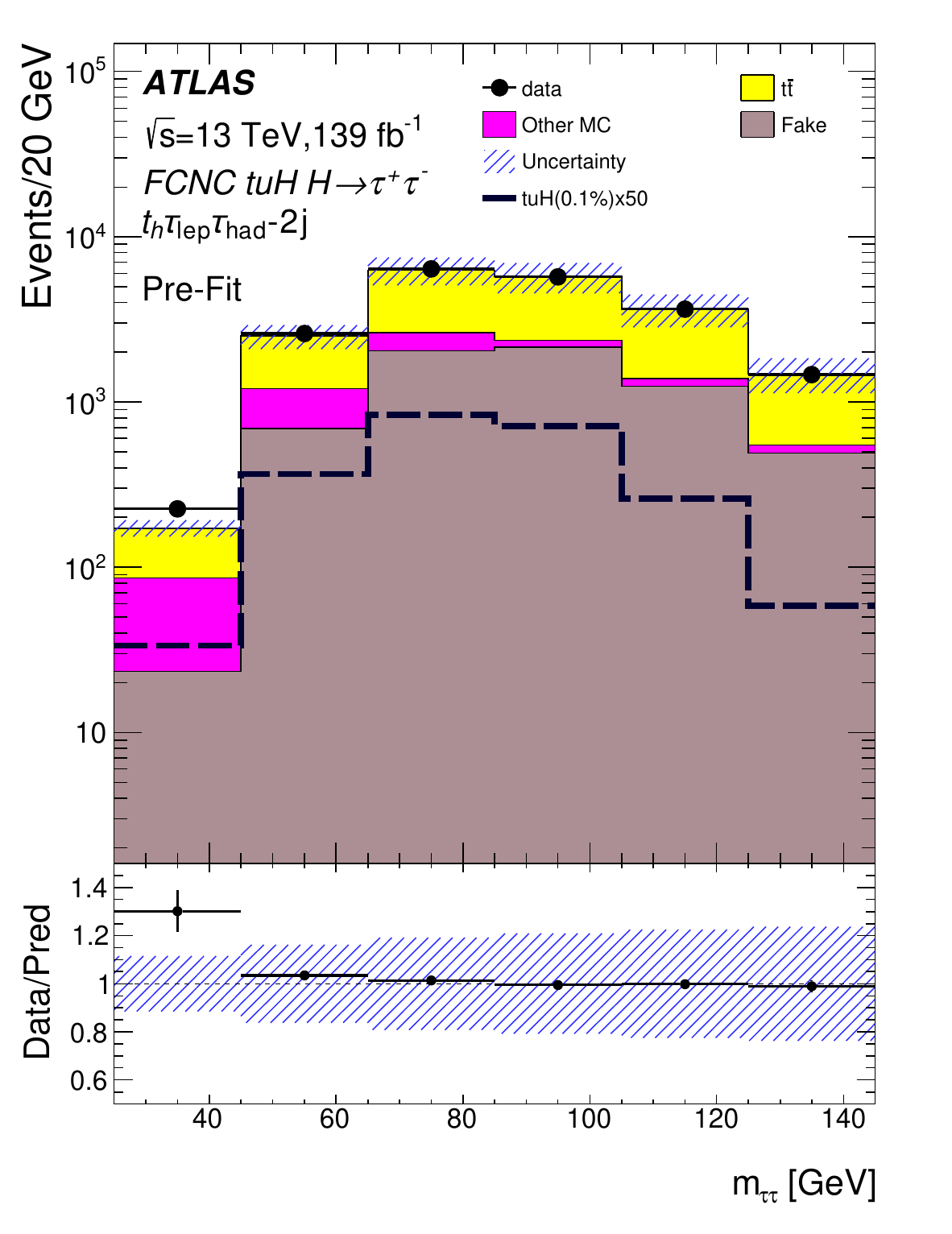}&
\includegraphics[page=1,width=0.28\textwidth]{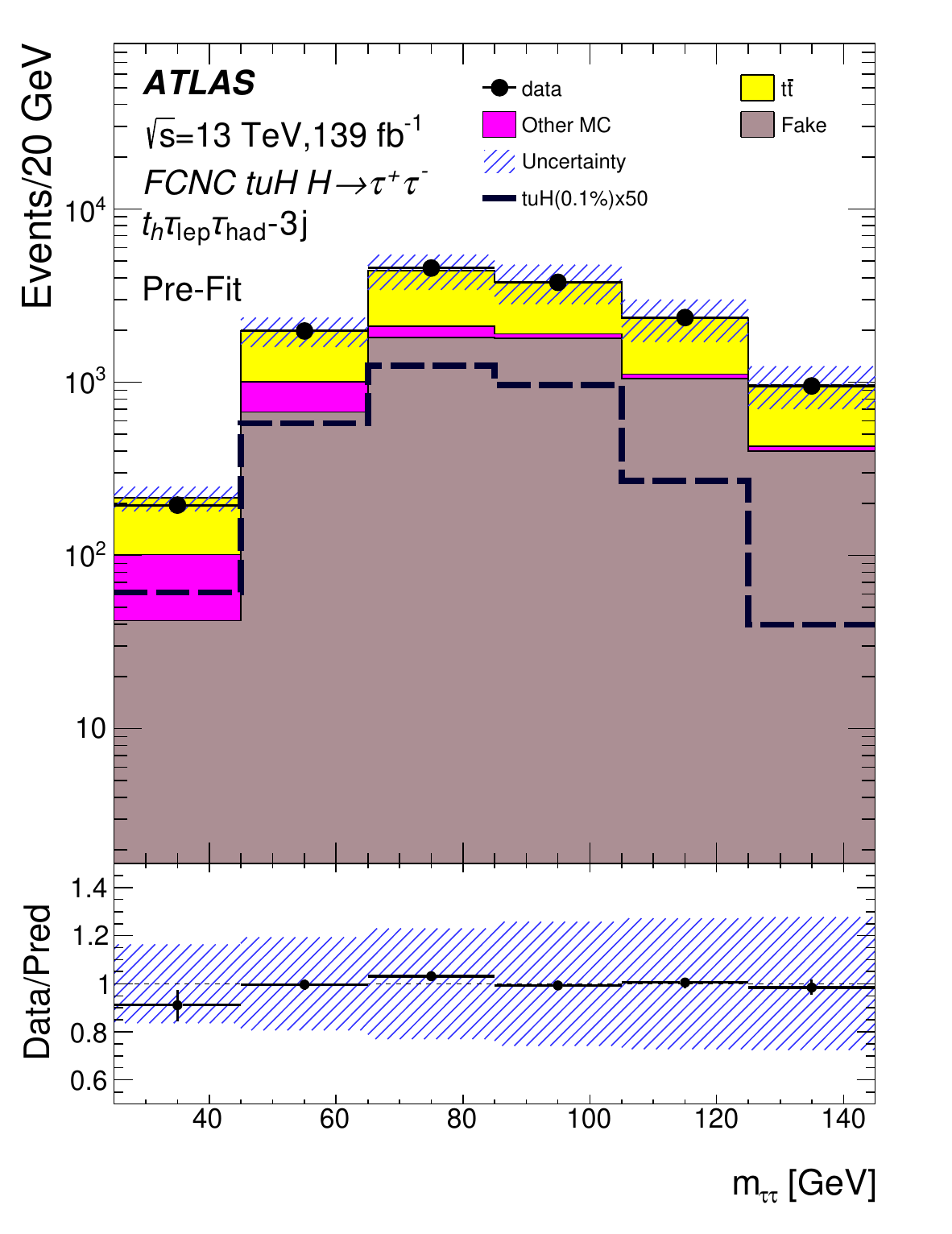}&
\includegraphics[page=1,width=0.28\textwidth]{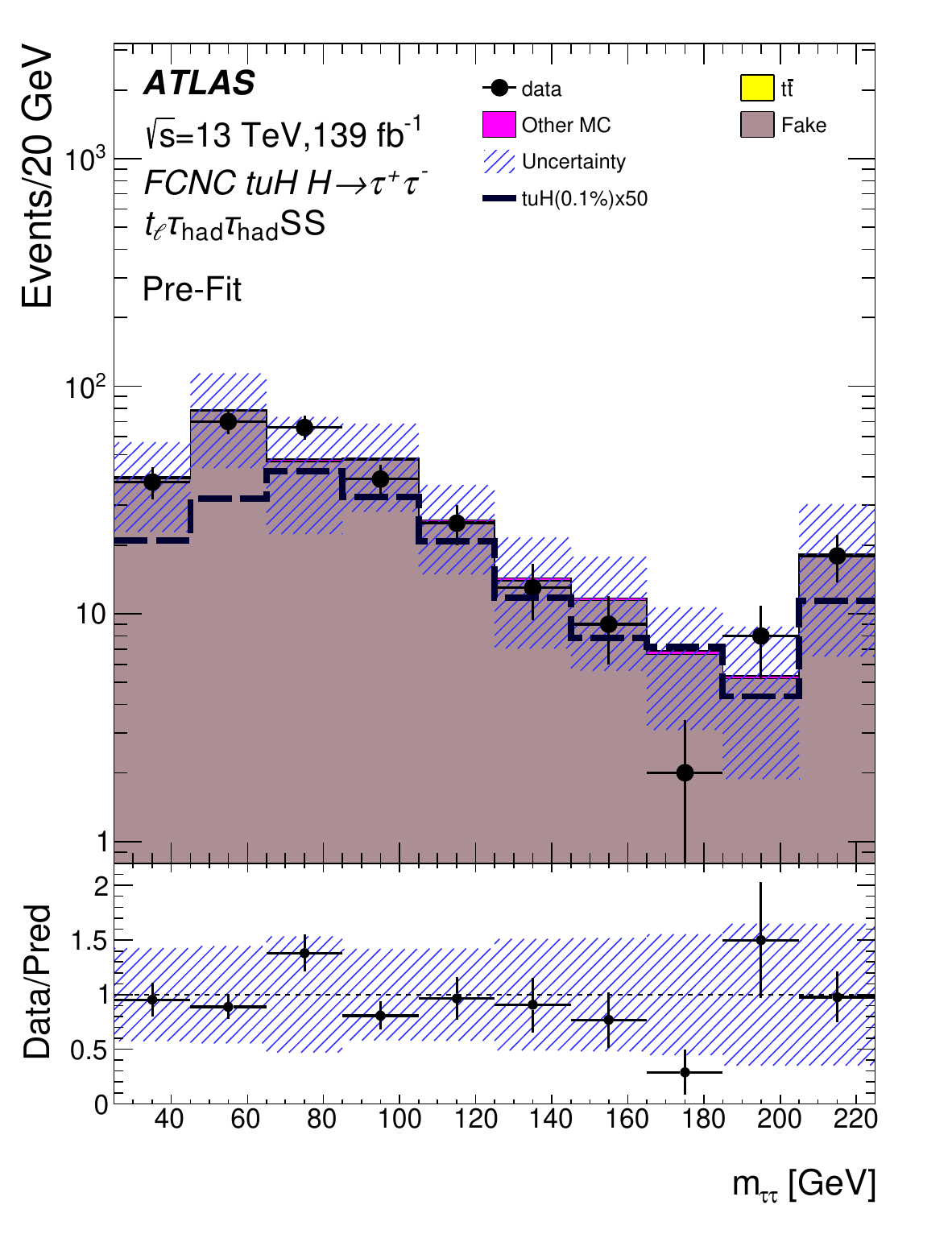}\\
(d) & (e) & (f)\\
\includegraphics[page=1,width=0.28\textwidth]{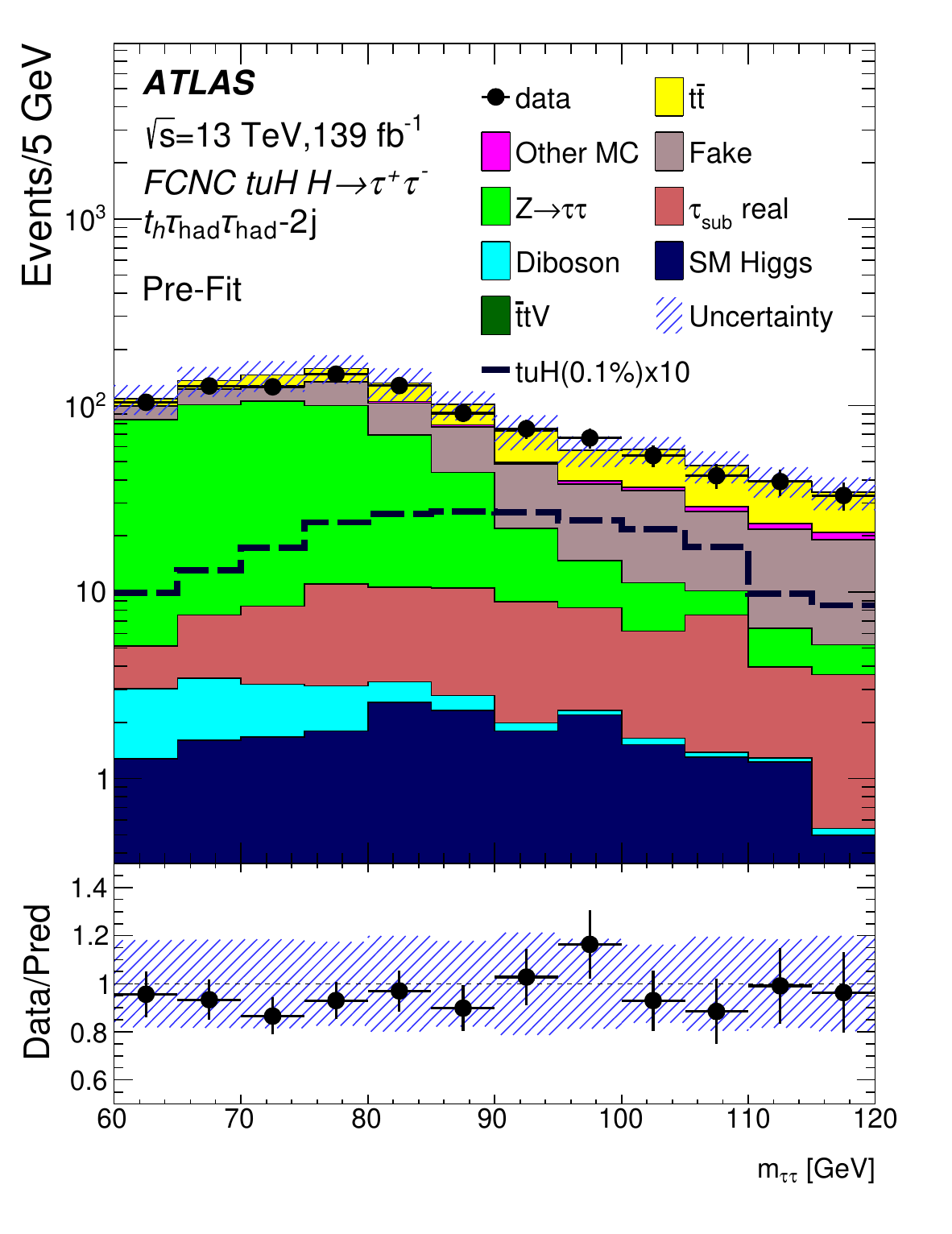}&
\includegraphics[page=1,width=0.28\textwidth]{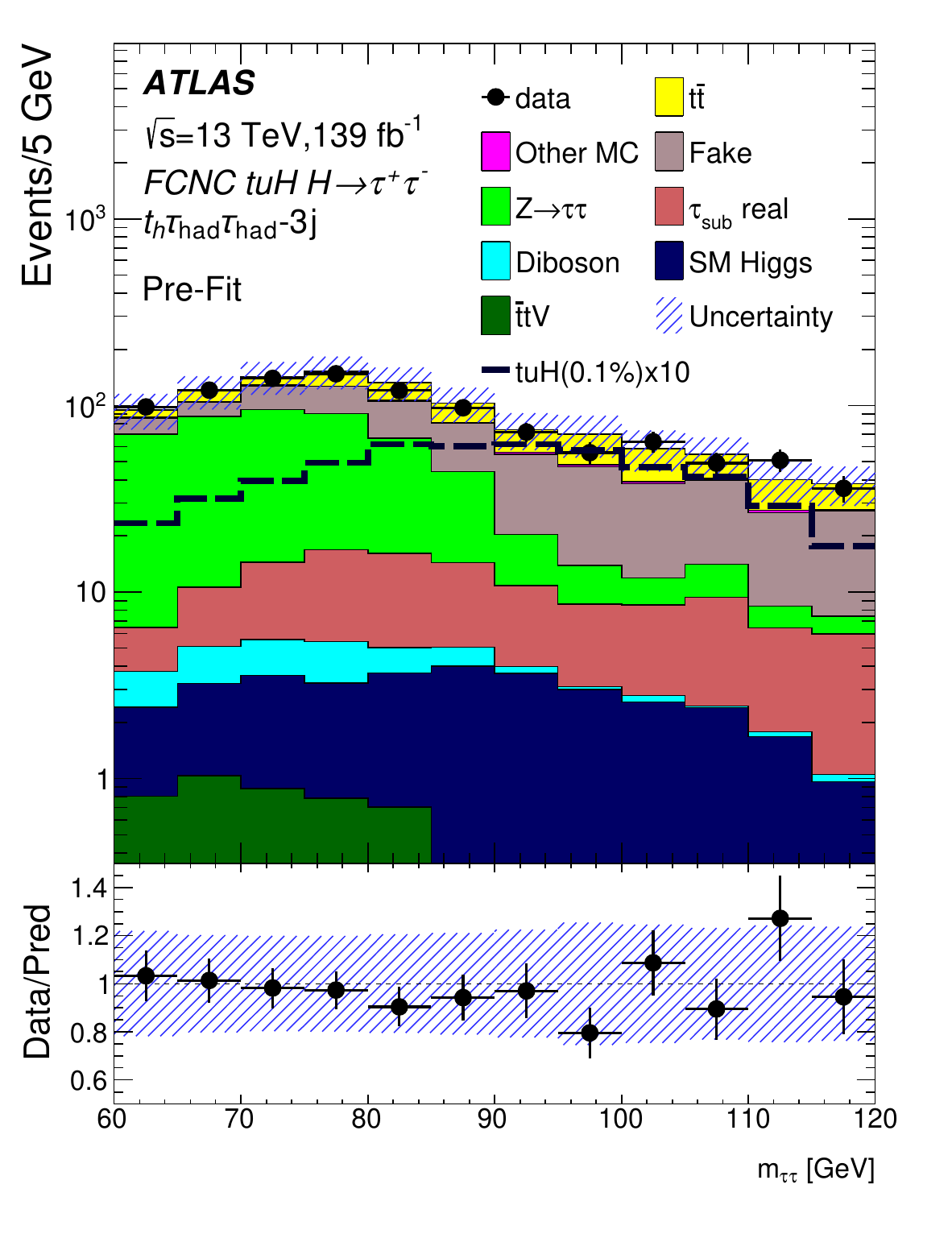}&
\includegraphics[page=1,width=0.28\textwidth]{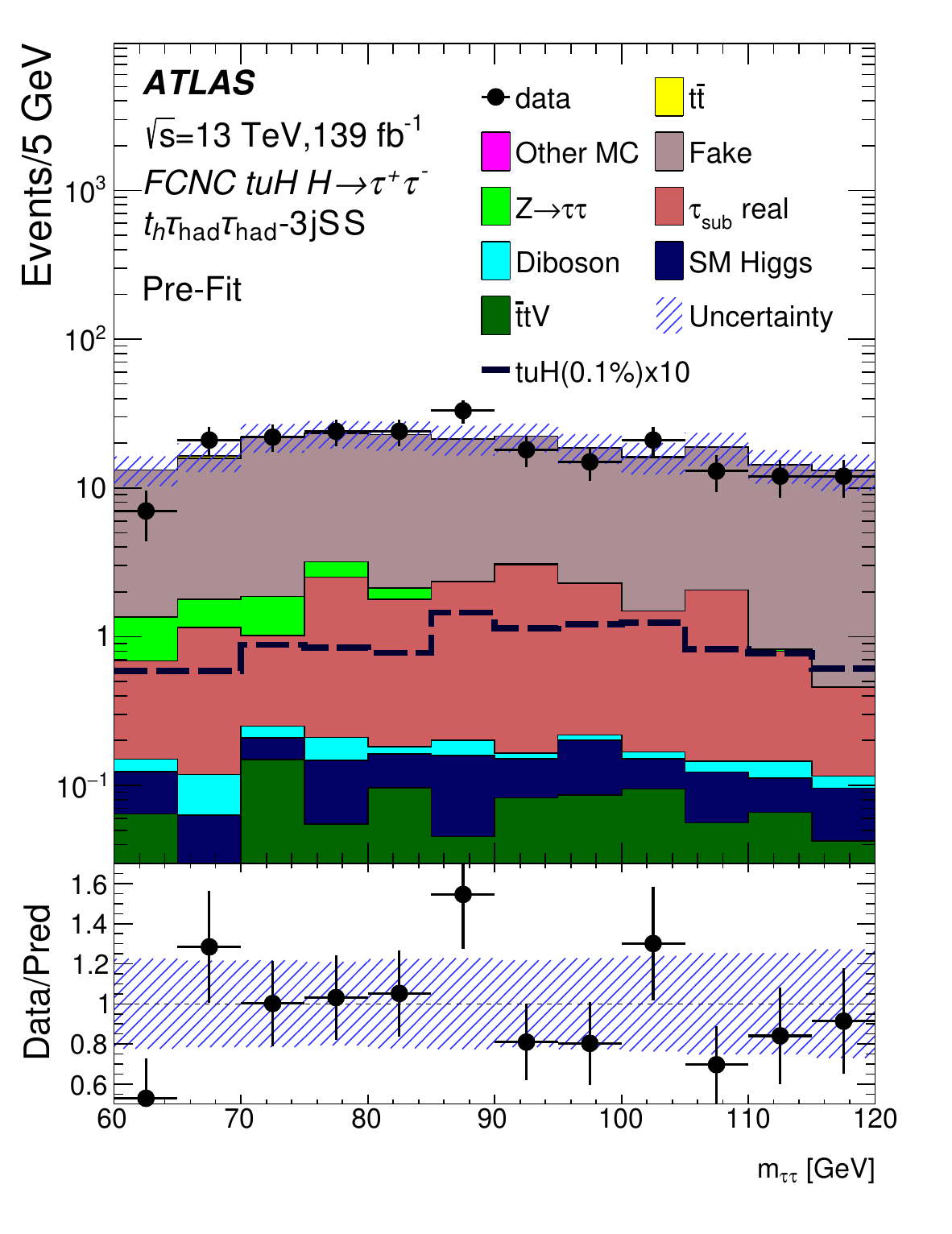}\\
(g) & (h) & (i)\\
\end{tabular}
\caption{Distributions of the visible invariant mass of the di-$\tau$ system ($m_{\tau\tau}$) obtained before the fit to data (`Pre-Fit') showing
the expected background and $tuH$ signals after applying fake factors in the following regions: (a) $t_{\ell}\thadhad$,
(b) $t_{\ell}\tauhad$-1j,  (c) $t_{\ell}\tauhad$-2j, (d) $t_h\tlhad$-2j, (e) $t_h\tlhad$-3j, (f) $t_{\ell}\thadhad$-SS, (g) $t_h\thadhad$-2j, (h) $t_h\thadhad$-3j and (i) $t_h\thadhad$-3j SS.
The total statistical and systematic uncertainty of the background prediction is indicated by the hatched band.
Overflow events are included in the last bin. `Other MC' includes single-top, and $V$+jets and other small backgrounds in the leptonic and hadronic channels. The $tuH$ signal is scaled by a normalisation factor of either 2, 10, or 50.
The lower panels show the ratio of data to prediction.}
\label{fig:mtt}
\end{figure}


\section{Systematic uncertainties}
\label{sec:systematics}
 
Several sources of systematic uncertainty that can affect the normalisation of signal
and background and/or the shape of their corresponding discriminant distributions are considered.
Each source is considered to be uncorrelated with the other sources.
Correlations of a given systematic uncertainty are maintained across processes and channels
as appropriate. The following sections describe the systematic uncertainties considered.
Table~\ref{tab:had_sys_impact} shows a summary of the dominant systematic uncertainties in the measured $\BR(t\to qH)$ branching ratio
resulting from the fits to the data in the signal regions as described in Section~\ref{sec:result}.
 
\subsection{Luminosity}
 
The uncertainty in the integrated luminosity is 1.7\%, affecting the overall normalisation of all processes estimated from the simulation.
It is derived, following a methodology similar to that detailed in Ref.~\cite{Aaboud:2016hhf} and using the LUCID-2 detector
for the baseline luminosity measurements \cite{Avoni:2018iuv}, from a calibration of the luminosity scale using $x$--$y$ beam-separation scans.
 
\subsection{Reconstructed objects}
 
Uncertainties associated with electrons, muons, and $\had$ candidates arise from the trigger, reconstruction,
identification and, in the case of electrons and muons, isolation efficiencies, as well as the momentum scale and resolution.
These are measured using $Z\to \ell^+\ell^-$ and $J/\psi\to \ell^+\ell^-$ events ($\ell =e, \mu$)~\cite{PERF-2017-01,Aad:2016jkr}
in the case of electrons and muons, and using $Z\to \tau^+\tau^-$ events in the case of $\had$ candidates~\cite{ATLAS-CONF-2017-029}.
 
Uncertainties associated with jets arise from the jet energy scale (JES)
and resolution (JER), and the efficiency to pass the JVT requirements.
The largest contribution comes from the jet energy scale, whose uncertainty dependence on jet $\pt$ and $\eta$, jet flavour, and pile-up treatment
is split into 43 uncorrelated components that are treated independently~\cite{Aaboud:2017jcu}. The total JES uncertainty is
below 5\% for most jets and below 1\% for central jets with $\pt$ between 300~\GeV\ and 2~\TeV. The difference between the JER values
in data and MC events is represented by one nuisance parameter (NP). It is applied to the MC events by smearing the jet $\pt$ within the prescribed uncertainty.
 
Uncertainties associated with energy scales and resolutions of leptons and jets
are propagated to $\met$. Additional uncertainties originating from the modelling
of the underlying event, in particular its impact on the $\pt$ scale and resolution
of unclustered energy, are negligible.
 
Efficiencies to tag $b$-jets and $c$-jets in the simulation are corrected by $\pt$-dependent factors to match the efficiencies in data,
while the light-jet efficiency is scaled by $\pt$- and $\eta$-dependent factors.
The $b$-jet efficiency is measured in a data sample enriched in $\ttbar$ events~\cite{FTAG-2018-01}, 
while the $c$-jet efficiency is measured
using $\ttbar$ events~\cite{ATLAS-CONF-2018-001} or $W$+$c$-jet events~\cite{Aad:2015ydr}.
The light-jet efficiency is measured in a multi-jet data sample enriched in light-flavour jets~\cite{ATLAS-CONF-2018-006}.
The uncertainties in these scale factors come from a total of 44 independent sources affecting $b$-jets, 19 sources affecting $c$-jets, and 19 sources affecting light jets.
These systematic uncertainties are taken to be uncorrelated between $b$-jets, $c$-jets, and light jets.
Scale factors are applied to reweight simulated events in order to obtain the pile-up distribution corresponding to the data.
An uncertainty in these reweighting scale factors is estimated by reweighting the profile in data while keeping it within its uncertainties.
 
\subsection{Background modelling}
\label{sec:syst_bkgmodeling}
 
A number of sources of systematic uncertainty affecting the modelling of $t\bar{t}$+jets are considered: the choice of renormalisation and factorisation scale in the matrix-element calculation, the choice of matching scale when matching the matrix elements to the parton shower generator, the uncertainty in the value of $\alpha_\text{s}$ when modelling initial-state radiation (ISR), and
the choice of renormalisation scale when modelling final-state radiation (FSR).
 
The $h_{\mathrm{damp}}$ parameter, which controls the amount of radiation produced by the parton shower in
\powheg\,v2, is set to 1.5$m_t$ in the $t\bar t$ sample. An alternative sample was generated with $h_{\mathrm{damp}}=3m_t$. The
difference between the two samples is treated as a systematic uncertainty called `$t\bar{t}$ $h_{\mathrm{damp}}$'.
 
The uncertainty due to the choice of parton shower and hadronisation (PS \& Had) model is derived
by comparing the predictions from {\powheg} interfaced either to \textsc{Pythia\,8} or \textsc{Herwig\,7}.
The latter uses the \textsc{MMHT2014lo}~\cite{Harland-Lang:2014zoa} PDF set in combination with the H7UE tune~\cite{Bellm:2015jjp}.
The uncertainty in the modelling of additional radiation from the PS is assessed by
varying the corresponding parameter of the A14 set~\cite{ATL-PHYS-PUB-2016-004} and by varying the radiation renormalisation and factorisation scales
by a factor of 2.0 and 0.5, respectively.

Another significant background in the hadronic channel stems from the $Z\rightarrow \tau\tau$ samples. Several sources of uncertainty are
considered for these samples: the PDF variation, which is evaluated by considering the standard deviation
produced by the event weights from the 100 NNPDF replicas for the \textsc{NNPDF3.0nnlo}~\cite{Ball:2015NNPDF} PDF set used in \sherpa,
the renormalisation ($\mu_\text{r}$) and factorisation ($\mu_\text{f}$) scales, the jet-to-parton matching uncertainty, the resummation scale uncertainty,
the variation in the choice of $\alpha_\text{s}$, and the use of alternative PDFs, which is evaluated by comparing predictions from the \textsc{NNPDF3.0nnlo} PDF set (nominal)
with those from the \textsc{MMHT2014nnlo68cl} and \textsc{CT14nnlo}~\cite{Lai:2010vv,Gao:2013xoa} PDF sets.
 
Uncertainties affecting the normalisation of the $V$+jets background are estimated separately for $V$+light-jets, $V$+$\geq$1$c$+jets,
and $V$+$\geq$1$b$+jets subprocesses. The total normalisation uncertainty of $V$+jets processes is estimated to be approximately 30\% by taking the maximum
difference between the data and the total background prediction in the different analysis regions considered, but requiring exactly zero, one, and two $b$-jets.
This is driven mainly by differences between MC simulations in the $V$+jets regions with a high multiplicity of jets.

Uncertainties affecting the modelling of the single-top-quark background include a
$+5\%$/$-4\%$ uncertainty in the total cross section, which is estimated as a weighted average
of the theoretical uncertainties in $t$-, $tW$- and $s$-channel production~\cite{Kidonakis:2011wy,Kidonakis:2010ux,Kidonakis:2010tc}.
Additional uncertainties associated with the parton shower, hadronisation and ISR/FSR are also considered by using the same procedure
as for $t\bar t$. Uncertainties in the diboson background normalisation include those estimated from variations of the renormalisation and factorisation scales,  \textsc{NNPDF3.0nnlo} variations and $\alpha_\text{s}$ variations. Uncertainties in the $\ttbar V$ and $\ttbar H$ cross sections are estimated to be 12\% and [$+5.8\%/-9.2\%$], respectively,
from the uncertainties in their respective NLO theoretical cross sections~\cite{TOPQ-2018-08}.

The statistical uncertainties of the fake-$\tau$-background calibration in the leptonic channels are applied with uncorrelated uncertainties for
different sources of fake $\tau$-leptons and different $\pt$ slices. The uncertainties in the MC modelling of the various processes used for
evaluation of the fake factors in the CRtt are treated as fully correlated in the fit.
The uncertainty in the ABCD method is applied to the normalisation factors for muons and electrons, including both
its statistical fluctuation and the differences between the normalisation factors for the various leptonic channels.
The uncertainties in the fake-factor method applied to the hadronic channel includes the statistical uncertainty of each fake factor and
differences between the sets of fake factors derived from different signal-depleted CRs.
 
\subsection{Signal modelling}
 
Several normalisation and shape uncertainties are taken into account and are treated as fully correlated for the $\Hq$ and $pp\to tH$ signals.
Uncertainties in the Higgs boson branching ratios are taken
into account by following the recommendation in Ref.~\cite{deFlorian:2016spz}.
The uncertainties due to ISR, FSR, the scales, and the PDFs, are considered and are treated as fully correlated in all SRs. The parton shower uncertainties are estimated by comparing
the nominal sample with an alternative sample interfaced with \textsc{Herwig\,7}.


\section{Statistical analysis}
\label{sec:stat_analysis}

The final discriminant distributions are the BDT outputs from all the considered analysis regions. They are jointly analysed for the
presence of a signal. The BDT distributions are binned to maximise sensitivity to the signal.
The statistical analysis uses a binned likelihood function ${\cal L}(\mu,\theta)$ constructed as
a product of Poisson probability terms over all bins considered in the search. This function depends
on the signal-strength parameter $\mu$, defined as a factor multiplying the expected yield of $tH$ and $tt(qH)$ signal events
normalised to a reference branching ratio $\BR_{\mathrm{ref}}(t\to qH)=0.1\%$,
and on $\theta$, a set of nuisance parameters that encode the effect of systematic uncertainties on the signal and background expectations.
The $pp\rightarrow tH$ production cross section is related to $\BR(t\rightarrow qH)$ through their dimension-6 operators~\cite{fcnc_production_theory}.
It follows that the expected total number of events in a given bin depends on $\mu$ and $\theta$.
All nuisance parameters are subject to Gaussian constraints in the likelihood function.
For a given value of $\mu$, the nuisance parameters $\theta$ allow variations of the expectations for signal and background
consistent with the corresponding systematic uncertainties, and their fitted values result in the deviations from
the nominal expectations that globally provide the best fit to the data.
This procedure reduces the impact of systematic uncertainties on
the search's sensitivity by taking advantage of the highly populated background-dominated bins included in the likelihood fit.
Statistical uncertainties in each bin of the predicted final discriminant distributions are taken into account through dedicated parameters in the fit.
The best-fit $\BR(t\to qH)$ is obtained by performing a binned likelihood fit to the data under the signal-plus-background
hypothesis, and maximising the likelihood function ${\cal L}(\mu,\theta)$ over $\mu$ and $\theta$.
 
The fitting procedure was initially validated through extensive studies using pseudo-data (which is defined as the sum of all predicted backgrounds
plus an injected signal of variable strength) as well as by performing fits to real data where bins of the final discriminating variable with
an expected signal contamination above 10\%, assuming $\BR_{\mathrm{ref}}(t\to qH)=0.1\%$, are excluded (referred to as `blinding' requirements).
In both cases, the robustness of the model with respect to systematic uncertainties is established by verifying the stability of the fitted background
when varying assumptions about some of the leading sources of uncertainty.
After this, the blinding requirements
are removed in the data and a fit under the signal-plus-background hypothesis is performed. Further checks involve the comparison of the fitted
nuisance parameters before and after removal of the blinding requirements, and their values are found to be consistent. In addition, it is verified that the
fit is able to determine the strength of a simulated signal injected into the real data.
 
The test statistic $q_\mu$ is defined as the profile likelihood ratio,
$q_\mu = -2\ln({\cal L}(\mu,{\hat{\hat{\theta}}}_\mu)/{\cal L}(\hat{\mu},\hat{\theta}))$,
where $\hat{\mu}$ and $\hat{\theta}$ are the values of the parameters that
maximise the likelihood function (subject to the constraint $\hat{\mu} \geq 0$), and the ${\hat{\hat{\theta}}}_\mu$ are the values of the
nuisance parameters that maximise the likelihood function for a given value of $\mu$.
The test statistic $q_\mu$ is evaluated with the {\textsc RooFit} package~\cite{Verkerke:2003ir,RooFitManual}.
 
Exclusion limits are set on $\mu$ and thus on
$\BR(t\to qH)$, derived by using $q_\mu$ in the CL$_{\textrm{s}}$ method~\cite{Junk:1999kv,Read:2002hq}.
For a given signal scenario, values of $\BR(t\to qH)$ yielding CL$_{\textrm{s}} < 0.05$,
where CL$_{\textrm{s}}$ is computed using the asymptotic approximation~\cite{Cowan:2010js}, are excluded with at least 95\% confidence.


\section{Results}
\label{sec:result}
 
This section presents the results obtained from the individual channels, as well as their combination,
by following the statistical analysis discussed in Section~\ref{sec:stat_analysis}.
 
A binned likelihood fit under the signal-plus-background hypothesis is performed on the BDT discriminant distributions in the seven
signal regions. The unconstrained parameter of the fit is the signal strength.
No significant pulls or constraints are obtained for the fitted nuisance parameters, resulting in a post-fit background prediction in each analysis region that is
very close to the pre-fit prediction, albeit with reduced uncertainties
resulting from the fit.
Figures~\ref{fig:asimov_postfitbdtHu} and~\ref{fig:asimov_postfitbdtHc} show the BDT output distributions after the signal-plus-background fit to the
data for the $tcH$ and $tuH$ searches, respectively.
The observed and predicted yields after a background-only fit to the data are summarised in Table~\ref{tab:HtautauPostfitYieldsUnblind}.
A slight excess of data events, with a significance of 2.3$\sigma$, is observed above the expected background. This is mainly in the high BDT-score region of the most sensitive channel, $t_{\ell}\thadhad$, as shown in Figures~\ref{fig:asimov_postfitbdtHu}(a) and~\ref{fig:asimov_postfitbdtHc}(a).
The kinematic distributions for the observed excess in
the high BDT-score region were checked. Within the large statistical uncertainty, the observed distributions are compatible with the background shapes, but also with a small signal contribution. There is no indication that the excess is from a specific data period.
The background modelling in this signal region was also checked using the VR in which both $\thad$ candidates have the same charge.
In this VR the signal contribution is negligible in the highest BDT bins and the background shape is well reproduced by the data.

\begin{figure}[H]
\centering

\begin{tabular}{@{}ccc@{}}
\includegraphics[width=0.29\textwidth]{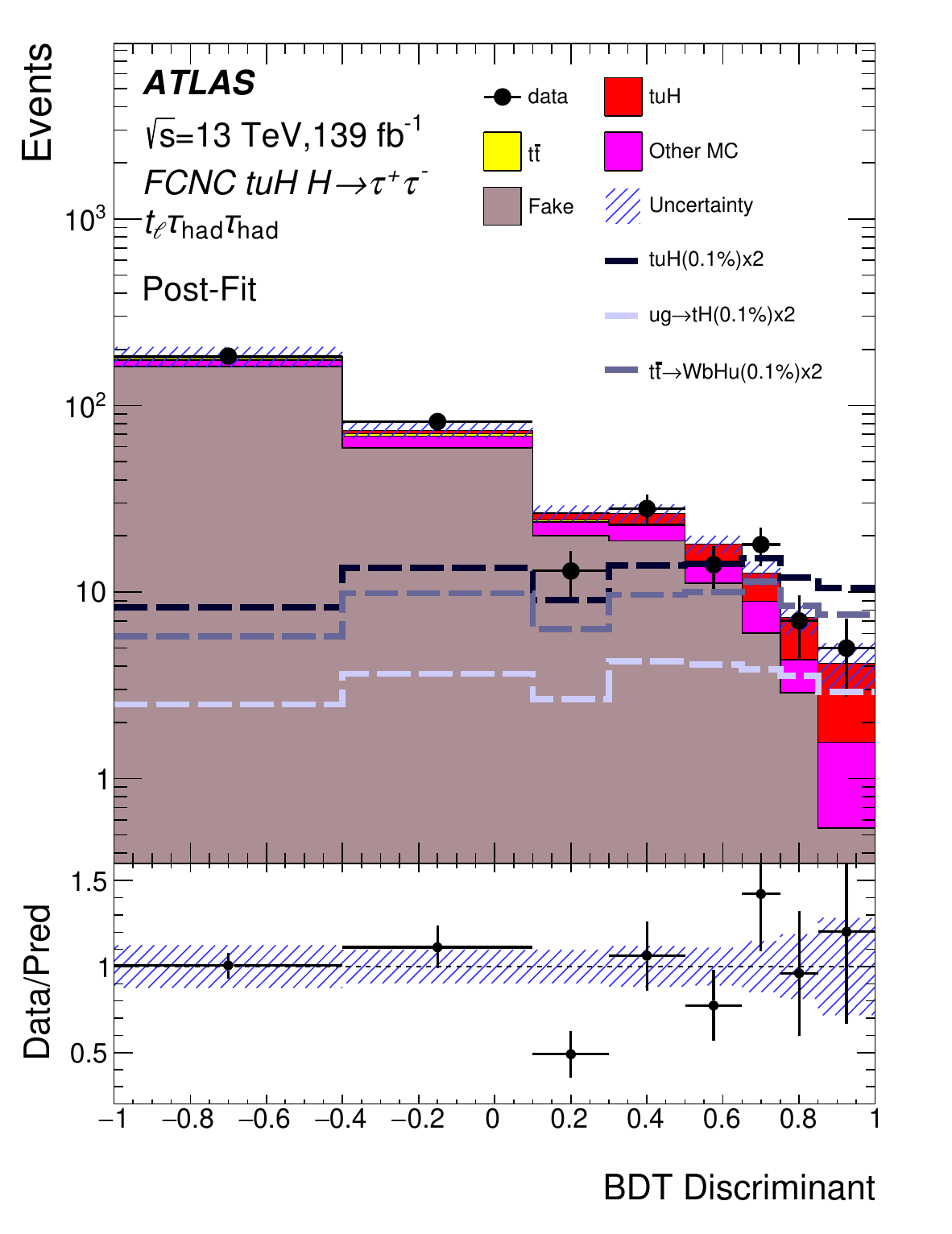}&
\includegraphics[width=0.29\textwidth]{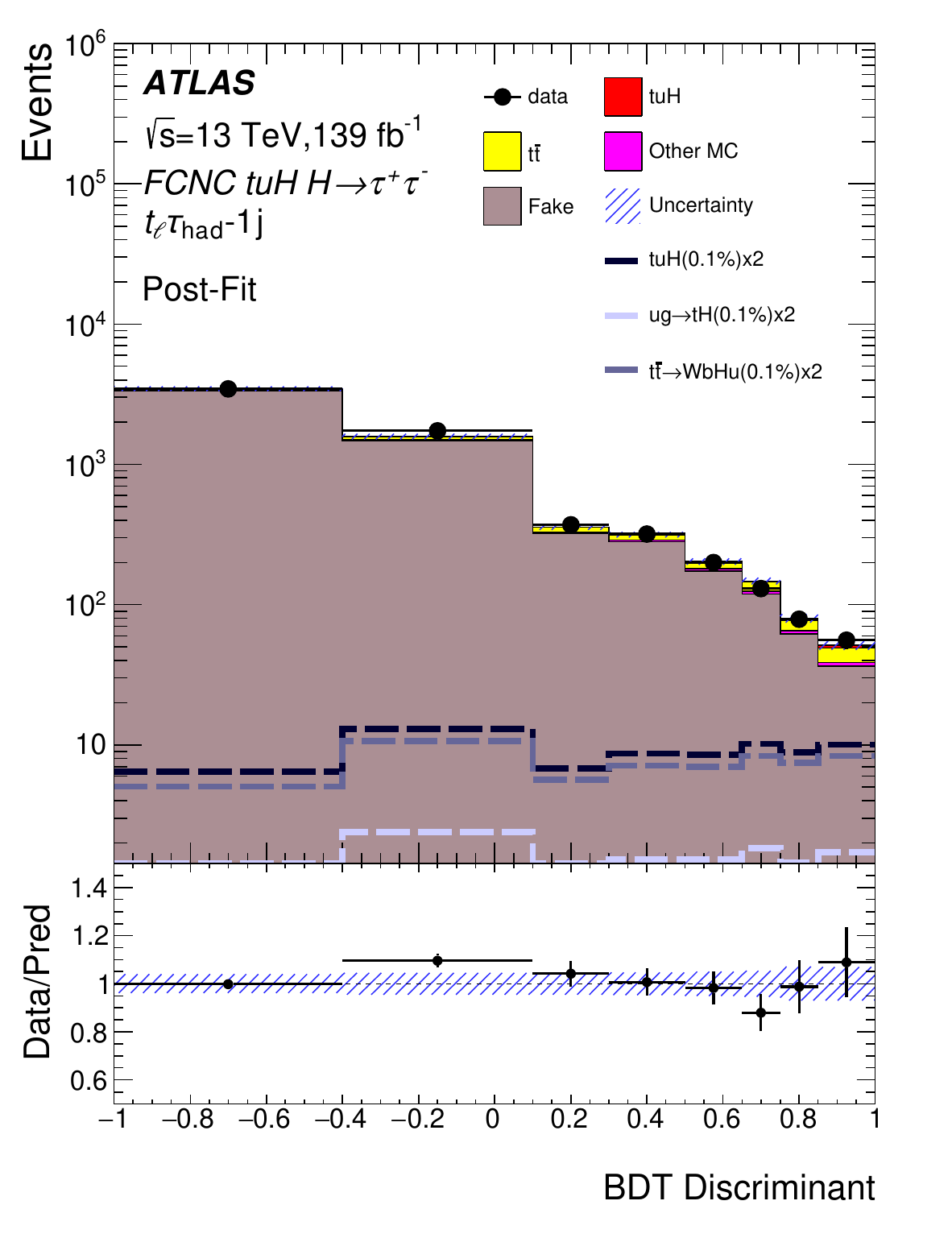}&
\includegraphics[width=0.29\textwidth]{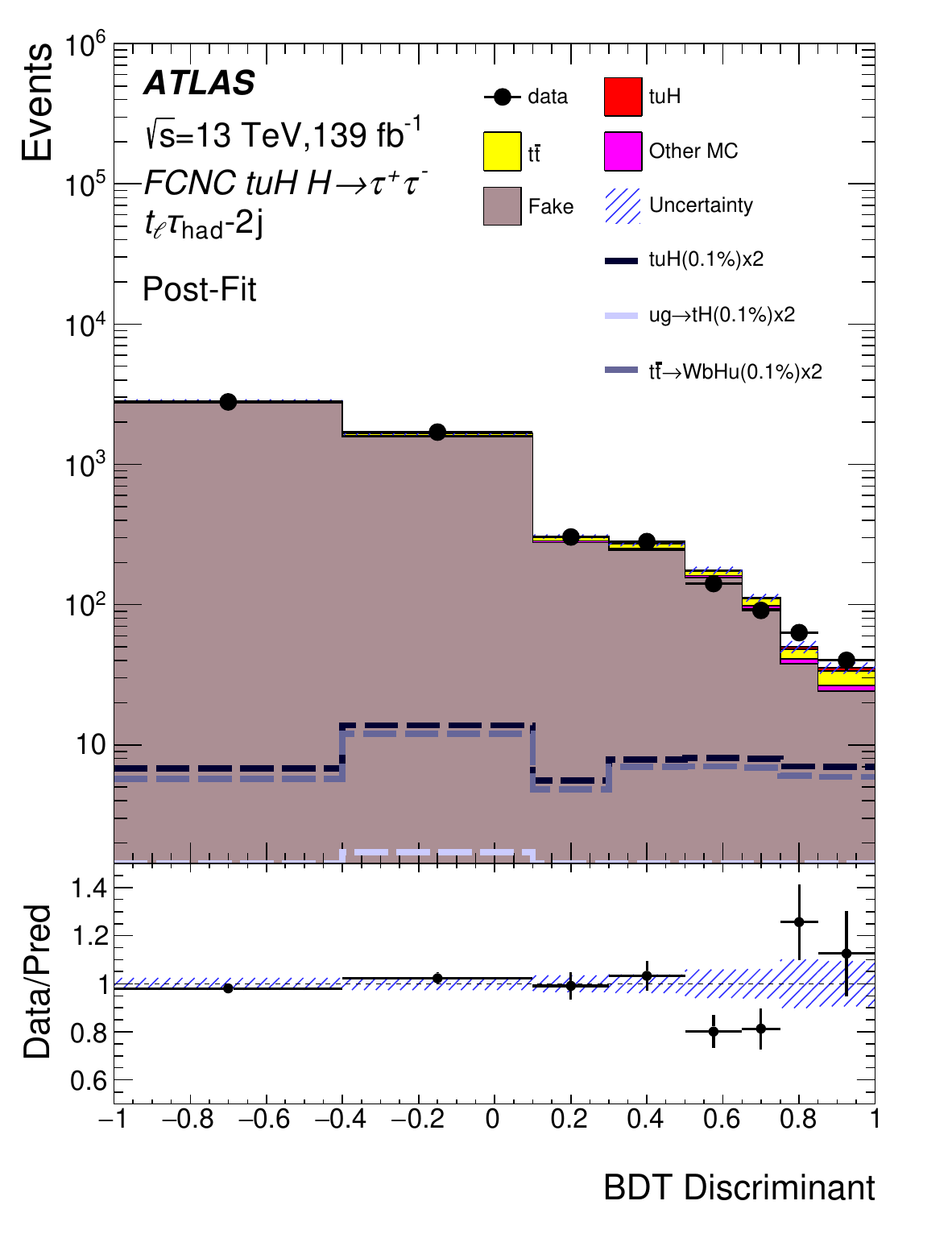}\\
(a)  & (b) & (c) \\
\includegraphics[width=0.29\textwidth]{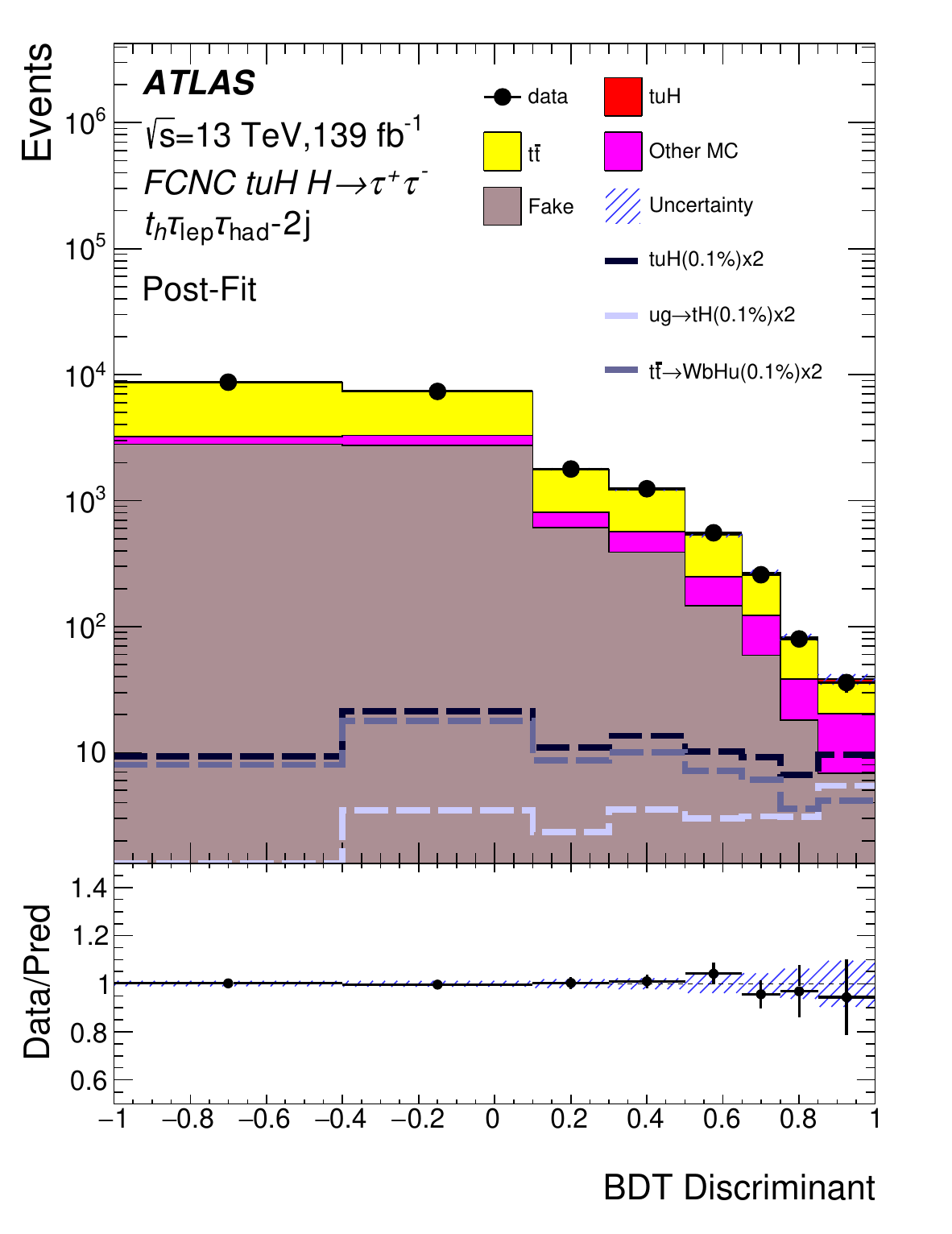}&
\includegraphics[width=0.29\textwidth]{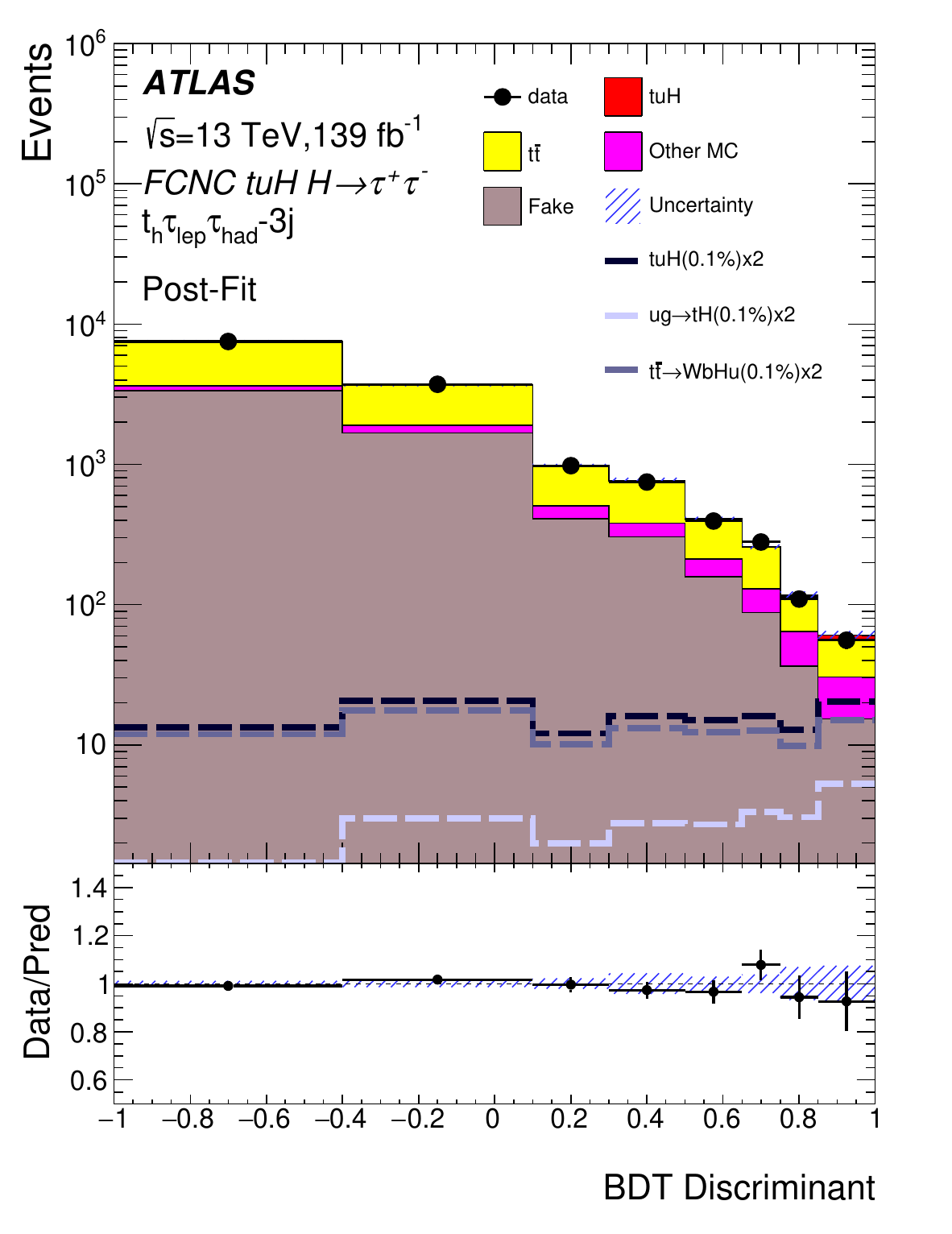}&
\includegraphics[width=0.29\textwidth]{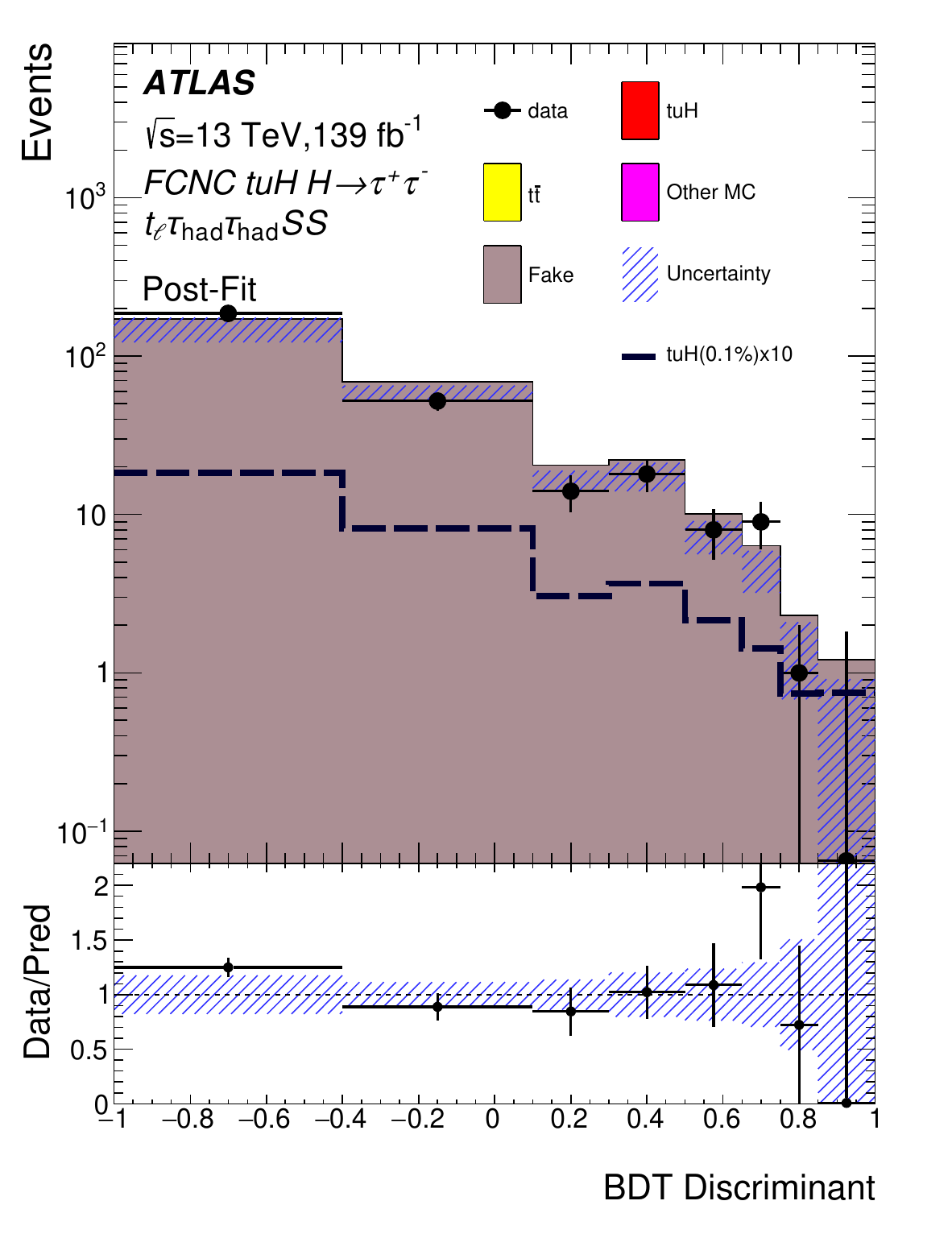}\\
(d) & (e)  & (f) \\
\includegraphics[width=0.29\textwidth]{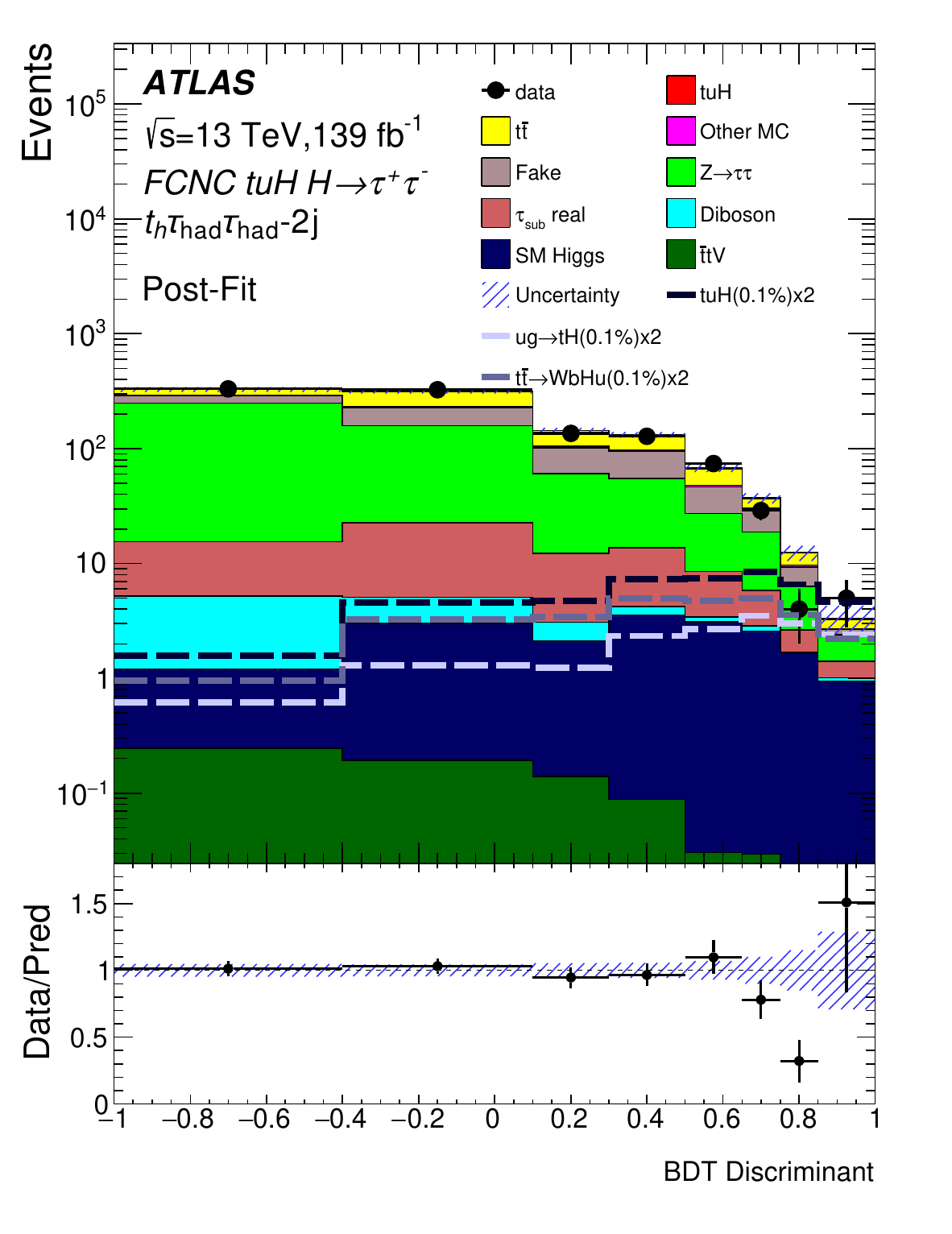}&
\includegraphics[width=0.29\textwidth]{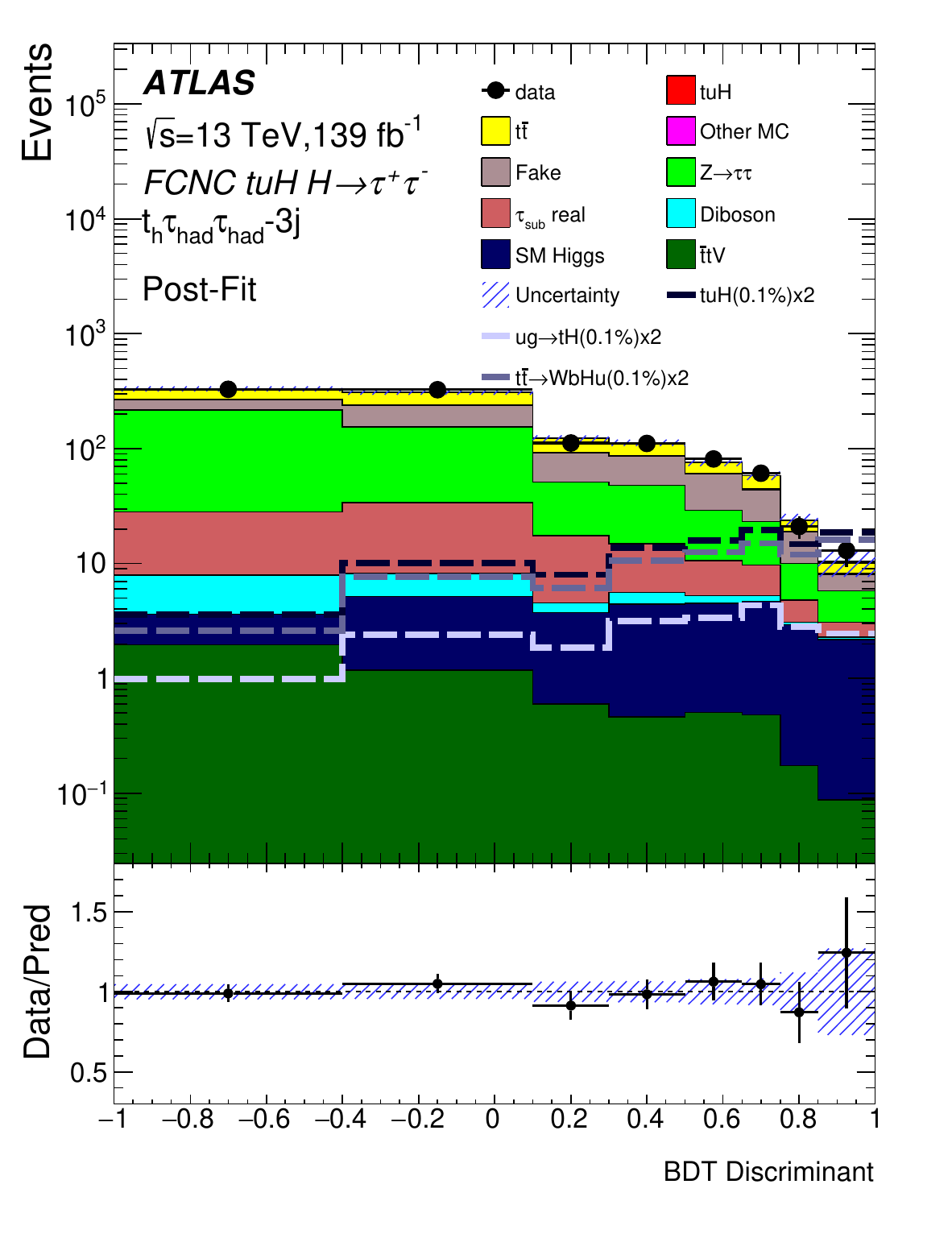}&
\includegraphics[width=0.29\textwidth]{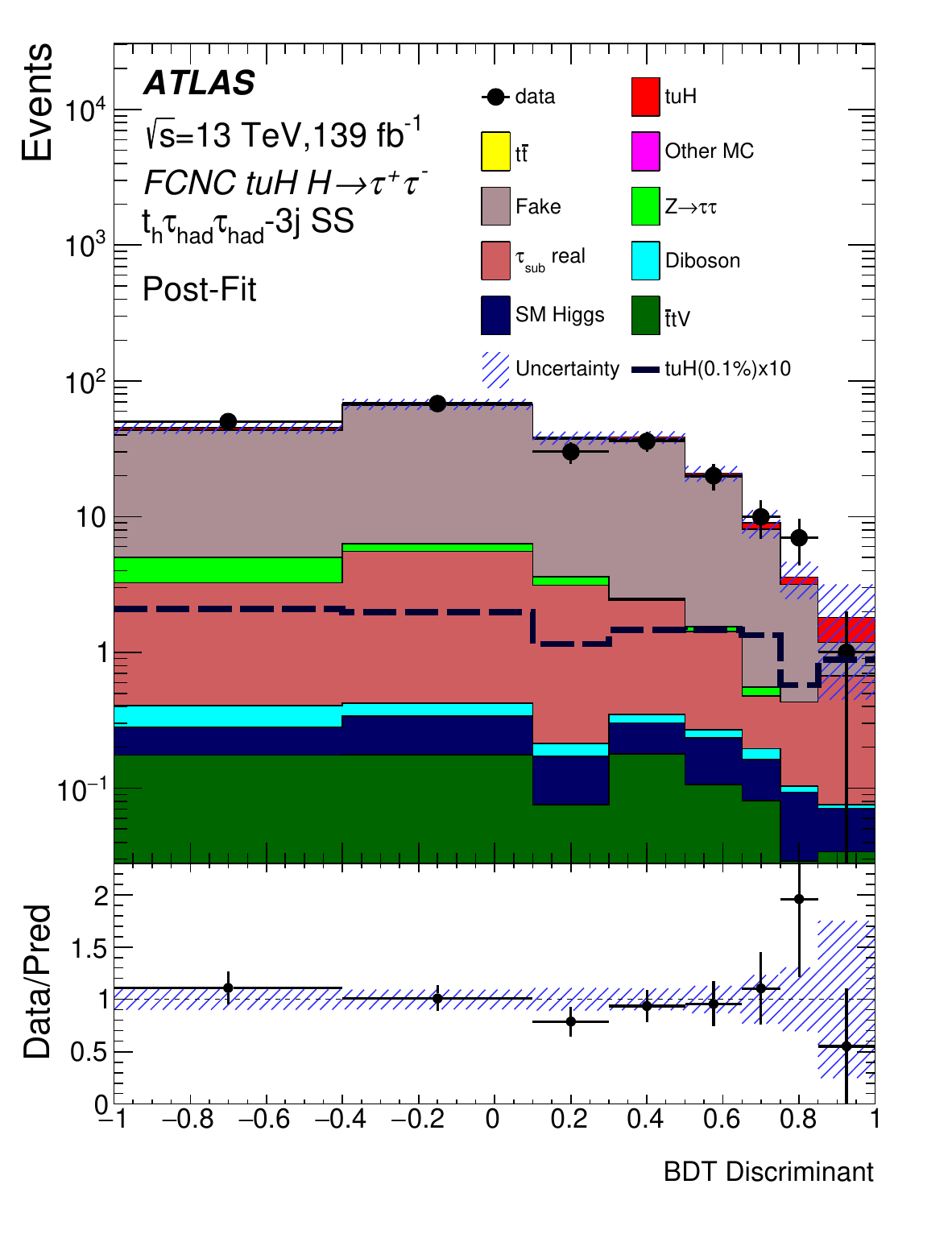}\\
(g) & (h)  & (i) \\
\end{tabular}
\caption{ BDT output distributions obtained from a signal+background fit to the data for the $tuH$ search:
(a) $t_{\ell}\thadhad$, (b) $t_{\ell}\tauhad$-1j,  (c) $t_{\ell}\tauhad$-2j, (d) $t_h\tlhad$-2j, (e) $t_h\tlhad$-3j, (f) $t_{\ell}\thadhad$-SS, (g) $t_h\thadhad$-2j, (h) $t_h\thadhad$-3j and (i) $t_h\thadhad$-3j SS.
The total statistical and systematic uncertainty is indicated by the hatched band. The signal shapes of $tt(uH)$, $tH$, and their sum are also shown using a normalisation of $2 \times\BR(t\to uH)$ of 0.1\%.
}
\label{fig:asimov_postfitbdtHu}
\end{figure}

\begin{figure}[H]
\centering
\begin{tabular}{@{}ccc@{}}
\includegraphics[width=0.29\textwidth]{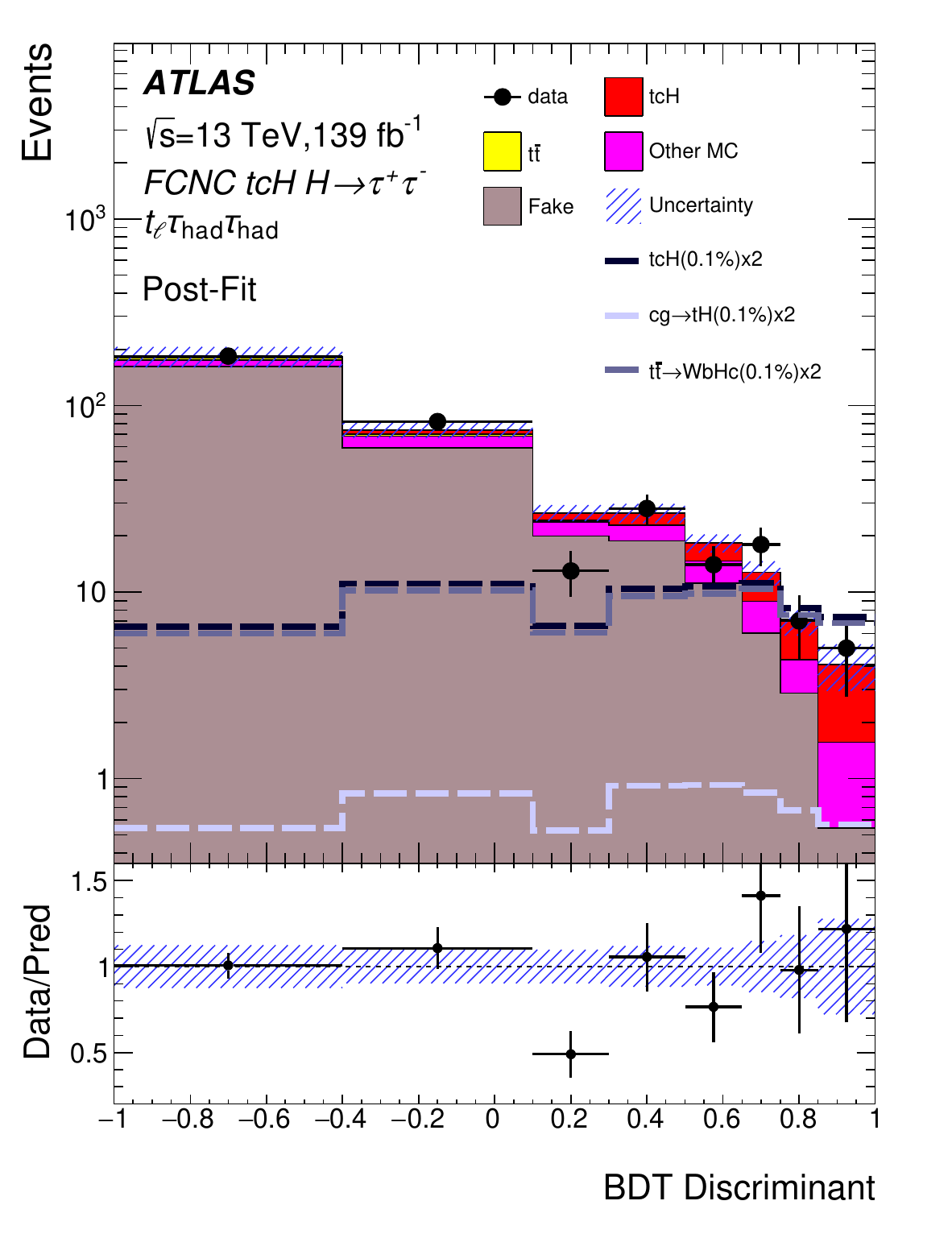}&
\includegraphics[width=0.29\textwidth]{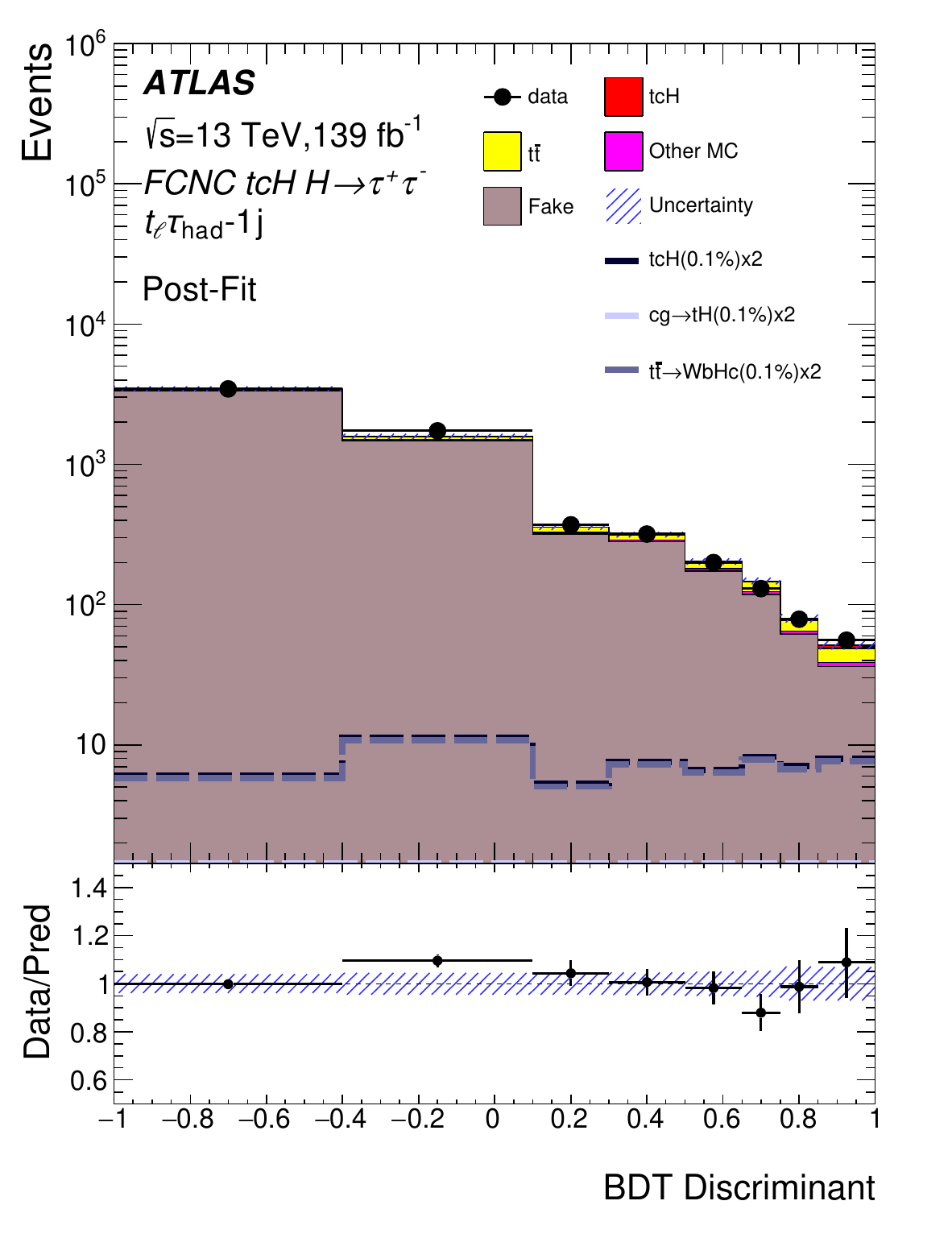}&
\includegraphics[width=0.29\textwidth]{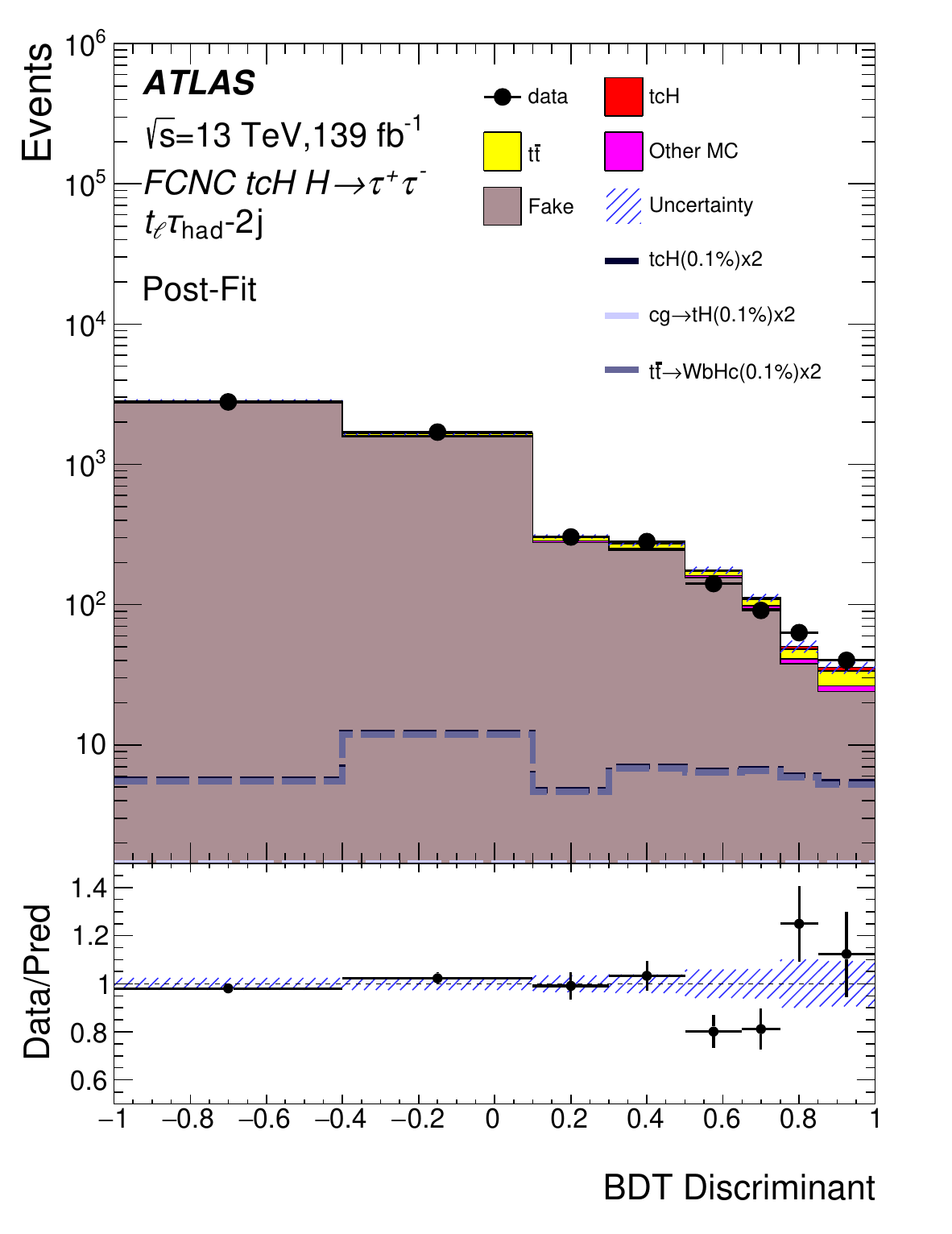}\\
(a)  & (b) & (c) \\
\includegraphics[width=0.29\textwidth]{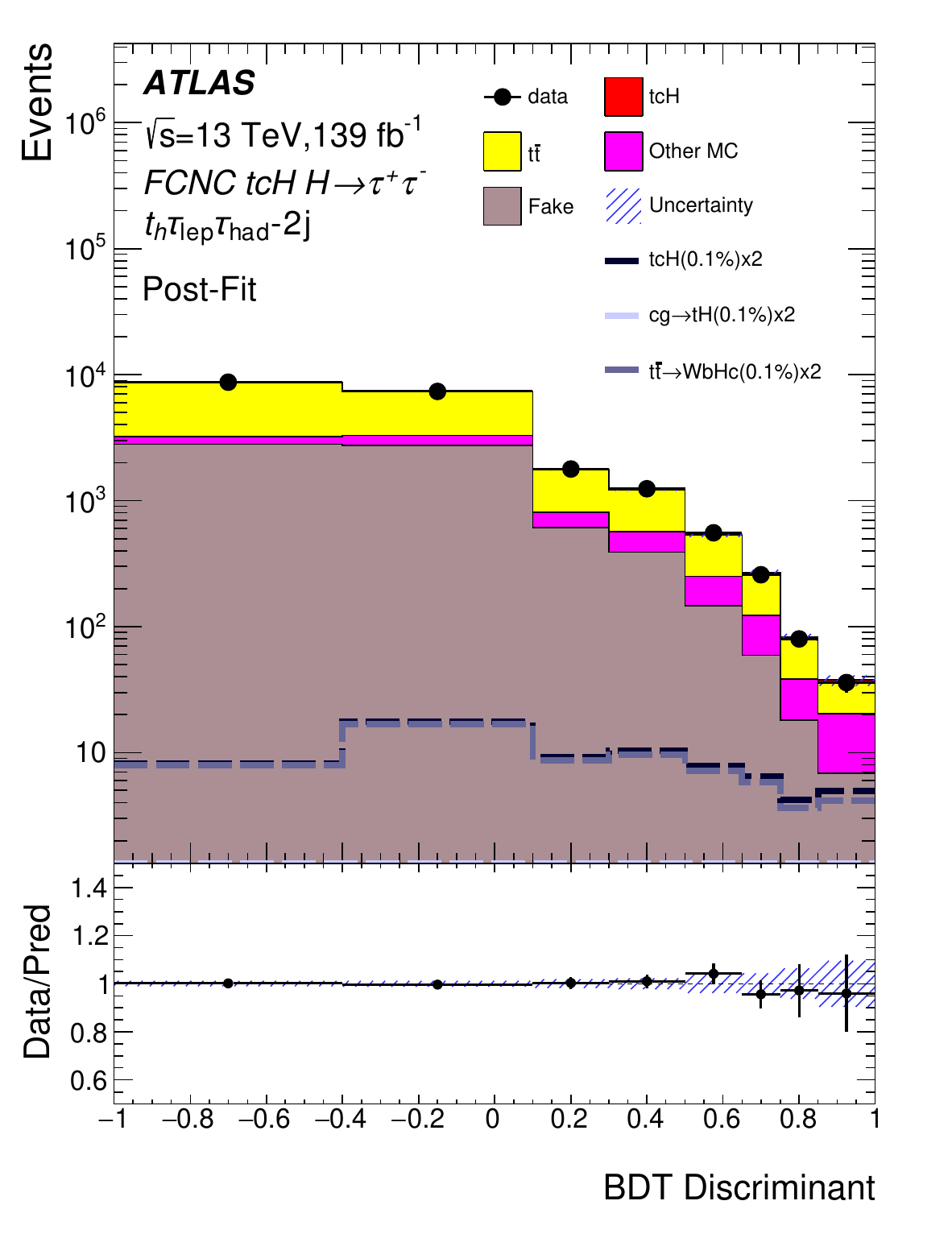}&
\includegraphics[width=0.29\textwidth]{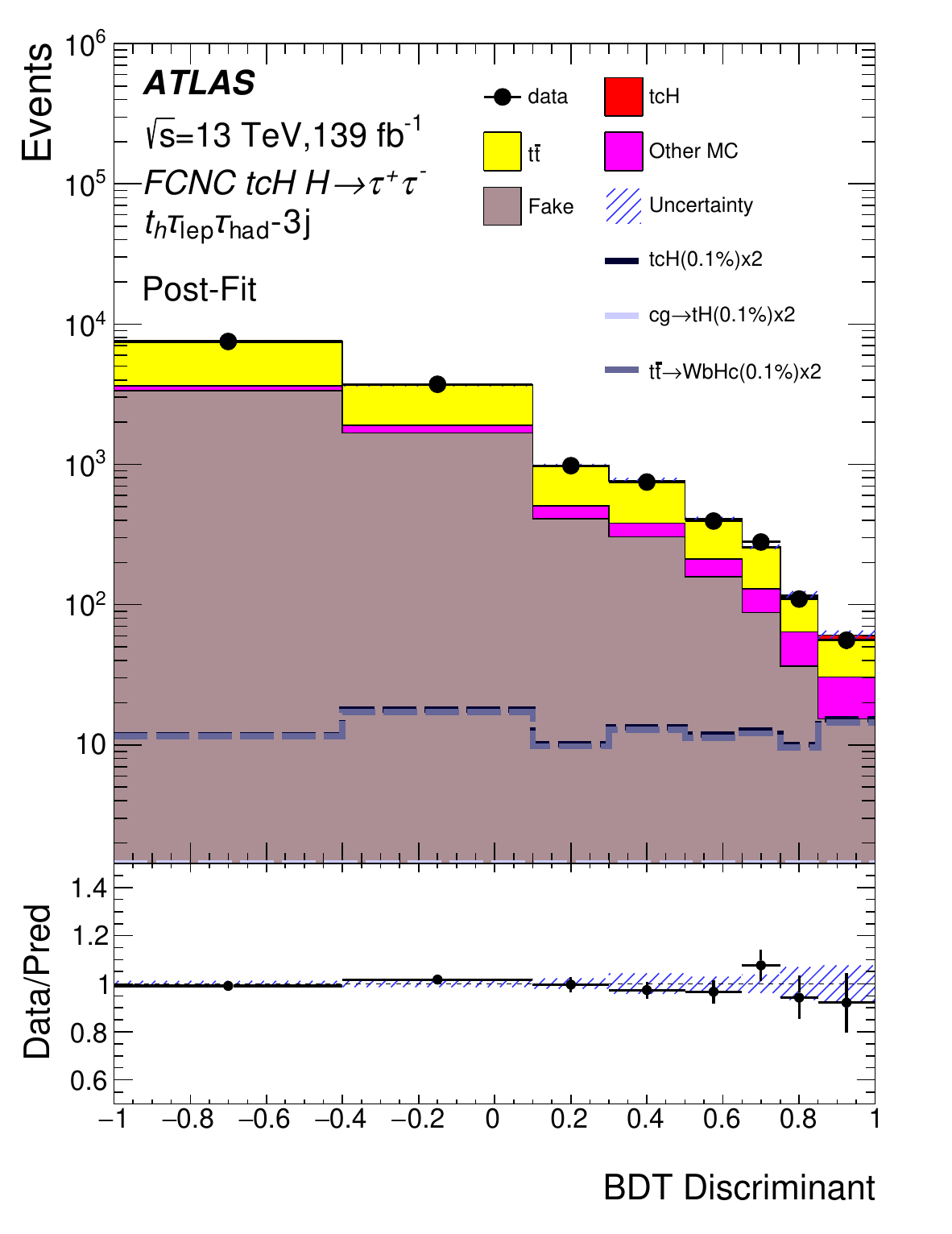}&
\includegraphics[width=0.29\textwidth]{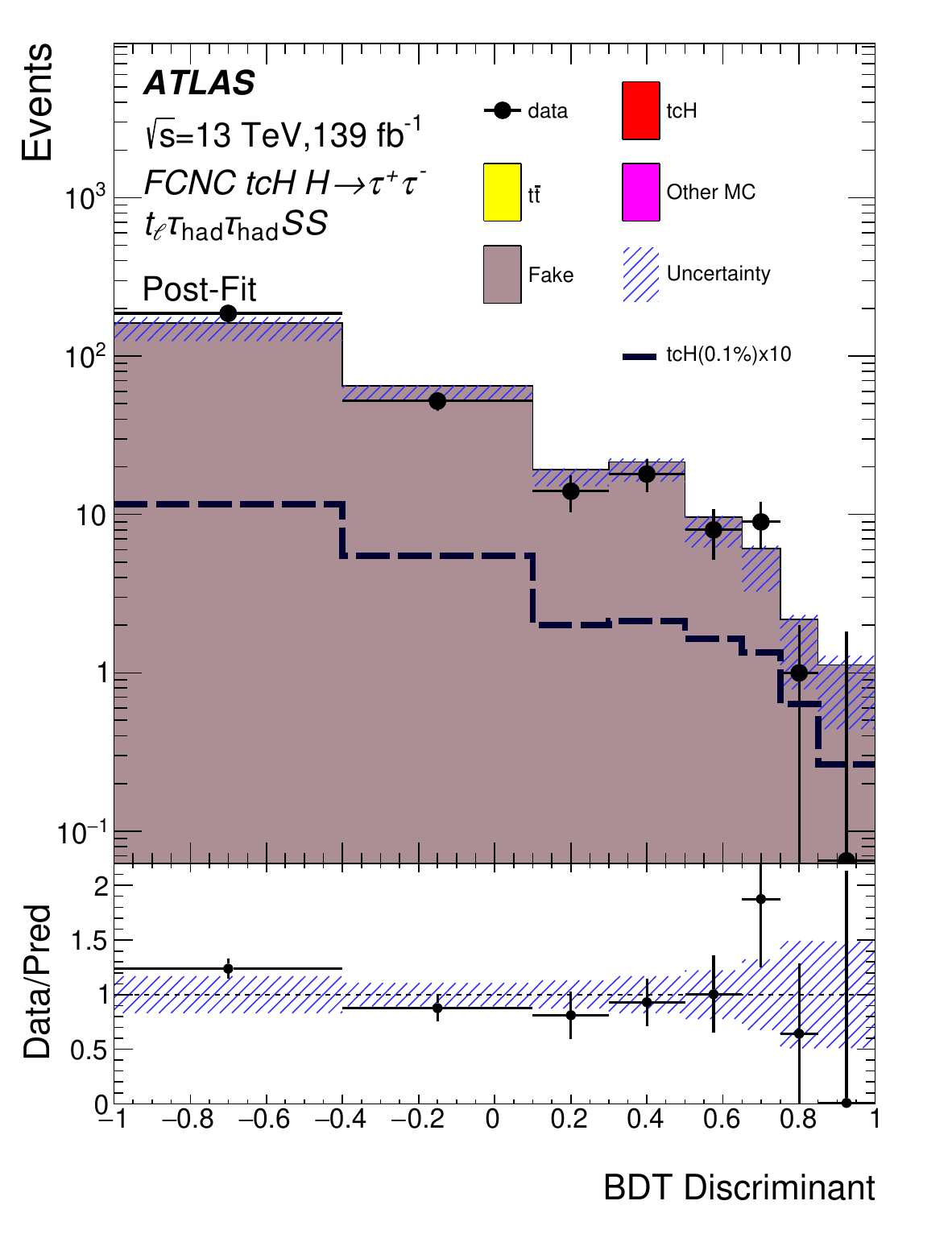}\\
(d) & (e)  & (f) \\
\includegraphics[width=0.29\textwidth]{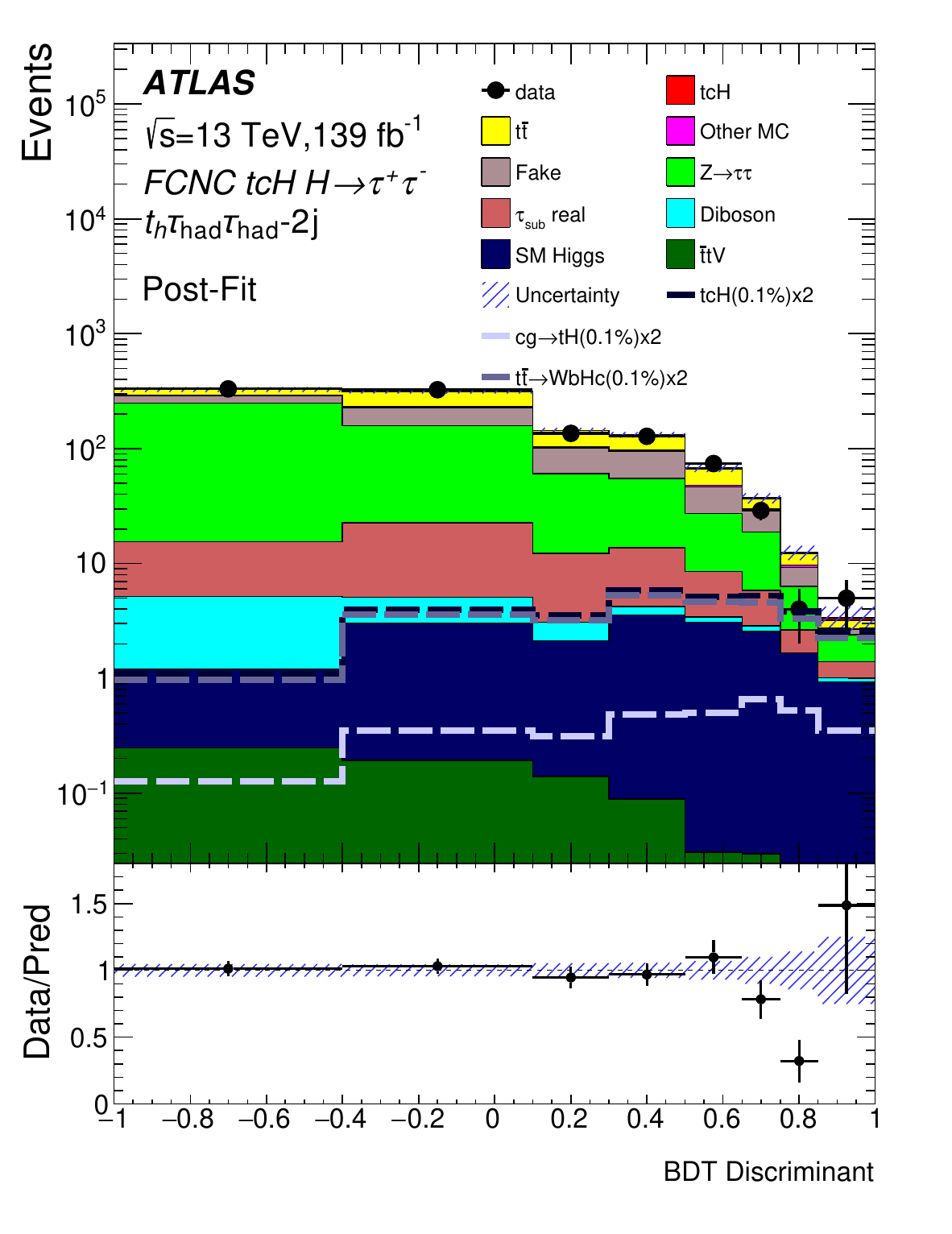}&
\includegraphics[width=0.29\textwidth]{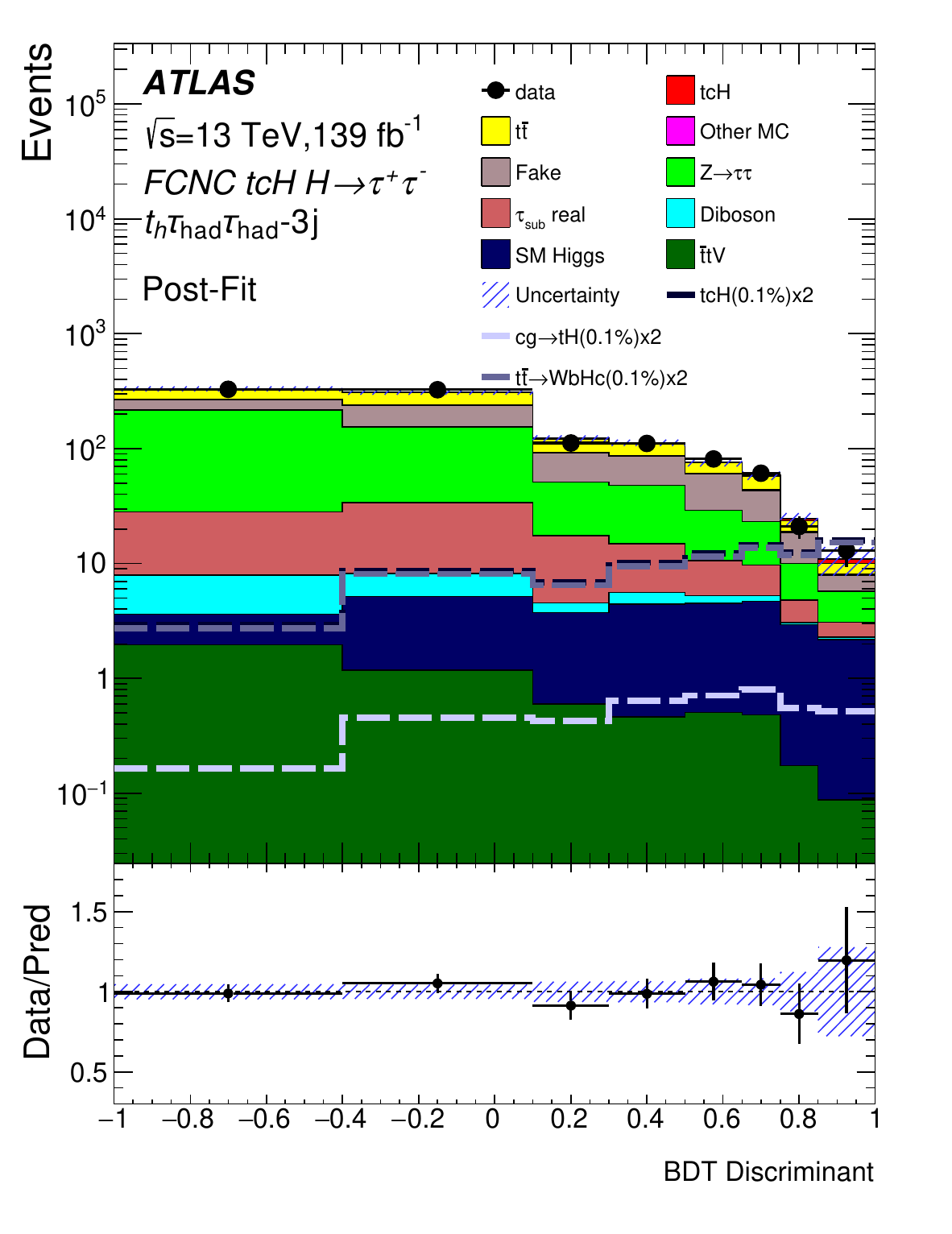}&
\includegraphics[width=0.29\textwidth]{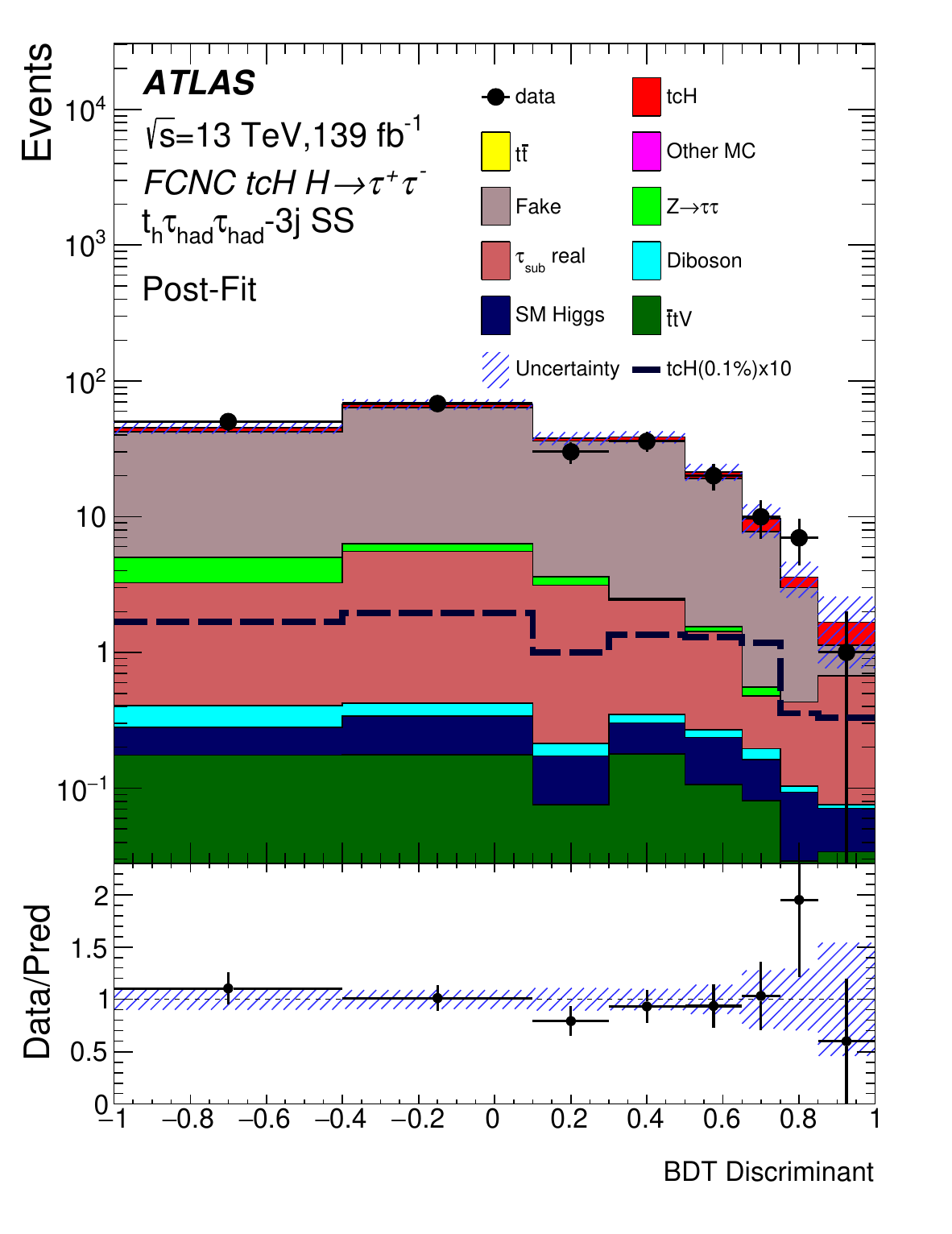}\\
(g) & (h)  & (i) \\
\end{tabular}
\caption{ BDT output distributions obtained from a signal+background fit to the data for the $tcH$ search:
(a) $t_{\ell}\thadhad$, (b) $t_{\ell}\tauhad$-1j,  (c) $t_{\ell}\tauhad$-2j, (d) $t_h\tlhad$-2j, (e) $t_h\tlhad$-3j, (f) $t_{\ell}\thadhad$-SS, (g) $t_h\thadhad$-2j, (h) $t_h\thadhad$-3j and (i) $t_h\thadhad$-3j SS.
The total statistical and systematic
uncertainty is indicated by the hatched band. The signal shapes of $tt(cH)$, $tH$, and their sum are also shown using a normalisation of $2 \times\BR(t\to cH)$ of 0.1\%.
}
\label{fig:asimov_postfitbdtHc}
\end{figure}

\begin{table}[htbp]
\caption{
Predicted and observed yields in each of the analysis regions considered. The background prediction is shown after a background-only fit is applied to the data.
Also shown are the expected signals for $\Hc$ and
$\Hu$ assuming $\BR(t\to cH)=0.1\%$ and $\BR(t\to uH)=0.1\%$ respectively. The contributions with real $\had$ candidates from $\ttbar$ and  $Z\to \ell^+\ell^-$ ($\ell = e, \mu$),
diboson, $\ttbar V$, $\ttbar H$, single-top-quark, and other small backgrounds are combined into a single background source referred to as `Other MC' in the leptonic channels,
whereas single-top-quark and the small contributions are combined into `Rare' in the hadronic channels.
The quoted uncertainties are the sum in quadrature of the statistical and systematic uncertainties of the yields.}
\small
\centering
 
\begin{tabular}{cccccc} \toprule\toprule
& $t_\ell\tauhad$-1j & $t_\ell\tauhad$-2j & $t_h\tlhad$-3j &$t_h\tlhad$-2j  & $t_\ell\thadhad$ \\\midrule
Double Fake            & $--       $  & $--       $  & $--       $  &  $--       $  & ~~$73 \pm 24    $ \\
$\bar{t}tV$            & ~~$9.3 \pm 1.2 $  & $22.6 \pm 2.8$  & $23.5 \pm 3.0$  &  $13.7 \pm 1.7$  & ~~$2.57 \pm 0.35$ \\
SM Higgs               & ~~$5.8 \pm 0.8 $  & $13.7 \pm 1.7$  & $32.8 \pm 3.5$  &  $13.5 \pm 2.5$  & $16.7 \pm 1.9 $ \\
Diboson                & $32.6 \pm 3.4$  & $19.9 \pm 2.1$  & $36 \pm 4    $  &  $46 \pm 5    $  & $13.2 \pm 1.4 $ \\
Other MC               & $35.6 \pm 3.1$  & $15.9 \pm 1.7$  & $226 \pm 21  $  &  $620 \pm 40  $  & ~~$6.7 \pm 0.6  $  \\
$Z\rightarrow\tau\tau$ & ~~$0 \pm 6     $  & ~~$9.1 \pm 2.2 $  & $500 \pm 60  $  &  $880 \pm 90  $  & ~~$2.1 \pm 0.7  $ \\
Lep.\ Fake               & $212 \pm 30  $  & ~~$80 \pm 10   $  & $292 \pm 26  $  &  $490 \pm 70  $  & ~~$0.9 \pm 0.4  $ \\
QCD Fake               & ~~$670 \pm 200 $  & $310 \pm 90  $  & $180 \pm 70  $  &  ~~$330 \pm 110 $  & $--        $  \\
$b$ Fake                 & ~~$960 \pm 140 $  & $1250 \pm 230$  & ~~$710 \pm 140 $  &  ~~$710 \pm 130 $  & ~~$82 \pm 13    $ \\
$W$-jet Fake             & ~~$970 \pm 200 $  & $1090 \pm 240$  & $3300 \pm 500$  &  $3800 \pm 600$  & ~~$5.5 \pm 1.8  $ \\
Other Fake             & $3020 \pm 260$  & $2470 \pm 160$  & $1420 \pm 220$  &  $1320 \pm 320$  & $129 \pm 14   $ \\
$\bar{t}t$             & $281 \pm 14  $  & $195 \pm 24  $  & $7100 \pm 400$  &  $11800 \pm 500$~~ & ~~$7.7 \pm 2.7  $ \\ \midrule
Total background       & $6200 \pm 170$  & $5480 \pm 100$  & $13820 \pm 140$~~ &  $20000 \pm 170$~~ & $339 \pm 27   $ \\  \midrule
$tcH$                    & $30 \pm 5    $  & $27 \pm 4    $  & $51 \pm 8    $  &  $34 \pm 6    $  & $36 \pm 5     $ \\
$tuH$                    & $36 \pm 8    $  & $32 \pm 5    $  & ~~$63 \pm 10   $  &  $45 \pm 7    $  & $48 \pm 7     $ \\ \midrule
Data                   & $6353        $  & $5410        $  & $13804       $  &  $20000       $  & $351          $ \\
\bottomrule\bottomrule
\end{tabular}\\

\begin{tabular}{ccc} \toprule\toprule
& $t_{h}\thadhad$-2j & $t_{h}\thadhad$-3j\\\midrule
$t\bar{t}V$              & ~~$0.7 \pm 0.4 $ & ~~$5.5 \pm 1.0 $  \\
Diboson                  & ~~$8.4 \pm 1.6 $ & $10.8 \pm 1.5$  \\
Rare                     & $17.9 \pm 3.1$ & $10.2 \pm 2.6$  \\
SM Higgs                 & $17.4 \pm 2.5$ & $25.9 \pm 3.1$  \\
only $\tau_\text{sub}$ real   & ~~$56 \pm 30   $ & ~~$80 \pm 50   $  \\
$t\bar{t}$               & $221 \pm 28  $ & $220 \pm 40  $  \\
Fake $\tau$              & $220 \pm 70  $ & $270 \pm 70  $  \\
$Z\rightarrow\tau\tau$   & $490 \pm 50  $ & $420 \pm 50  $  \\ \midrule
Total background         & $1040 \pm 35 $~~ & $1040 \pm 40 $~~  \\ \midrule
$tcH$                      & $15.6 \pm 2.5$ & $42 \pm 8    $  \\
$tuH$                      & $23 \pm 4    $ & ~~$52 \pm 10   $  \\ \midrule
Data                     & $1033       $& $1052 $       \\
\bottomrule\bottomrule
\end{tabular}
\label{tab:HtautauPostfitYieldsUnblind}
\end{table}

\begin{table}[h!]
\caption{Absolute uncertainties on $\BR(t\to qH)$ ($q=u,c$) obtained from the combined fit to the data. The uncertainties are symmetrised
and grouped into the categories described in the Section~\ref{sec:systematics}.}
\label{tab:had_sys_impact}
\begin{center}
\begin{tabular}{
@{}l
S
S
@{}
}
\toprule\toprule
\multirow{2}{*}{Source of uncertainty}      & \multicolumn{2}{c}{$\Delta\BR$ [$10^{-5}$]} \\
& \multicolumn{1}{c}{$t\rightarrow uH$} & \multicolumn{1}{c}{$t\rightarrow cH$} \\\midrule
Lepton ID                               & 0.6           &0.8         \\
$\met$                                  & 0.7           &0.7         \\
Fake lepton  modeling                   & 1.2           &1.7         \\
JES and JER                             & 2.5           &3.3         \\
Flavour tagging                         & 2.7           &3.7         \\
$t\bar{t}$ modeling                     & 2.6           &3.9         \\
Other MC modeling                       & 2.1           &3.0         \\
Fake $\tau$ modeling                    & 3.3           &4.7         \\
Signal modeling including Br($H\to\tau\tau$)&1.8        &1.5         \\
$\tau$ ID                               & 3.3           &4.4         \\
Luminosity and Pileup                   & 1.7           &2.4         \\
MC statistics                           & 5.1           &7.1         \\\midrule
Total systematic uncertainty            & 10.1          &14.1        \\
Data statistical uncertainty            & 14.9          &19.4        \\\midrule
Total uncertainties                     & 18            &24          \\
\bottomrule\bottomrule
\end{tabular}
\end{center}
\end{table}

Upper limits on $\BR(t\to cH)$ and $\BR(t\to uH)$ are derived using the CL$_{\textrm{s}}$ method~\cite{Junk:1999kv,Read:2002hq}, and
the observed (expected) 95\% CL limits are
$\BR(t\to cH)<9.4 \times 10^{-4}\,(4.8^{+2.2}_{-1.4}\times10^{-4})$, assuming $\BR(t\to uH)=0$, and $\BR(t\to uH)<6.9 \times 10^{-4}\,(3.5^{+1.5}_{-1.0}\times10^{-4})$, assuming $\BR(t\to cH)=0$.
These results are dominated by the leptonic channels, whose sensitivity is a factor of two better than that of the hadronic channels.
The expected sensitivity is a factor of five better than that of the previous ATLAS search, which was based on 36 fb$^{-1}$ of data and used $H\to \tau\tau$ decays~\cite{TOPQ-2017-07}.
A factor of~2 improvement in sensitivity comes from the larger dataset, and a further factor of~2.5 comes from including
additional leptonic channels, $tH$ production, and improved techniques.
 
In both cases, the results are dominated by the statistical uncertainty.
The main contributions to the total systematic uncertainty arise from  the size of the MC samples,
the uncertainties in the $\thad$ identification efficiency,
the renormalisation and factorisation scales, the $b$-tagging efficiency, the choice of parton shower and hadronisation schemes for $t\bar t$ modelling, and
the fake-$\had$ background estimation in the hadronic channels. Their absolute impacts on the signal strength are summarised in Table~\ref{tab:had_sys_impact}.
A summary of the upper limits, significance and best-fit values of the branching ratios obtained by the individual searches, as well as their combination, is given
in Table~\ref{tab:limits_summary} and in Figures~\ref{fig:limits_combo_1D_hc}(a) and~\ref{fig:limits_combo_1D_hc}(b).
 
Upper limits on the branching ratios $\BR(t\to qH)$ ($q=u,\, c$) can be translated into upper limits on the dimension-6 (D6) operator Wilson coefficients ($C^{i3}_{u\phi}$,
$C^{3i}_{u\phi}$) appearing in the effective field theory Lagrangian for the $tqH$ interaction~\cite{fcnc_production_theory}:
\begin{equation*}
\mathcal{L}_{EFT} = \frac{C^{i3}_{u\phi}}{\Lambda^{2}}(\phi^{\dagger}\phi)(\bar{q_{i}}t)\tilde{\phi} + \frac{C^{3i}_{u\phi}}{\Lambda^{2}}(\phi^{\dagger}\phi)(\bar{t}q_{i})\tilde{\phi}
\end{equation*}
where the subscript $i=1,\, 2$ represents the generation of the light-quark fields ($q=u,\, c$).
The branching ratio $\BR(t\to qH)$ is estimated as the ratio of its partial width to the SM $t \to Wb$ partial width including next-to-leading-order QCD corrections. The coefficients can be extracted as $C_{q\phi} = \sqrt{1946.6~\BR(t\to qH)}$~\cite{fcnc_production_theory}. The $C_{q\phi}$ coefficient corresponds to the sum in quadrature of the coefficients relative to the two possible chirality combinations of the quark fields,
$C_{q\phi} =\sqrt{(C^{i3}_{q\phi})^2 + (C^{3i}_{q\phi})^2}$~\cite{fcnc_production_theory}. The observed (expected) upper limits on the D6 Wilson coefficients from the combination of the search
results are $C_{c\phi}<1.35\,(0.97)$ and $C_{u\phi}<1.16\,(0.82)$ for a new-physics scale $\Lambda=1$~\TeV.

\begin{table}[t!]
\caption{\small{Summary of 95\% CL upper limits on $\BR(t \to cH)$ and $\BR(t \to uH)$, significance and best-fit branching ratio in the signal regions with a
benchmark branching ratio of $\BR(t \to qH)=0.1\%$}. The expected significance is obtained from an Asimov fit~\cite{Cowan:2010js} with a signal injection corresponding to a branching ratio of 0.1\%. }
\begin{center}
\resizebox{\textwidth}{!}{
\begin{tabular}{lcccccc}
\toprule\toprule
 
\multirow{3}{*}{Signal Region} & \multicolumn{3}{c}{$t\to cH$}                                & \multicolumn{3}{c}{$t\to uH$}  \\
&  95\% CL upper limit\,[$10^{-3}$]  & Significance             &  $\BR\,[10^{-3}]$ &     95\% CL upper limit\,[$10^{-3}$]  & Significance   &  $\BR\,[10^{-3}]$  \\
&  \multicolumn{2}{c}{Observed (Expected)}                     &        &     \multicolumn{2}{c}{Observed (Expected)}& \\
\midrule
$t_h\thadhad$-2j                & $1.80$\,($2.72^{+1.18}_{-0.76}$)&$-0.96$\,($0.78$) & $-1.03^{+1.03}_{-1.03}$&$1.07$\,($1.60^{+0.71}_{-0.45}$)&$-0.90$\,($1.31$) &$-0.55^{+0.58}_{-0.58}$ \\
$t_h\thadhad$-3j                & $1.14$\,($1.02^{+0.45}_{-0.29}$)& \,~~$0.34$\,($1.87$)  & \,~~$0.16^{+0.47}_{-0.47}$ &$0.97$\,($0.86^{+0.38}_{-0.24}$)& \,~~$0.36$\,($2.25$)  & \,~~$0.14^{+0.40}_{-0.40}$  \\ \midrule
Hadronic combination            & $1.00$\,($0.95^{+0.42}_{-0.27}$)& \,~~$0.26$\,($1.99$) & \,~~$0.11^{+0.43}_{-0.43}$ &$0.76$\,($0.76^{+0.33}_{-0.21}$)& \,~~$0.12$\,($2.52$)  & \,~~$0.04^{+0.34}_{-0.34}$  \\
\midrule
$t_\ell\tauhad$-2j                 &$4.77$\,($4.23^{+1.72}_{-1.18}$) & \,~~$0.41$\,($0.47$)   & \,~~$0.85^{+2.06}_{-2.06}$  & $3.84$\,($3.48^{+1.42}_{-0.97}$) & \,~~$0.36$\,($0.58$)  & \,~~$0.61^{+1.68}_{-1.68}$\\
$t_\ell\tauhad$-1j                 &$3.80$\,($3.56^{+1.51}_{-0.99}$) & \,~~$0.22$\,($0.58$)   & \,~~$0.36^{+1.70}_{-1.70}$  & $2.98$\,($2.78^{+1.17}_{-0.78}$) & \,~~$0.22$\,($0.73$)  & \,~~$0.29^{+1.33}_{-1.33}$\\
$t_h\tlhad$-2j                  &$4.71$\,($5.71^{+2.68}_{-1.60}$) &$-0.52$\,($0.38$)  &$-1.36^{+2.56}_{-2.56}$ & $2.50$\,($2.97^{+1.25}_{-0.83}$) & $-0.47$\,($0.70$) &$-0.66^{+1.38}_{-1.38}$\\
$t_h\tlhad$-3j                  &$2.71$\,($2.71^{+1.25}_{-0.76}$) &$-0.03$\,($0.77$)  &$-0.03^{+1.26}_{-1.26}$ & $2.02$\,($2.03^{+0.86}_{-0.57}$) & $-0.05$\,($0.99$) &$-0.03^{+0.98}_{-0.98}$\\
$t_\ell\thadhad$                   &$1.35$\,($0.61^{+0.27}_{-0.17}$) & \,~~$2.64$\,($3.31$)   & \,~~$0.74^{+0.33}_{-0.33}$  & $0.97$\,($0.44^{+0.19}_{-0.12}$) & \,~~$2.64$\,($4.38$)  & \,~~$0.53^{+0.24}_{-0.24}$\\ \midrule
Leptonic combination            &$1.25\,$($0.58^{+0.25}_{-0.16}$) & \,~~$2.61$\,($3.46$)   & \,~~$0.69^{+0.31}_{-0.31}$  & $0.88$\,($0.41^{+0.18}_{-0.11}$) & \,~~$2.60$\,($4.62$)  & \,~~$0.49^{+0.22}_{-0.22}$\\
\midrule
Combination                     &$0.94$\,($0.48^{+0.20}_{-0.14}$)& \,~~$2.34$\,($4.02$)   &~~$0.51^{+0.24}_{-0.24}$ & $0.69$\,($0.35^{+0.15}_{-0.10}$)& \,~~$2.31$\,($5.18$)& \,~~$0.37^{+0.18}_{-0.18}$  \\
\bottomrule\bottomrule
\end{tabular}
}
\label{tab:limits_summary}
\end{center}
\end{table}

\begin{figure*}[h!]
\begin{center}
\begin{tabular}{@{}cc@{}}
\includegraphics[width=0.49\textwidth]{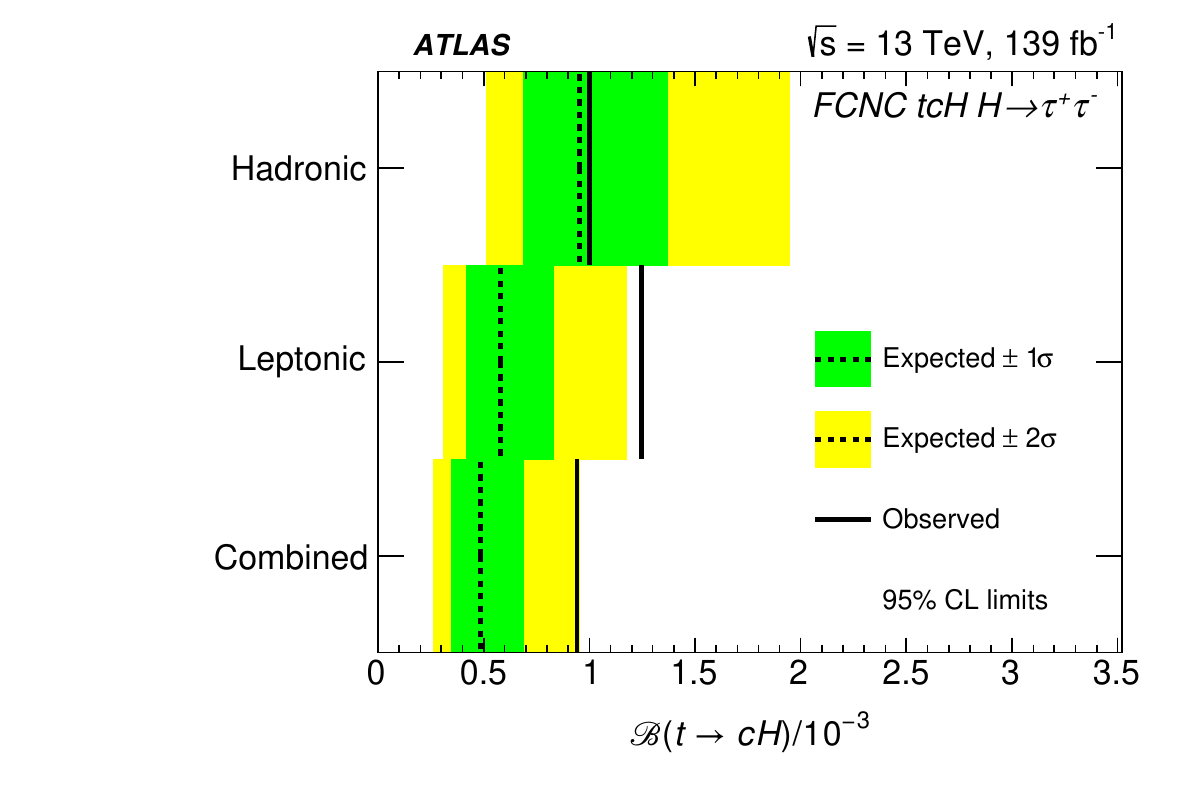}&
\includegraphics[width=0.49\textwidth]{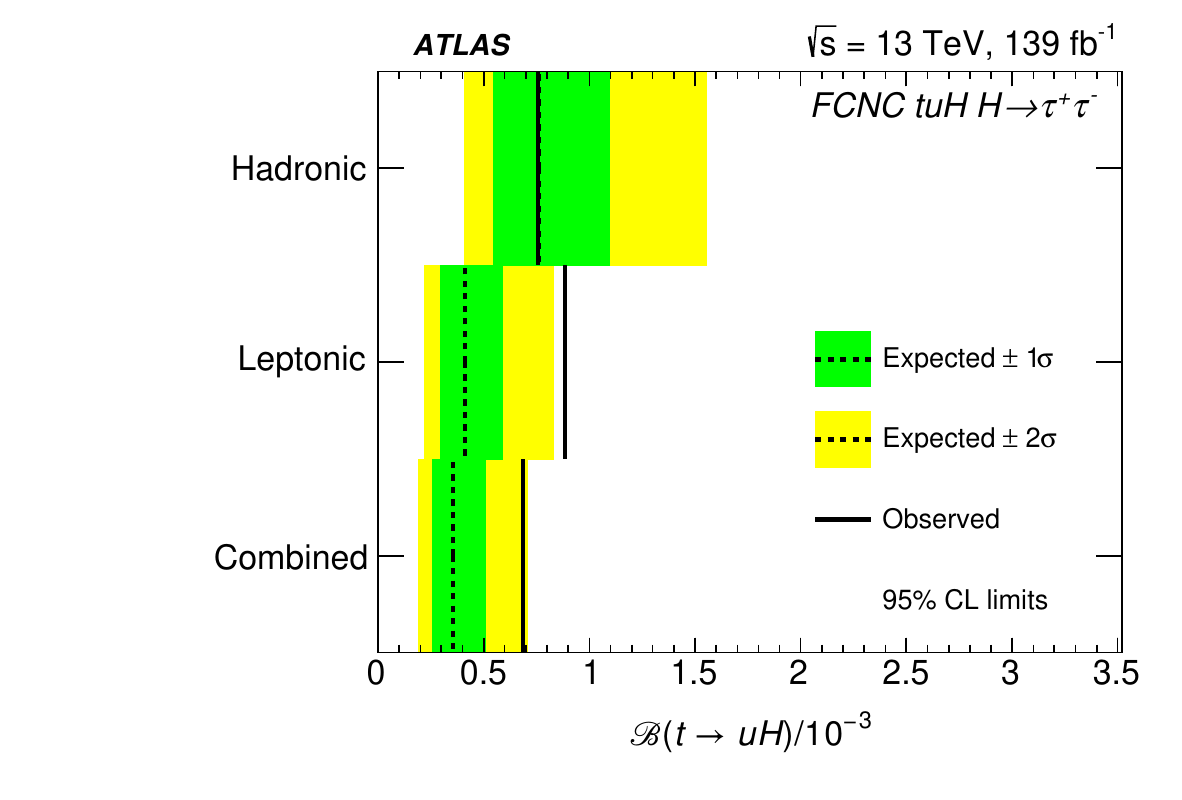}\\
(a) $tcH$ & (b) $tuH$ \\
\end{tabular}
\caption{\small{(a) This figure shows 95\% CL upper limits on $\BR(t\to cH)$ for the individual searches as well as their
combination, assuming $\BR(t\to uH)=0$. (b) This figure shows 95\% CL upper limits on $\BR(t\to uH)$ for the individual searches as well as their
combination, assuming $\BR(t\to cH)=0$. The observed limits (solid lines) are compared with the
expected (median) limits under the background-only hypothesis (dotted lines). The surrounding shaded bands correspond to the 68\% and 95\% CL intervals around the expected limits,
denoted by $\pm 1\sigma$ and $\pm 2\sigma$ respectively.
}}
\label{fig:limits_combo_1D_hc}
\end{center}
\end{figure*}

A similar set of results can be obtained by simultaneously varying both branching ratios in the likelihood function.
Figure~\ref{fig:limits_combo_2D}(a) shows the 95\% CL upper limits on the branching ratios in the $\BR(t\to uH)$ versus $\BR(t\to cH)$ plane.
The corresponding upper limits on the D6 Wilson coefficient couplings in the $C_{u\phi}$ versus $C_{c\phi}$ plane are shown in Figure~\ref{fig:limits_combo_2D}(b).
 
\begin{figure*}[t!]
\begin{center}
\subfloat[]{\includegraphics[width=0.49\textwidth]{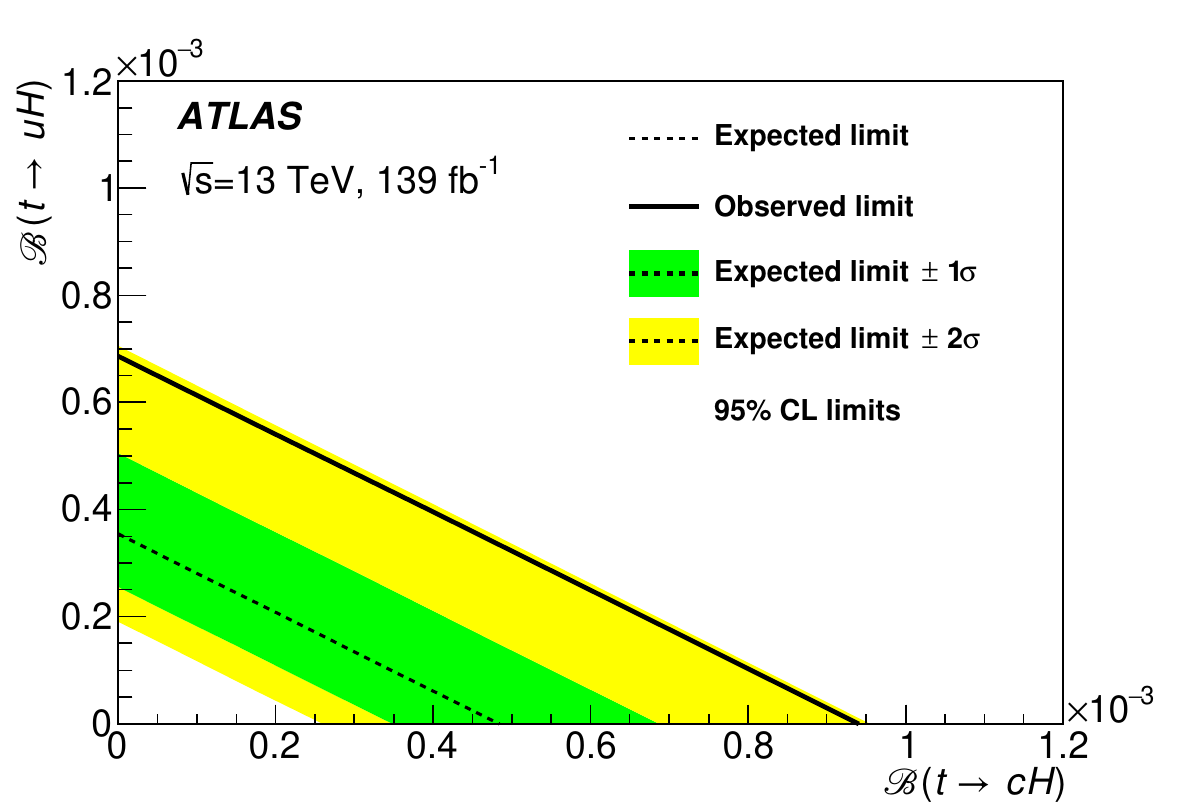}}
\subfloat[]{\includegraphics[width=0.49\textwidth]{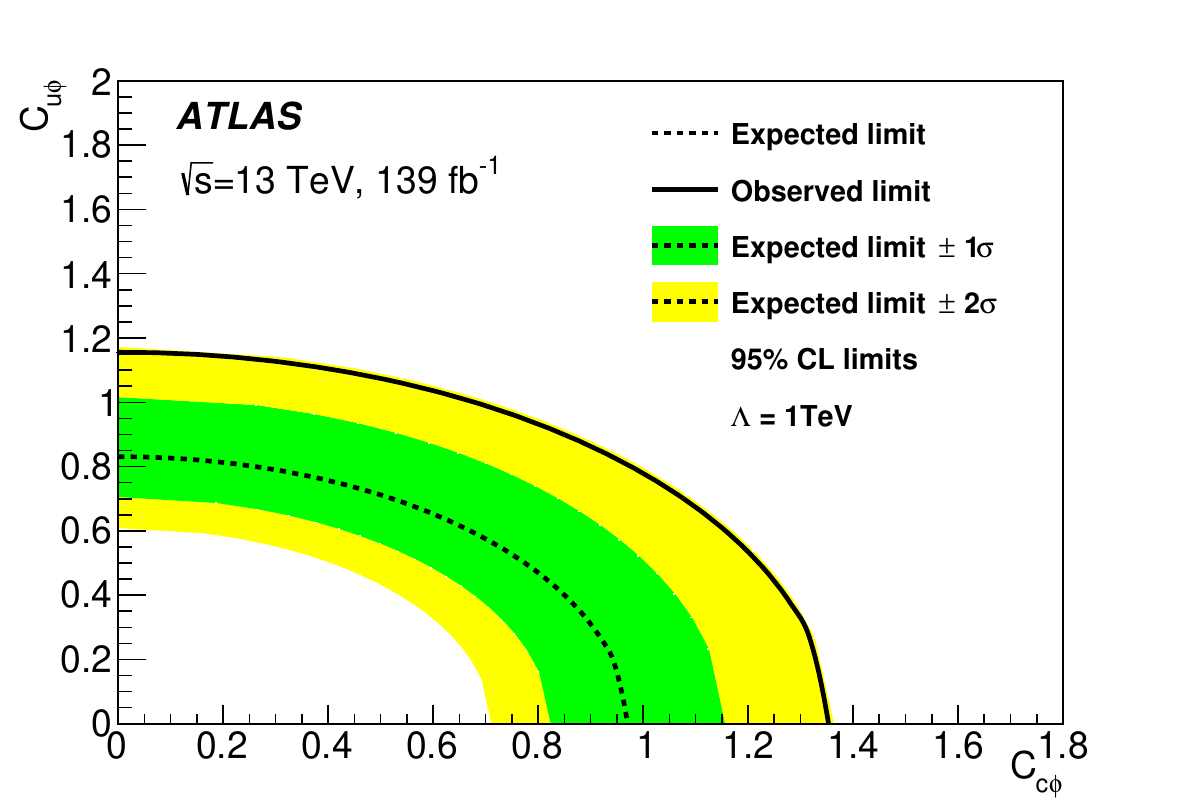}}
\caption{\small{95\% CL upper limits (a) in the $\BR(t\to cH)$ versus $\BR(t\to uH)$ plane and (b) in the
$C_{c\phi}$ versus $C_{u\phi}$ plane for the combination of the searches. The observed limits (solid lines) are compared with the expected (median) limits under the background-only hypothesis (dotted lines). The surrounding shaded bands correspond to the 68\% and 95\% CL intervals around the expected limits,
denoted by $\pm 1\sigma$ and $\pm 2\sigma$ respectively.}}
\label{fig:limits_combo_2D}
\end{center}
\end{figure*}


\FloatBarrier

\section{Conclusion}
A search for flavour-changing neutral current processes involving a top quark, another up-type quark ($q=u, c$), and a SM Higgs boson is presented.
The search uses 139 fb$^{-1}$ of 13~\TeV\ $pp$ collisions recorded by the ATLAS detector at the LHC.
Evidence of FCNC $tqH$ interactions is sought in both the $t\bar{t}$ decay mode, where one top quark decays via SM processes and the other one decays through $t\rightarrow qH$, and the production mode ($pp\rightarrow tH$), where a single top quark produced via the
FCNC interaction decays as $t\to Wb$.
A slight excess of data is observed above background, with a significance of 2.3$\sigma$. Upper limits at the
95\% confidence level are set on the $t\to qH$ branching ratios and the corresponding
dimension-6 operator Wilson coefficients in the effective $tqH$ couplings.
The observed (expected) 95\% CL upper limits set on the $t\to cH$ and $t\to uH$ branching ratios are $9.4\times10^{-4}$ ($4.8^{+2.2}_{-1.4}\times10^{-4}$) and $6.9\times10^{-4}$ ($3.5^{+1.5}_{-1.0}\times10^{-4}$), respectively.
The corresponding combined observed (expected) upper limits on the dimension-6 operator Wilson coefficients in
the effective $tqH$ couplings, for a new-physics scale $\Lambda$ of 1~\TeV, are $C_{c\phi} <1.35\, (0.97)$ and $C_{u\phi} <1.16\, (0.82)$.
These results improve significantly upon the previously published ATLAS studies in this channel and provide more stringent limits than the previous combination of ATLAS  $t\rightarrow qH$ results.


\section*{Acknowledgements}


We thank CERN for the very successful operation of the LHC, as well as the
support staff from our institutions without whom ATLAS could not be
operated efficiently.
 
We acknowledge the support of
ANPCyT, Argentina;
YerPhI, Armenia;
ARC, Australia;
BMWFW and FWF, Austria;
ANAS, Azerbaijan;
CNPq and FAPESP, Brazil;
NSERC, NRC and CFI, Canada;
CERN;
ANID, Chile;
CAS, MOST and NSFC, China;
Minciencias, Colombia;
MEYS CR, Czech Republic;
DNRF and DNSRC, Denmark;
IN2P3-CNRS and CEA-DRF/IRFU, France;
SRNSFG, Georgia;
BMBF, HGF and MPG, Germany;
GSRI, Greece;
RGC and Hong Kong SAR, China;
ISF and Benoziyo Center, Israel;
INFN, Italy;
MEXT and JSPS, Japan;
CNRST, Morocco;
NWO, Netherlands;
RCN, Norway;
MEiN, Poland;
FCT, Portugal;
MNE/IFA, Romania;
MESTD, Serbia;
MSSR, Slovakia;
ARRS and MIZ\v{S}, Slovenia;
DSI/NRF, South Africa;
MICINN, Spain;
SRC and Wallenberg Foundation, Sweden;
SERI, SNSF and Cantons of Bern and Geneva, Switzerland;
MOST, Taiwan;
TENMAK, T\"urkiye;
STFC, United Kingdom;
DOE and NSF, United States of America.
In addition, individual groups and members have received support from
BCKDF, CANARIE, Compute Canada and CRC, Canada;
PRIMUS 21/SCI/017 and UNCE SCI/013, Czech Republic;
COST, ERC, ERDF, Horizon 2020 and Marie Sk{\l}odowska-Curie Actions, European Union;
Investissements d'Avenir Labex, Investissements d'Avenir Idex and ANR, France;
DFG and AvH Foundation, Germany;
Herakleitos, Thales and Aristeia programmes co-financed by EU-ESF and the Greek NSRF, Greece;
BSF-NSF and MINERVA, Israel;
Norwegian Financial Mechanism 2014-2021, Norway;
NCN and NAWA, Poland;
La Caixa Banking Foundation, CERCA Programme Generalitat de Catalunya and PROMETEO and GenT Programmes Generalitat Valenciana, Spain;
G\"{o}ran Gustafssons Stiftelse, Sweden;
The Royal Society and Leverhulme Trust, United Kingdom.
 
The crucial computing support from all WLCG partners is acknowledged gratefully, in particular from CERN, the ATLAS Tier-1 facilities at TRIUMF (Canada), NDGF (Denmark, Norway, Sweden), CC-IN2P3 (France), KIT/GridKA (Germany), INFN-CNAF (Italy), NL-T1 (Netherlands), PIC (Spain), ASGC (Taiwan), RAL (UK) and BNL (USA), the Tier-2 facilities worldwide and large non-WLCG resource providers. Major contributors of computing resources are listed in Ref.~\cite{ATL-SOFT-PUB-2021-003}.


\clearpage
\appendix
\part*{Appendix}
\addcontentsline{toc}{part}{Appendix}

 
\section{Fake-$\thad$ scale factor calibration in the CRtt}
\label{sec:tauFF_appendix}
 
Figure~\ref{fig:ttCR_mtt} presents the post-fit distributions of the leading $\thad$ $\pt$ in the CRtt. These show good agreement between the data and the fitted background model.
Tables~\ref{tab:ff1_summary} and~\ref{tab:ff2_summary} summarise the scale factors for 1-prong and 3-prong fake $\tau$ decays
derived from the CRtt including both the statistical and systematic uncertainties. However, the interpretation of these scale factors is difficult due to their large correlations.
 
\begin{figure}[H]
\centering
\begin{tabular}{@{}ccc@{}}
\includegraphics[page=1,width=0.29\textwidth]{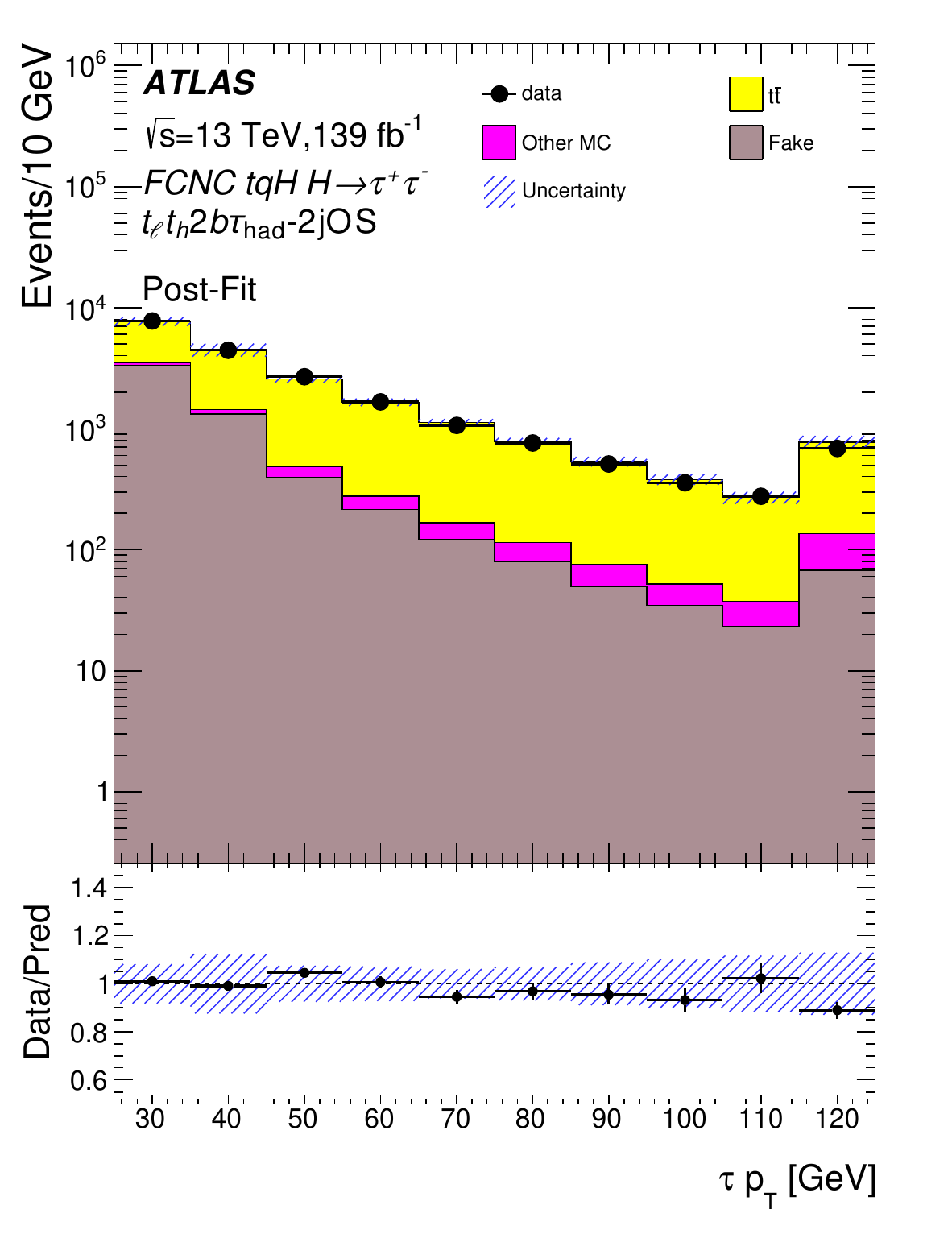} &
\includegraphics[page=1,width=0.29\textwidth]{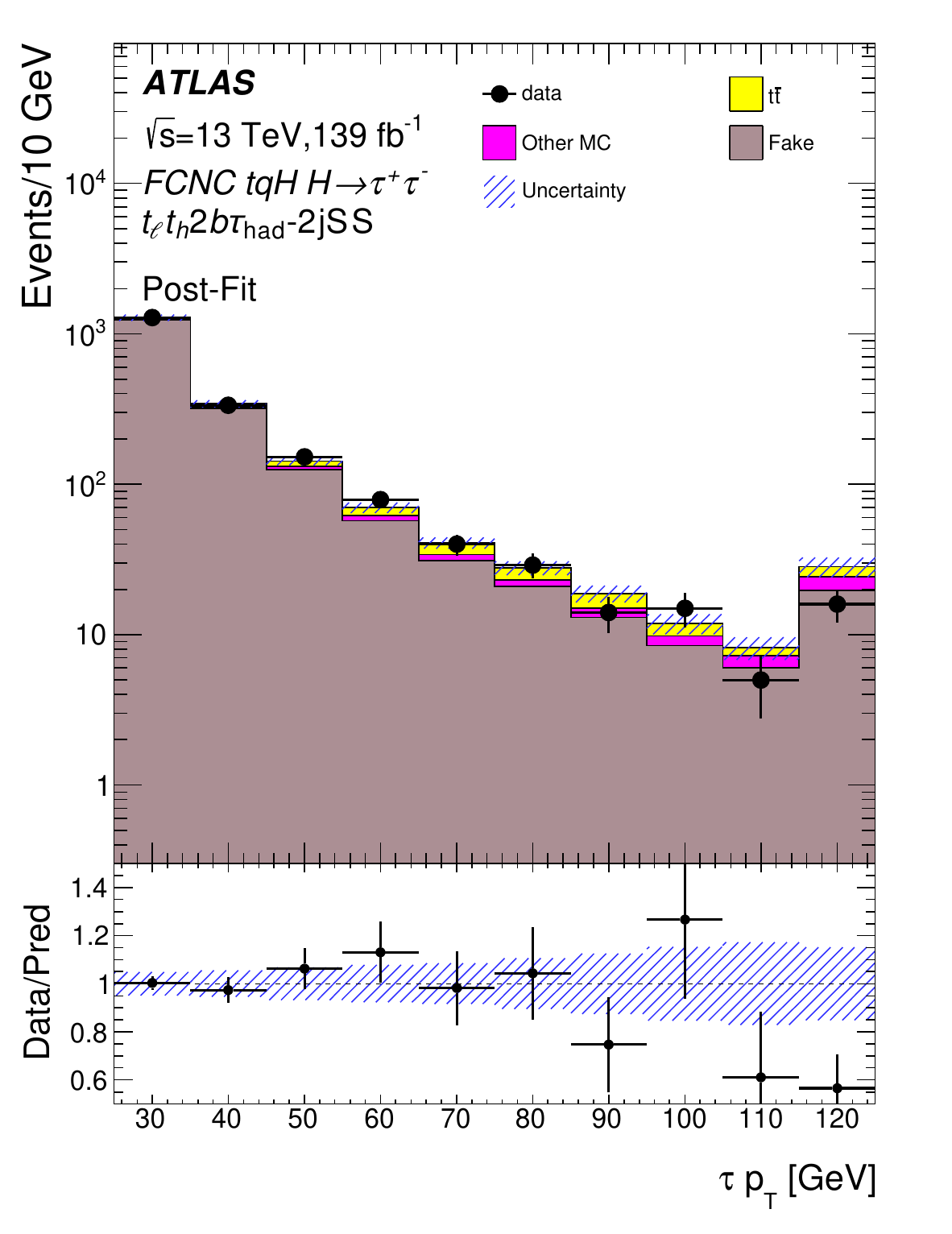}&
\includegraphics[page=1,width=0.29\textwidth]{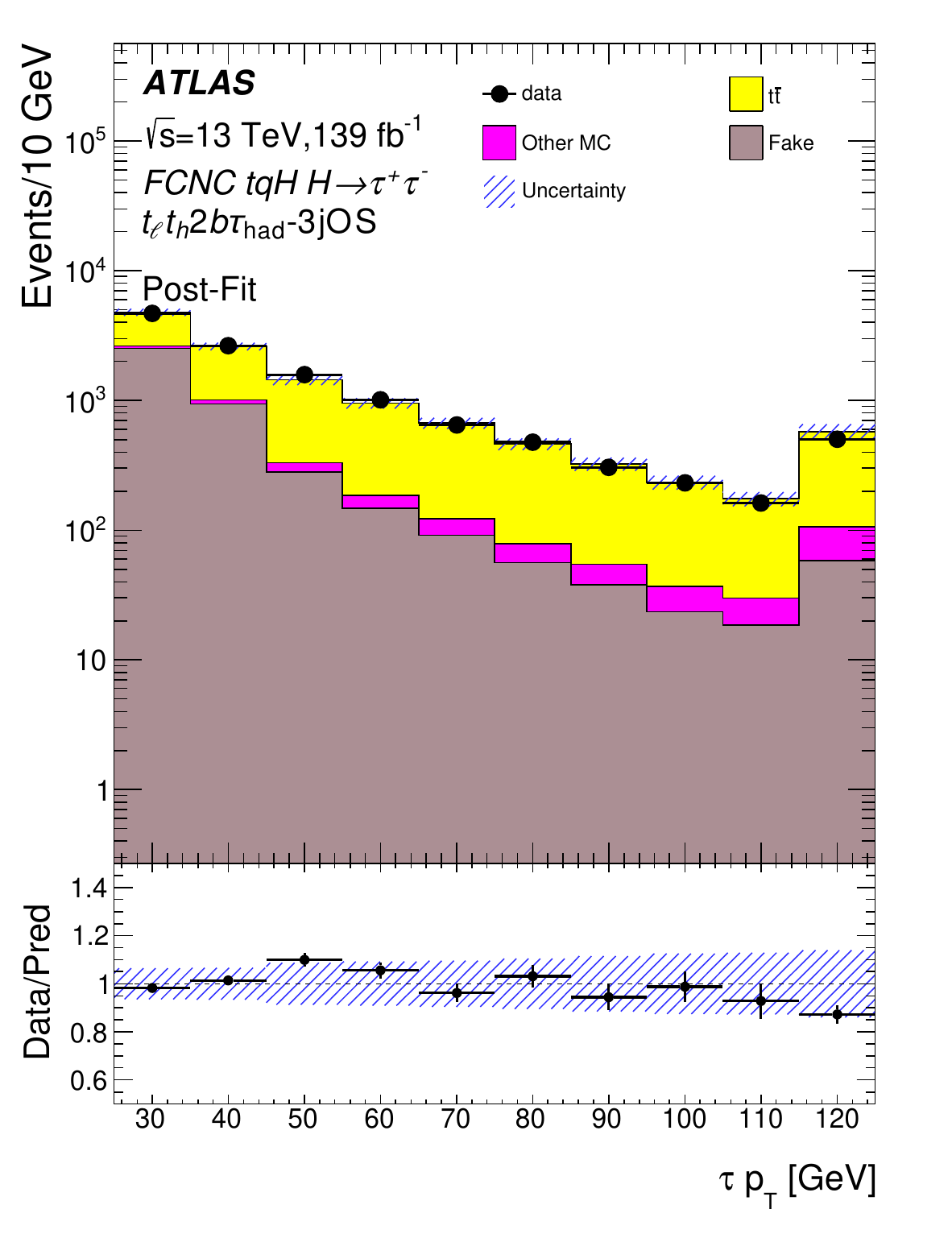}\\
(a) & (b) & (c) \\
\includegraphics[page=1,width=0.29\textwidth]{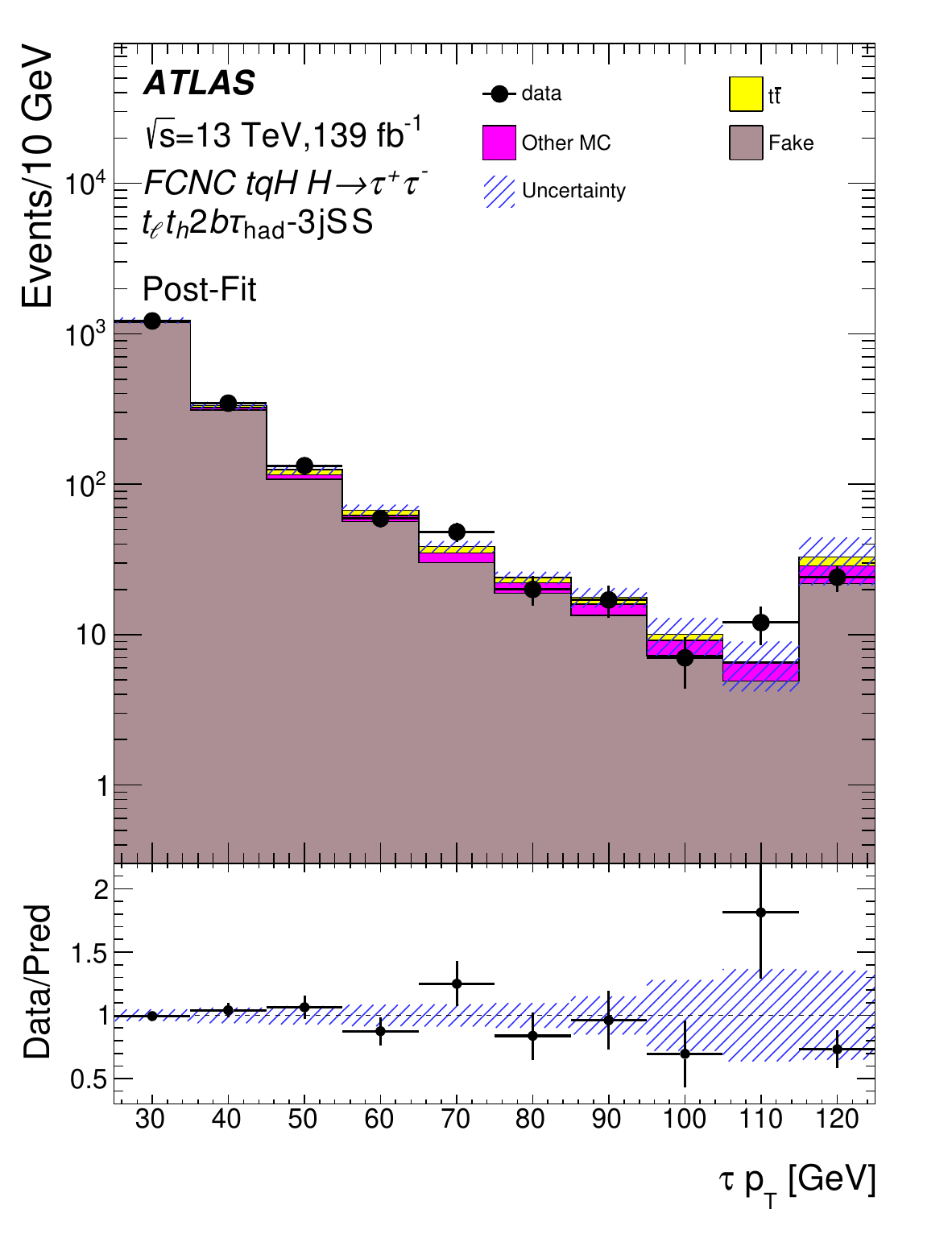}&
\includegraphics[page=1,width=0.29\textwidth]{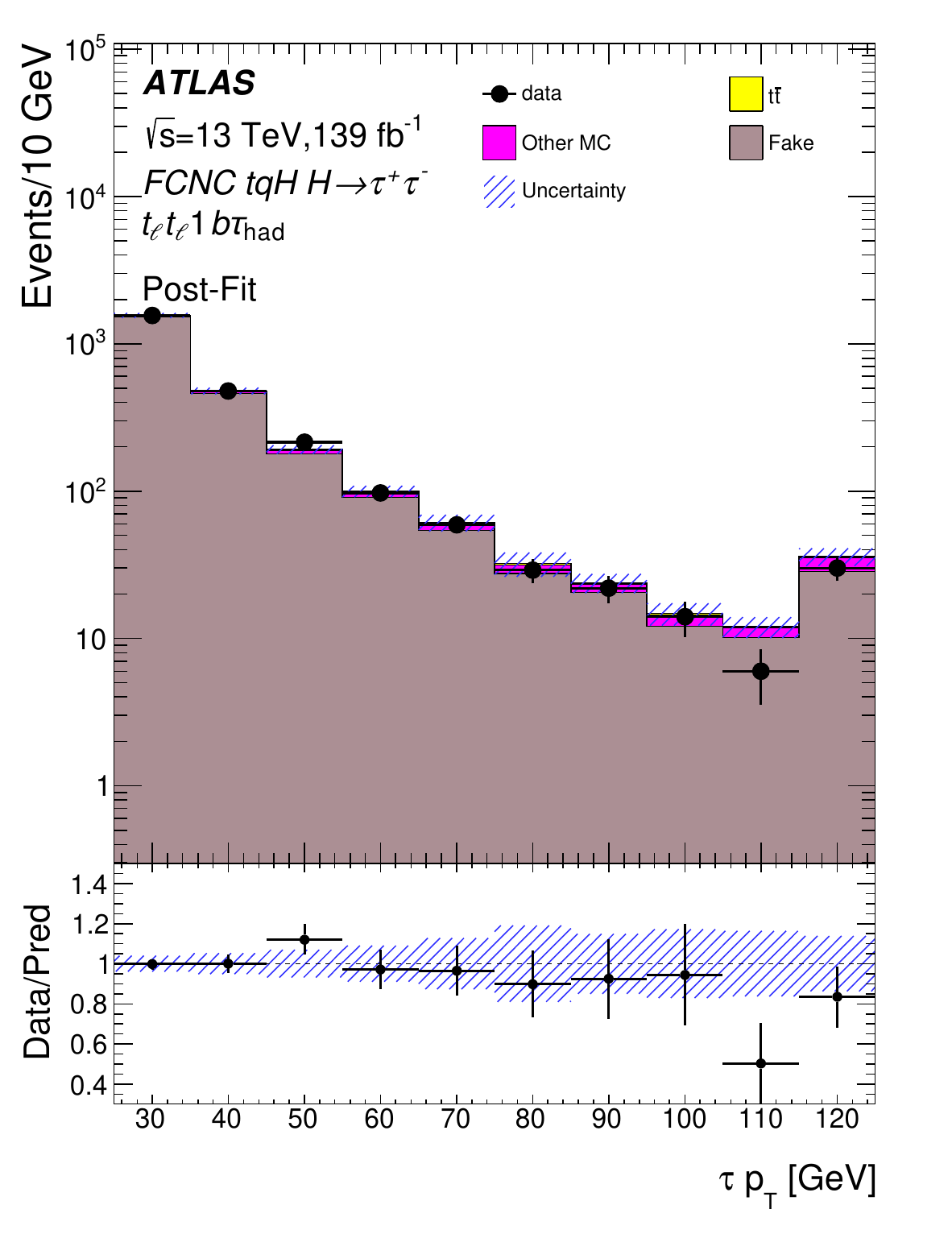}&
\includegraphics[page=1,width=0.29\textwidth]{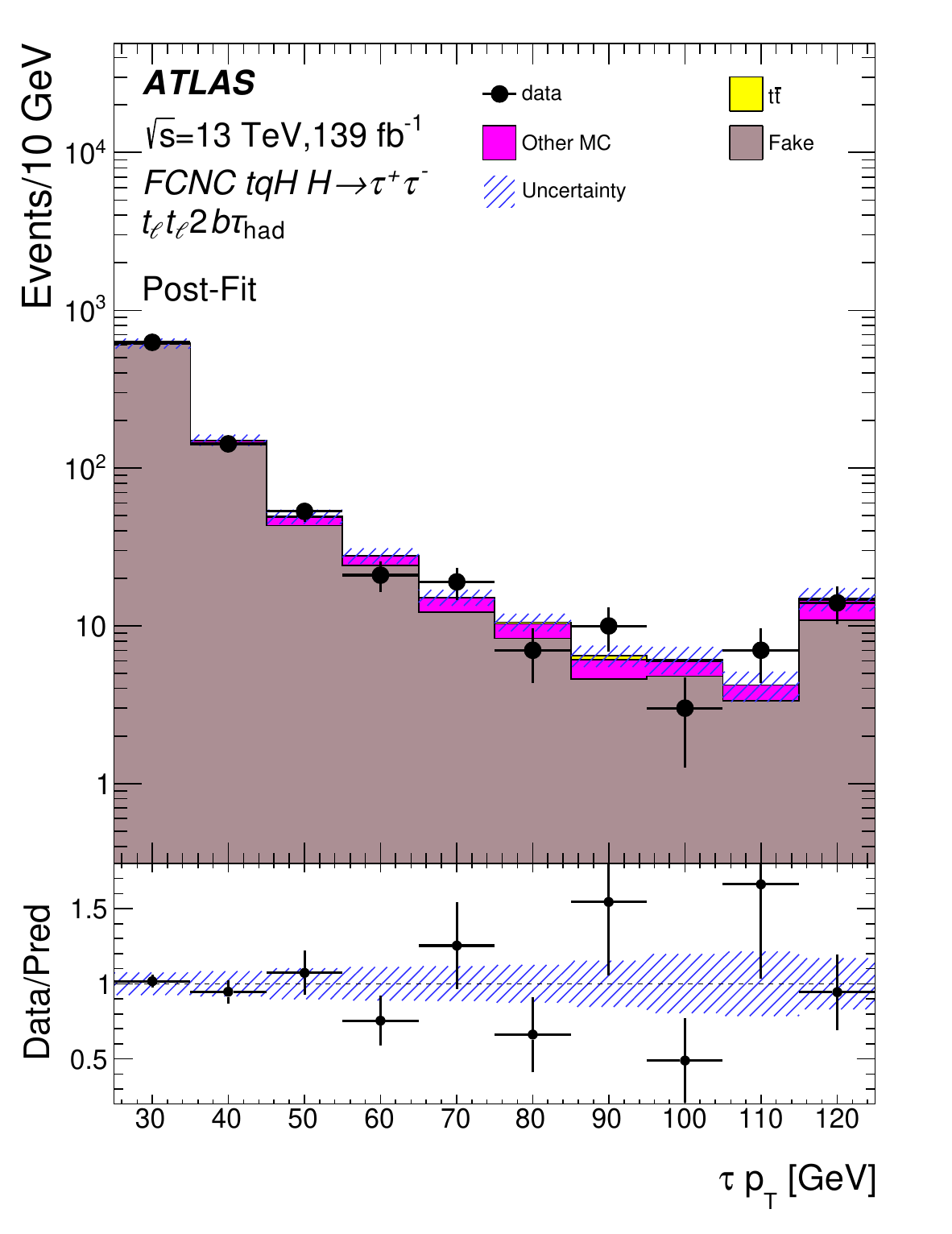}\\
(d) & (e) & (f)\\
\end{tabular}
\caption{Leading $\thad$ $\pt$  distributions obtained after the fit to data (`Post-Fit') for the fake-$\thad$ scale factors in the following CRtt:
(a) $t_{\ell}t_{h}$2b$\thad$-2j OS, (b) $t_{\ell}t_{h}$2b$\thad$-2j SS, (c) $t_{\ell}t_{h}$2b$\thad$-3j OS, (d) $t_{\ell}t_{h}$2b$\thad$-3j SS,
(e) $t_{\ell}t_{\ell}$1b$\thad$-nj and (f) $t_{\ell}t_{\ell}$2b$\thad$-nj.
The total statistical and systematic uncertainty of the background prediction is indicated by the hatched band.
Overflow events are included in the last bin. The lower panels show the ratio of data to prediction.}
\label{fig:ttCR_mtt}
\end{figure}

\begin{table}[H]
\caption{ Summary of fake-$\tau$ (1-prong) scale factors derived in the CRtt. The numbers are shown as central values, statistical uncertainties, and systematics uncertainties. }
\begin{center}
\begin{tabular}{lcccccc}
\toprule\toprule
 
Fake-$\tau$ types & \multicolumn{3}{c}{1-prong $\tau$ decay}  \\ \midrule
$\thad$ $\pt$                                   &  25--35~\GeV  & 35--45~\GeV  &  $>45$~\GeV            \\
\midrule
Type-1                          &$0.71 \pm 0.01 \pm 0.03 $ &$0.61 \pm 0.02 \pm 0.04 $ &$0.38 \pm 0.02 \pm 0.05 $  \\
Type-2                          &$0.76 \pm 0.06 \pm 0.04 $ & $0.37 \pm 0.08 \pm 0.02$ & $0.74 \pm 0.08 \pm 0.02 $ \\
Type-3                          &$0.62 \pm 0.10 \pm 0.03 $ &$0.83 \pm 0.09 \pm 0.03 $ &$0.94 \pm 0.07 \pm 0.02 $  \\
Type-4                          &$1.20 \pm 0.02 \pm 0.01 $ & $1.01 \pm 0.04 \pm 0.02 $ &$0.76 \pm 0.03 \pm 0.03 $  \\
\bottomrule\bottomrule
\end{tabular}
\label{tab:ff1_summary}
\end{center}
\end{table}

\begin{table}[H]
\caption{ Summary of fake-$\tau$ (3-prong) scale factors derived in the CRtt. The numbers are shown as central values, statistical uncertainties, and systematics uncertainties. }
\begin{center}
\begin{tabular}{lcccccc}
\toprule\toprule
 
Fake-$\tau$ types        & \multicolumn{3}{c}{3-prong $\tau$ decays}  \\ \midrule
$\thad$ $\pt$                             &  25--35~\GeV  & 35--45~\GeV       &  $>45$~\GeV  \\
\midrule
Type-1                                    & $1.01 \pm 0.03 \pm 0.04 $ & $1.09 \pm 0.04 \pm 0.05 $ & $0.30 \pm 0.05 \pm 0.07 $ \\
Type-2                                    & $0.93 \pm 0.10 \pm 0.04 $ & $1.05 \pm 0.09 \pm 0.03 $ & $0.79 \pm 0.09 \pm 0.04 $ \\
Type-3                                    & $1.07 \pm 0.13 \pm 0.03 $ &$1.39 \pm 0.12 \pm 0.03 $ &$1.26 \pm 0.10 \pm 0.04 $  \\
Type-4                                    &$1.28 \pm 0.07 \pm 0.02 $ &$0.66 \pm 0.08 \pm 0.01 $ & $0.71 \pm 0.07 \pm 0.02 $ \\
\bottomrule\bottomrule
\end{tabular}
\label{tab:ff2_summary}
\end{center}
\end{table}


\clearpage
 
\printbibliography

\clearpage
 
\begin{flushleft}
\hypersetup{urlcolor=black}
{\Large The ATLAS Collaboration}

\bigskip

\AtlasOrcid[0000-0002-6665-4934]{G.~Aad}$^\textrm{\scriptsize 101}$,
\AtlasOrcid[0000-0002-5888-2734]{B.~Abbott}$^\textrm{\scriptsize 119}$,
\AtlasOrcid[0000-0002-7248-3203]{D.C.~Abbott}$^\textrm{\scriptsize 102}$,
\AtlasOrcid[0000-0002-1002-1652]{K.~Abeling}$^\textrm{\scriptsize 55}$,
\AtlasOrcid[0000-0002-8496-9294]{S.H.~Abidi}$^\textrm{\scriptsize 29}$,
\AtlasOrcid[0000-0002-9987-2292]{A.~Aboulhorma}$^\textrm{\scriptsize 35e}$,
\AtlasOrcid[0000-0001-5329-6640]{H.~Abramowicz}$^\textrm{\scriptsize 150}$,
\AtlasOrcid[0000-0002-1599-2896]{H.~Abreu}$^\textrm{\scriptsize 149}$,
\AtlasOrcid[0000-0003-0403-3697]{Y.~Abulaiti}$^\textrm{\scriptsize 116}$,
\AtlasOrcid[0000-0003-0762-7204]{A.C.~Abusleme~Hoffman}$^\textrm{\scriptsize 136a}$
% ...
\bigskip
\\

% ...
$^{29}$Physics Department, Brookhaven National Laboratory, Upton NY; United States of America.\\
% ...
$^{35}$$^{(a)}$Facult\'e des Sciences Ain Chock, R\'eseau Universitaire de Physique des Hautes Energies - Universit\'e Hassan II, Casablanca;$^{(b)}$Facult\'{e} des Sciences, Universit\'{e} Ibn-Tofail, K\'{e}nitra;$^{(c)}$Facult\'e des Sciences Semlalia, Universit\'e Cadi Ayyad, LPHEA-Marrakech;$^{(d)}$LPMR, Facult\'e des Sciences, Universit\'e Mohamed Premier, Oujda;$^{(e)}$Facult\'e des sciences, Universit\'e Mohammed V, Rabat;$^{(f)}$Institute of Applied Physics, Mohammed VI Polytechnic University, Ben Guerir; Morocco.\\
% ...
$^{55}$II. Physikalisches Institut, Georg-August-Universit\"{a}t G\"ottingen, G\"ottingen; Germany.\\
% ...
$^{101}$CPPM, Aix-Marseille Universit\'e, CNRS/IN2P3, Marseille; France.\\
$^{102}$Department of Physics, University of Massachusetts, Amherst MA; United States of America.\\
% ...
$^{116}$Department of Physics, New York University, New York NY; United States of America.\\
% ...
$^{119}$Homer L. Dodge Department of Physics and Astronomy, University of Oklahoma, Norman OK; United States of America.\\
% ...
$^{136}$$^{(a)}$Departamento de F\'isica, Pontificia Universidad Cat\'olica de Chile, Santiago;$^{(b)}$Millennium Institute for Subatomic physics at high energy frontier (SAPHIR), Santiago;$^{(c)}$Instituto de Investigaci\'on Multidisciplinario en Ciencia y Tecnolog\'ia, y Departamento de F\'isica, Universidad de La Serena;$^{(d)}$Universidad Andres Bello, Department of Physics, Santiago;$^{(e)}$Instituto de Alta Investigaci\'on, Universidad de Tarapac\'a, Arica;$^{(f)}$Departamento de F\'isica, Universidad T\'ecnica Federico Santa Mar\'ia, Valpara\'iso; Chile.\\
% ...
$^{149}$Department of Physics, Technion, Israel Institute of Technology, Haifa; Israel.\\
$^{150}$Raymond and Beverly Sackler School of Physics and Astronomy, Tel Aviv University, Tel Aviv; Israel.\\
% ...
$^{a}$ Also Affiliated with an institute covered by a cooperation agreement with CERN.\\
$^{b}$ Also at An-Najah National University, Nablus; Palestine.\\
$^{c}$ Also at Borough of Manhattan Community College, City University of New York, New York NY; United States of America.\\
$^{d}$ Also at Bruno Kessler Foundation, Trento; Italy.\\
$^{e}$ Also at Center for High Energy Physics, Peking University; China.\\
$^{f}$ Also at Centro Studi e Ricerche Enrico Fermi; Italy.\\
% ...

\end{flushleft}


\end{document}